\documentclass{jfm}
\usepackage{graphicx}
\usepackage{epstopdf, epsfig}
\usepackage{subfig}
\usepackage[final]{changes}

\shorttitle{Non-equilibrium turbulence scalings and self-similarity in  turbulent planar jets} 
\shortauthor{G. Cafiero and J.C. Vassilicos}

\title{Non-equilibrium turbulence scalings and self-similarity in turbulent planar jets}

\author{{G. Cafiero} \corresp{\email{g.cafiero@imperial.ac.uk}} \and
{J.C. Vassilicos} \corresp{\email{j.c.vassilicos@imperial.ac.uk}}}

\affiliation{Turbulence, Mixing and Flow Control Group, Department of Aeronautics, Imperial College London, London SW7 2AZ, United Kingdom}

\begin{document}

\maketitle

\begin{abstract}
We study the self-similarity and dissipation scalings of a turbulent
planar jet and the theoretically implied mean flow scalings. Unlike
turbulent wakes where such studies have already been carried out
\citep{dairay2015,obligado2016}, this is a boundary-free turbulent
shear flow where the local Reynolds number increases with distance
from inlet. The Townsend-George theory revised by \cite{dairay2015}
is applied to turbulent planar jets. Only a few profiles need to
be self-similar in this theory. The self-similarity of mean flow,
turbulence dissipation, turbulent kinetic energy and Reynolds stress
profiles is supported by our experimental results from 18 to at least
54 nozzle sizes, the furthermost location investigated in this
work. Furthermore, the non-equilibrium dissipation scaling found in
turbulent wakes, decaying grid-generated turbulence, various instances
of periodic turbulence and turbulent boundary layers
(\citealp{vassilicos2015}, \citealp{dairay2015},
\citealp{goto&vassilicos2015}, \citealp{Nedic2017}) is also observed
in the present  turbulent planar jet and in the  turbulent planar jet of
\cite{Antonia1980}. Given
these observations, the theory implies new mean flow and jet width
scalings which are found to be consistent with our data and the data
of \cite{Antonia1980}. In particular, it implies a hitherto unknown
entrainment behaviour: the ratio of characteristic cross-stream to
centreline streamwise mean flow velocities decays as the -1/3 power of
streamwise distance in the region where the non-equilibrium
dissipation scaling holds.
\end{abstract}

\begin{keywords}
Authors should not enter keywords on the manuscript, as these must be
chosen by the author during the online submission process and will
then be added during the typesetting process (see
http://journals.cambridge.org/data/\linebreak[3]relatedlink/jfm-\linebreak[3]keywords.pdf
for the full list)
\end{keywords}

\section{Introduction}
Industrial and environmental applications of turbulent free shear
flows usually require knowledge of mean flow profiles. In the case of
turbulent jets one most often needs to know how the mean flow velocity
vector and the jet width evolve with downstream distance. The mean
flow velocity vector has a  cross-stream component which
relates to entrainment. In the aforementioned applications entrainment is of paramount importance, for example in the effectiveness of heating/cooling by
means of impinging jets \citep{Ianiro, cafieroetal2017}.

The modern theory of turbulent free shear flows has been initiated by
\cite{Townsend} and \cite{george1989}. It is based on hypotheses of
self-similar profiles and the equilibrium dissipation scaling whereby
the dissipation coefficient is constant. The dissipation coefficient
$C_{\varepsilon}$ is defined as the ratio of the turbulence
dissipation rate to the rate of non-linear energy losses by the
largest turbulent eddies. This latter rate is proportional to the 3/2
power of the turbulent kinetic energy $K$ divided by a length-scale
which characterizes the size of the largest turbulent
eddies.

Self-similarity is usually justified in terms of loss of memory of
inlet/initial conditions, which is why various previous investigations
have sought to find self-similar profiles quite far downstream
\citep{gutmarkwygnanski1976, kotsovinos1, kotsovinos2, everittrobins1978, deoetal2008,
  deoetal2013}. However, the studies of
axisymmetric turbulent wakes by \cite{Nedic2013}, \cite{dairay2015}
and \cite{obligado2016} found self-similar profiles starting from a
downstream distance as close as ten times the wake generator
size. Most industrial and even many environmental applications of
turbulent wakes and jets are not concerned with the extremely far
downstream flow. This makes the observation of self-similar profiles
at closer distances particularly relevant and these distances amenable
to theory.

Concerning the other hypothesis of the theory of \cite{Townsend} and
\cite{george1989}, the one about the turbulence dissipation scaling,
\cite{dairay2015} and \cite{obligado2016} did not find support for a
constant $C_{\varepsilon}$ in their experiments and numerical
simulations of axisymmetric turbulent wakes even at distances of the
order of 100 wake generator's size. In fact, the turbulent planar jet
investigations by \cite{gutmarkwygnanski1976} and \cite{Antonia1980}
did not find a constant turbulence dissipation coefficient either,
even though their measurements extended up to streamwise distances as
large as 160 nozzle widths. It may not have been fully clear at the
time, but it is becoming increasingly clear now, that deviations from
a constant $C_{\varepsilon}$ can imply deviations from current
textbook scalings of wake/jet widths and centreline mean flow
velocities. This is an important point which the present paper offers
support for in the particular case of the turbulent planar jet.

Evidence of a new non-equilibrium scaling for $C_{\varepsilon}$ in
flow regions where it is not constant has been found in turbulence
generated by various different types of grids and in axisymmetric
wakes \citep{vassilicos2015,dairay2015}, in both forced and
freely decaying periodic turbulence \citep{goto&vassilicos2015,goto&vassilicos2016} and, most recently, in
zero pressure gradient turbulent boundary layers \citep{Nedic2017}.
This non-equilibrium dissipation scaling appears to have some
universality as $C_{\varepsilon}$ is proportional to the ratio of a
global Reynolds number to a local Reynolds number in all these
cases. For example, in the axisymmetric turbulent wake case, the
global Reynolds number $Re_G$ is defined in terms of wake generator
size and incoming freestream velocity, and the local Reynolds number
$Re_{\delta}$ is defined in terms of local wake width $\delta$ and the
square root of the local centreline turbulent kinetic energy
$K_0$. Explanations for the use of the word "non-equilibrium" in this
context can be found in \cite{vassilicos2015} and
\cite{goto&vassilicos2016}.

\cite{dairay2015} modified the theory of \cite{Townsend} and
\cite{george1989} to take into account the non-equilibrium dissipation
scaling and to also make the other assumptions of the theory more
realistic and reduce them in number. They developed the theory for the
case of the axisymmetric turbulent wake and deduced streamwise
evolutions for the mean flow deficit and the wake width which differ
from the well-known textbook scalings
\citep{Townsend,tennekes&lumley} yet fit experimental measurements
well \citep{Nedic2013,dairay2015,obligado2016}.

In the present paper we start by describing the theory of turbulent
planar jets with particular emphasis on the theory's assumptions and
predictions which we then confront with experimental data. To be
assessed, the scaling predictions require data for the centreline mean
flow velocity, the jet width {\it and} the centreline turbulence
dissipation rate. To our knowledge the only previous study with
sufficient and reliable experimental measurements of all these three
quantities in a turbulent planar jet is the one by
\cite{Antonia1980}. We therefore use data from \cite{Antonia1980}
and we also use data from the experimental study of
\cite{deoetal2008} which are also relatively rare in that they report
streamwise profiles of both mean centreline velocity and turbulence
dissipation rate in a turbulent planar jet. However, the data of
\cite{deoetal2008} that we use to study dissipation were obtained for an inlet/global Reynolds number
that is six times smaller than that of \cite{Antonia1980} and this is
reflected in the results of our analysis. We carry out our own
experiment at an inlet Reynolds number that is about three times
larger than that of \cite{deoetal2008} with measurements that are
extensive enough to allow for assessments of various self-similar
profiles and various scalings, including the entrainment coefficient's
streamwise scaling which also turns out to be related to the
turbulence dissipation scaling.

Previous turbulent shear flow experiments where the non-equilibrium
dissipation scaling was observed were carried out in flows where the
local Reynolds number decreases with downstream distance. In turbulent
planar jets, the local Reynolds number $Re_{\delta}$ (defined on the
basis of the local jet width $\delta (x)$ and the square root of the
turbulent kinetic energy) increases with downstream distance $x$ from
the nozzle exit. It is therefore particularly interesting to see
whether the non-equilibrium dissipation scaling $C_{\varepsilon} \sim
(Re_{G}/Re_{\delta})^{m}$ with $m=1$ for high enough Reynods number,
and its consequences on the mean flow, also hold in a turbulent shear
flow with such "reversed" circumstances \citep{lumley1992, castro}. In
the turbulent planar jet flow, $Re_G$ is defined on the basis of the
inlet velocity $U_J$ and the size $h$ of the nozzle exit section (see
figure \ref{fig:planjet}a). As the paper shows, the theory also has
some important implications for the jet entrainment coefficient as
well as for the Reynolds shear stress scaling.

In section \ref{sec:theory} we present the self-similarity theory of
turbulent planar jets with particular attention to the assumptions and
deductions of the theory.
In section \ref{sec:origin} we revisit the experimental turbulent
planar jet data of \cite{deoetal2008} and \cite{Antonia1980}. In
section \ref{sec:expset} we describe our experimental apparatus and
validate our data against previous measurements and in sections
\ref{sec:dissipation}, \ref{sec:selfsimilarity} and \ref{sec:scalings}
we report the results from our experimental tests of the following
section's assumptions and predictions. We conclude in section
\ref{sec:concl}.

\section{Mean field theory of turbulent planar jet flow}  \label{sec:theory}

We apply to the turbulent planar jet flow the Townsend-George theory
of incompressible turbulent free shear flow (see \citealp{Townsend}
and \citealp{george1989}) as revised by \cite{dairay2015}. This theory is based
on the thin shear layer approximation of
the Reynolds-averaged streamwise momentum balance
\begin{equation}
U\frac{\partial U}{\partial x} +
V\frac{\partial U}{\partial y} =
-\frac{\partial}{\partial y}R_{xy}\label{eq:momentum0}
\end{equation}
and on the continuity equation
\begin{equation}
\frac{\partial U}{\partial x} +
\frac{\partial V}{\partial y} = 0
\label{eq:continuity}
\end{equation}
where $U$ and $V$ are the mean flow velocities in the streamwise ($x$)
and cross-stream ($y$) directions respectively (see Figure
\ref{fig:planjet}a) and $R_{xy}$ is the corresponding Reynolds shear
stress (average of the product of streamwise and cross-stream
fluctuating velocities obtained from a Reynolds decomposition
involving the mean flow velocities $U$ and $V$ respectively). These
two equations combined lead to re-writing the streamwise momentum
balance as follows
\begin{equation}
\frac{\partial U^{2}}{\partial x} = -\frac{\partial}{\partial y}(VU + R_{xy}).
\label{eq:mom}
\end{equation}

In all three versions of the Townsend-George theory (\citealp{Townsend}, \citealp{george1989}, \citealp{dairay2015})
one starts by making the assumption that $U(x,y)$ is self-similar, i.e.
\begin{equation}
U(x,y) = u_{0}(x)f_{1}(y/\delta)\label{eq:Uform}
\end{equation}
where $u_{0}(x)$ is the centreline and therefore maximum streamwise
mean flow velocity at streamwise location $x$, and $\delta = \delta
(x)$ is a measure of the jet width which we take to be
\begin{equation}
\delta(x)= \frac{1}{u_{0}(x)}\int_0^\infty \,U(x,y) \,\textup{d}y .   
\label{eq:delta}
\end{equation}
Integrating eq. (\ref{eq:mom}) over $y$ across the jet and using the
self-similar form of $U(x,y)$ (eq. \ref{eq:Uform}) leads to
\begin{equation}
 u_{0}^{2}(x)\delta (x) = U_{J}^{2} h .
 \label{eq:momFlux}
\end{equation}
The constancy of $u_{0}^{2}(x)\delta (x)$
(eq. (\ref{eq:momFlux})) in conjunction with the continuity
(eq. (\ref{eq:continuity})) of the planar mean flow, the self-similar
form of $U(x,y)$ (eq. \ref{eq:Uform}) and $V(x,0)=0$ imply that
$V(x,y)$ is also self-similar, i.e.
\begin{equation}
  V(x,y) = v_{0}(x)f_{2}(y/\delta),
\label{eq:Vform}
\end{equation}
and that $v_{0}$ and $u_0$ are related by 
\begin{equation}
v_{0} = \alpha u_{0} \sim \frac{d\delta}{dx} u_0
\label{eq:v0u0del}
\end{equation}
where $\alpha \sim d\delta/dx$ is the entrainment coefficient (\citealp{pope}).

Use of equation (\ref{eq:mom}), the constancy of $u_{0}^{2}(x)\delta
(x)$ (eq. (\ref{eq:momFlux})), the self-similar forms of both $U$ and
$V$ (eqns. (\ref{eq:Uform}), (\ref{eq:Vform})), and $R_{xy}(x,0)=0$
then imply that $R_{xy}$ is self-similar too, i.e.
\begin{equation}
R_{xy}(x,y) = R_{0}(x)g(y/\delta),\label{eq:Rform}
\end{equation}
and that the $x$-dependence of $R_0$ is given by $R_{0} \sim u_{0}^{2}
d\delta/dx$. 

To close the problem and obtain explicit $x$-dependencies of $u_0$,
$\delta$ and $v_0$, \cite{Townsend} and \cite{george1989} used the
equation for the turbulent kinetic energy $K$,
\begin{equation}
U\frac{\partial K}{\partial x}+V\frac{\partial K}{\partial y} = 
P+T-\varepsilon \label{eq:tke}
\end{equation}
where $P$, $T$ and $\varepsilon$ stand for turbulence production,
transport and dissipation respectively. At this point the approaches
of \cite{Townsend}, \cite{george1989} and \cite{dairay2015} diverge in
the detailed assumptions they make. A summary of the different
assumptions is given in table \ref{tab:theory}. We follow
\cite{dairay2015} and assume self-similarity of $\varepsilon$, $K$
  and $P+T$ and we write the first two terms as
\begin{eqnarray}
K(x,y) &=& K_0(x)h(y/\delta)\label{eq:functionalformsK}\\
\varepsilon(x,y) &=& D_0(x)e(y/\delta).\label{eq:functionalformsEps}
\end{eqnarray}
Use of eq. (\ref{eq:tke}) leads to
\begin{equation}
\frac{K_0\,u_0}{\delta} \frac{d\delta}{dx} \sim D_{0}, 
\label{eq:EnScal}
\end{equation}
a relation which was also obtained by \cite{Townsend} and
\cite{george1989}.
This procedure adds the extra constraint eq. (\ref{eq:EnScal}) and two
further unknowns ($K_0$ and $D_0$) to our already four unknowns $u_0$,
$v_0$, $\delta$ and $R_0$ and three constraints $R_{0} \sim
u_{0}^{2}d\delta/dx$, $u_{0}^{2}\delta = U_{J}^{2}h$ and $v_{0} \sim
u_{0} d\delta/dx$. We therefore have four constraints for six unknowns
and, in general, we cannot proceed without two additional constraints
to close the problem.

The one notable exception, as pointed out by \cite{dairay2015}, is
when the non-equilibrium dissipation scaling can be invoked, namely
$D_{0} = C_{\varepsilon} K_{0}^{3/2}/\delta \sim (Re_{G}/Re_{\delta})
K_{0}^{3/2}/\delta$ where $Re_{\delta} = \sqrt{K_{0}}\delta/\nu$, in
which case eq. (\ref{eq:EnScal}) implies $u_{0} d\delta/dx \sim U_{J}
h/\delta$ without interference from $K_0$. In this case the single
additional hypothesis $D_{0} \sim (Re_{G}/Re_{\delta})
K_{0}^{3/2}/\delta$ suffices to close the problem without further
additional assumptions ($m=1$ case in table \ref{tab:theory}) and one
obtains
\begin{eqnarray}
u_0(x)/U_J &=& A \bigl((x-x_{0})/h\bigr)^{-1/3} \label{eq:scalingNEQ_u_0}\\ 
\delta(x)/h &=& B
\bigl((x-x_{0})/h\bigr)^{2/3}
\label{eq:scalingNEQ_delta}
\end{eqnarray}
from $u_{0} d\delta/dx \sim U_{J} h/\delta$ and $u_{0}^{2} \delta =
U_{J}^{2}h$ in terms of two dimensionless coefficients $A$ and $B$ and
a unique virtual origin $x_0$. It follows that the entrainment
coefficient $\alpha$ is not constant but depends on $x$ as $\alpha \sim
d\delta/dx={2B\over 3}\bigl((x-x_{0})/h\bigr)^{-1/3}$. This is a very
different entrainment behaviour from the classical situation where
$\alpha$ is independent of $x$.

To retrieve both the classical and more general scalings we follow
\cite{dairay2015} and consider the general dissipation scaling
\begin{equation}
D_{0} \sim (Re_{G}/Re_{\delta})^{m} K_{0}^{3/2}/\delta 
\label{eq:neqScal}
\end{equation}
where the special case $m=0$ corresponds to the classical equilibrium
scaling used in the approaches of \cite{Townsend} and
\cite{george1989}. The theory is not conclusive without an additional
assumption when $m \not = 1$ so we adopt Townsend's assumption that
$K_0$ and $R_0$ have the same dependence on $x$, i.e. $K_{0} \sim
R_{0}$ \citep{Townsend}. This makes the theory conclusive and leads to
\begin{eqnarray}
u_0(x)/U_J &=& A \bigl((x-x_{0})/h\bigr)^{-a}\label{eq:scalingEQ1}\\ \delta(x)/h &=& B
\bigl((x-x_{0})/h\bigr)^{2a}\label{eq:scalingEQ2}\\
2a &=& \frac{m+1}{2m+1}\label{eq:avsm}
\label{eq:scalingEQ3}
\end{eqnarray}
which, in the classical equilibrium case $m=0$, leads to $2a = 1$ and
$\alpha$ independent of $x$ as predicted by \cite{Townsend} and
\cite{george1989} and as reported in textbooks
(e.g. \citealp{tennekes&lumley}, \citealp{davidson2004}).
\begin{table}
  \begin{center}
\def~{\hphantom{0}}
  \begin{tabular}{lccccc}
                    		 & \cite{Townsend}  	                     &	&     \cite{george1989} 	&& \cite{dairay2015} \\[3pt]
      Self-similarity  	 & $U, K, T, \varepsilon, u', v'$   		 &   &      $U, K, T, \varepsilon$	  & 			& $U, K, P+T, \varepsilon$ 	      \\
      Dissipation Scaling    &  $K_0^{3/2}/\delta$  	& &      $K_0^{3/2}/\delta$	    && $(Re_G/Re_\delta )^{m} K_{0}^{3/2}/\delta$	   	  \\
      Simplified production & no & & $P \approx -R_{xy} dU/dy$ & & no \\
      $R_{0}$        			 &  $\sim K_0$ 			&  &    no	    && no ($m=1$), $\sim K_{0}$ ($m\not = 1$)	     \\
  \end{tabular}
  \caption{Summary of the assumptions made in \cite{Townsend},
    \cite{george1989} and the present theory which has been adapted
    for planar jets from \cite{dairay2015}. The first row lists the
    quantities assumed to be self-similar. The second row gives the
    scaling assumed for the centreline turbulence dissipation
    rate. The third row states whether an approximation is or is not
    made for the production term. And the fourth row states if an
    assumption is or is not made concerning the centreline Reynolds
    shear stress. Notation: $u'$ is the rms streamwise turbulent
    velocity and $v'$ is the rms cross-stream (direction $y$)
    turbulent velocity.}
  \label{tab:theory}
  \end{center}
\end{table}

The scalings obtained for three different values of $m$ are summarized
in Table \ref{tab:scal}. The classical equilibrium scalings
(\cite{Townsend}, \cite{george1989}) correspond to $m=0$; the high
Reynolds number non-equilibrium scalings correspond to $m=1$. It is
worth pointing out that the entrainment coefficient $\alpha$ obeys
\begin{equation}
\alpha \sim 2aB \bigl((x-x_{0})/h\bigr)^{2a-1} 
\label{eq:entrain}
\end{equation}
and that it is constant only in the classical equilibrium case where
$m=0$. We stress that the virtual origin $x_0$ is the same in
equations (\ref{eq:scalingEQ1}), (\ref{eq:scalingEQ2}) and
(\ref{eq:entrain}). 

\begin{table}
  \begin{center}
\def~{\hphantom{0}}
  \begin{tabular}{lcc}
                    		     & \cite{Townsend}-\cite{george1989}  	               & Present ($m=1$) \\
      $u_0(x)$  	 			 & $\sim (x-x_0)^{-1/2}$   	 & $\sim (x-x_0)^{-1/3}$ \\
      $\delta(x)$   			 & $\sim (x-x_0)$  		     & $\sim (x-x_0)^{2/3}$ \\
      $\alpha \equiv v_0(x)/u_0(x)$ 		     & $\sim const$  	    & $\sim (x-x_0)^{-1/3}$ \\
  \end{tabular}
  \caption{Summary of the planar jet scalings obtained by
    \cite{Townsend} and \cite{george1989} and by the present version
    of the theory for $m=1$ (see eq. \ref{eq:neqScal})). The present
    version of the theory leads to the same scalings as
    \cite{Townsend} and \cite{george1989} if $m=0$.}
  \label{tab:scal}
  \end{center}
\end{table}    
In the next section we revisit the experimental turbulent planar jet
data of \citep{deoetal2008} and \citep{Antonia1980} by paying
particular attention to the fact that the virtual origin $x_0$ must be
the same in all power-law dependencies on streamwise distance.

\section{Centreline data from previous experiments} \label{sec:origin}

The turbulence dissipation scaling (eq. \ref{eq:neqScal}) is a pillar
of the mean flow scaling eqns. (\ref{eq:scalingEQ1}),
(\ref{eq:scalingEQ2}), (\ref{eq:scalingEQ3}) and
(\ref{eq:entrain}). From eqns. (\ref{eq:neqScal}),
(\ref{eq:scalingEQ1}), (\ref{eq:scalingEQ2}) and
(\ref{eq:scalingEQ3}),
\begin{equation}
D_{0} \sim (x-x_{0})^{-\gamma}
\label{eq:DissExp1}
\end{equation}
where the virtual origin $x_0$ must be  the same as the one in eqns.
(\ref{eq:scalingEQ1}) and (\ref{eq:scalingEQ2}) and 
\begin{equation}
\gamma = \bigl(2a - \frac{1}{2}\bigr)m +\bigl(\frac{3}{2}+2a\bigr). 
\label{eq:DissExp}
\end{equation}

Direct numerical simulations (DNS) of turbulent planar jets do not
reach sufficiently high Reynolds numbers and very few laboratory
studies report centreline turbulent dissipation profiles alongside
centreline profiles of $u_0$ and/or $\delta$ for turbulent planar
jets. The main exceptions seem to be the experimental data of
\cite{deoetal2008} who reported streamwise profiles of $D_0$ and
$u_0$ (as well as some values of $\delta$ but at very few points, not
enough for verifying eq. (\ref{eq:scalingEQ2})) and the experimental data
of \cite{Antonia1980} who reported streamwise profiles of $D_0$,
$u_0$ and $\delta$ at an inlet/global Reynolds number $Re_{G}=42800$
which is about 6 times larger that the value of $Re_{G}$ in
\cite{deoetal2008}. We now analyse these data by first identifying
the single virtual origin which returns best fits to the streamwise
scalings of the available quantities.

\begin{figure} 
\centering
{\includegraphics[width=0.7\columnwidth]{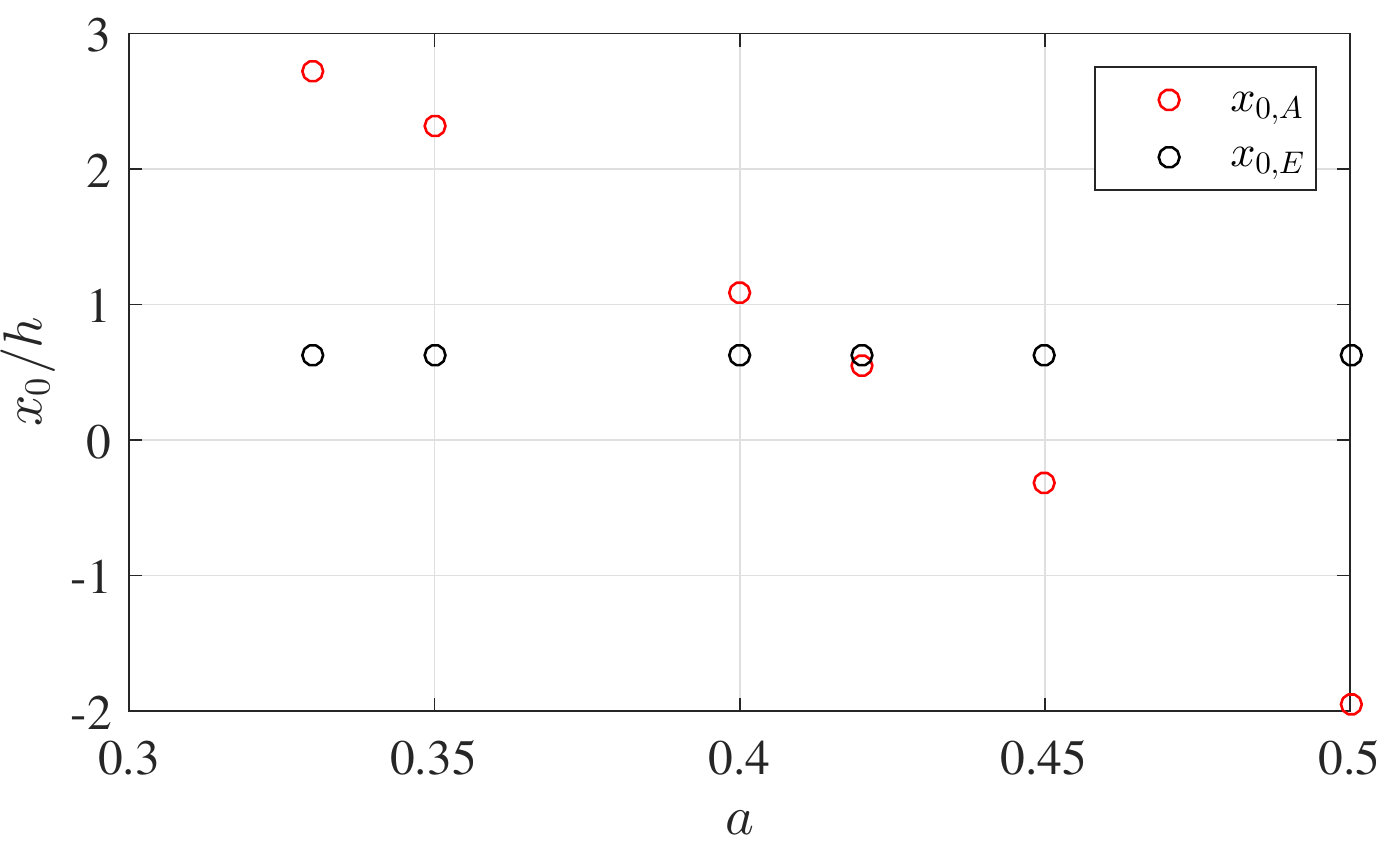}}
\caption{Virtual origins $x_{0,A}$ and $x_{0,E}$ obtained from the
  data of \cite{deoetal2008} ($Re_{G}=7000$) by applying different
  exponents to the power laws eqns. (\ref{eq:scalingEQ1}) and
  (\ref{eq:DissExp1}) respectively, with $0.33\le a \le 0.5$ (i.e. $0
  \le m \le 1$ from eq. (\ref{eq:scalingEQ3})). The optimal exponent
  $a=0.42$, corresponding to $m=0.2353$, is obtained for $x_{0} =
  x_{0,A} = x_{0,E}$.}
\label{fig:DeoVO}
\end{figure}
\begin{figure} 
\centering 
\subfloat[][] 
{\includegraphics[width = 0.5\columnwidth]{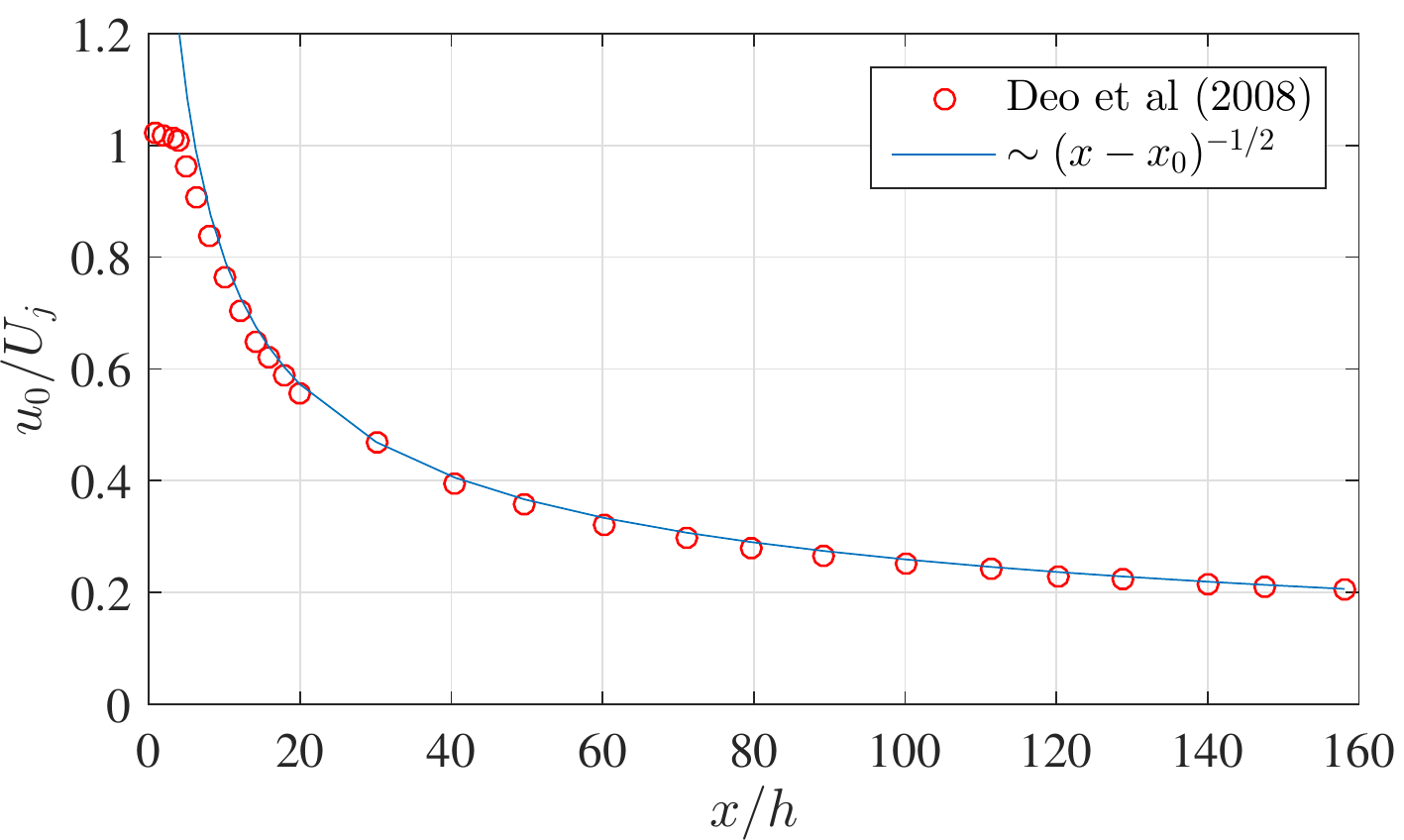}} 
\subfloat[][]
{\includegraphics[width =0.5\columnwidth]{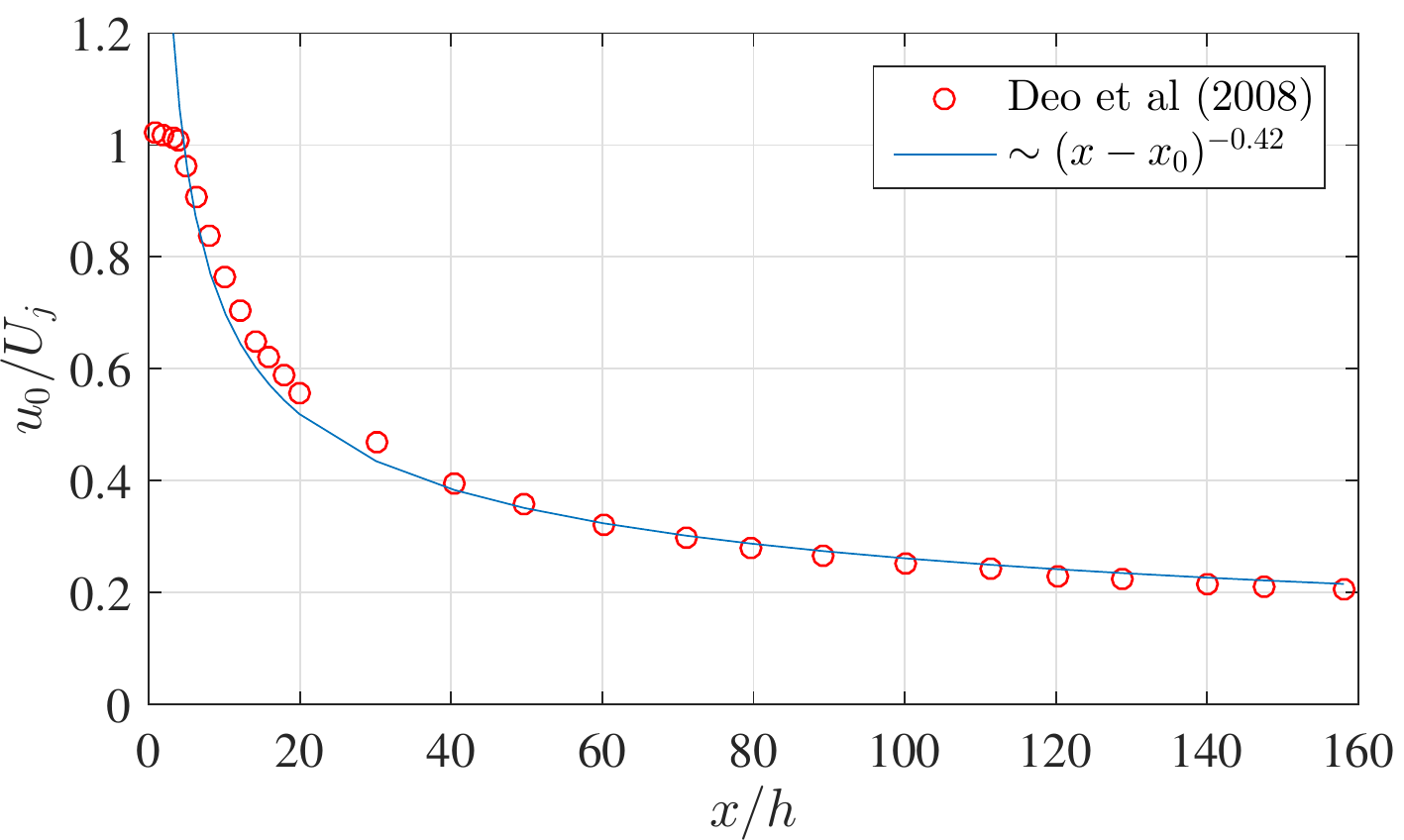}}\\
\caption{Normalised centreline streamwise mean velocity $u_0/U_J$
  versus $x/h$.  (a) Compared to eq. (\ref{eq:scalingEQ1}) with $a=0.5$
  ($m=0$) and $x_{0} = {x_{0,A} + x_{0,E}\over 2} \approx -0.65h$
  (which differs from $x_{0,A} \approx 0.65h$ and $x_{0,E} \approx
  -1.95h$); (b) Compared to eq. (\ref{eq:scalingEQ1}) with $a=0.42$
  ($m=0.2353$) and $x_{0} \approx 0.5h$ which is about the same as
  $x_{0,A}$ and $x_{0,E}$.}
\label{fig:Deou0}
\end{figure}
\begin{figure} 
\centering
\subfloat[][]
{\includegraphics[width = 0.5\columnwidth]{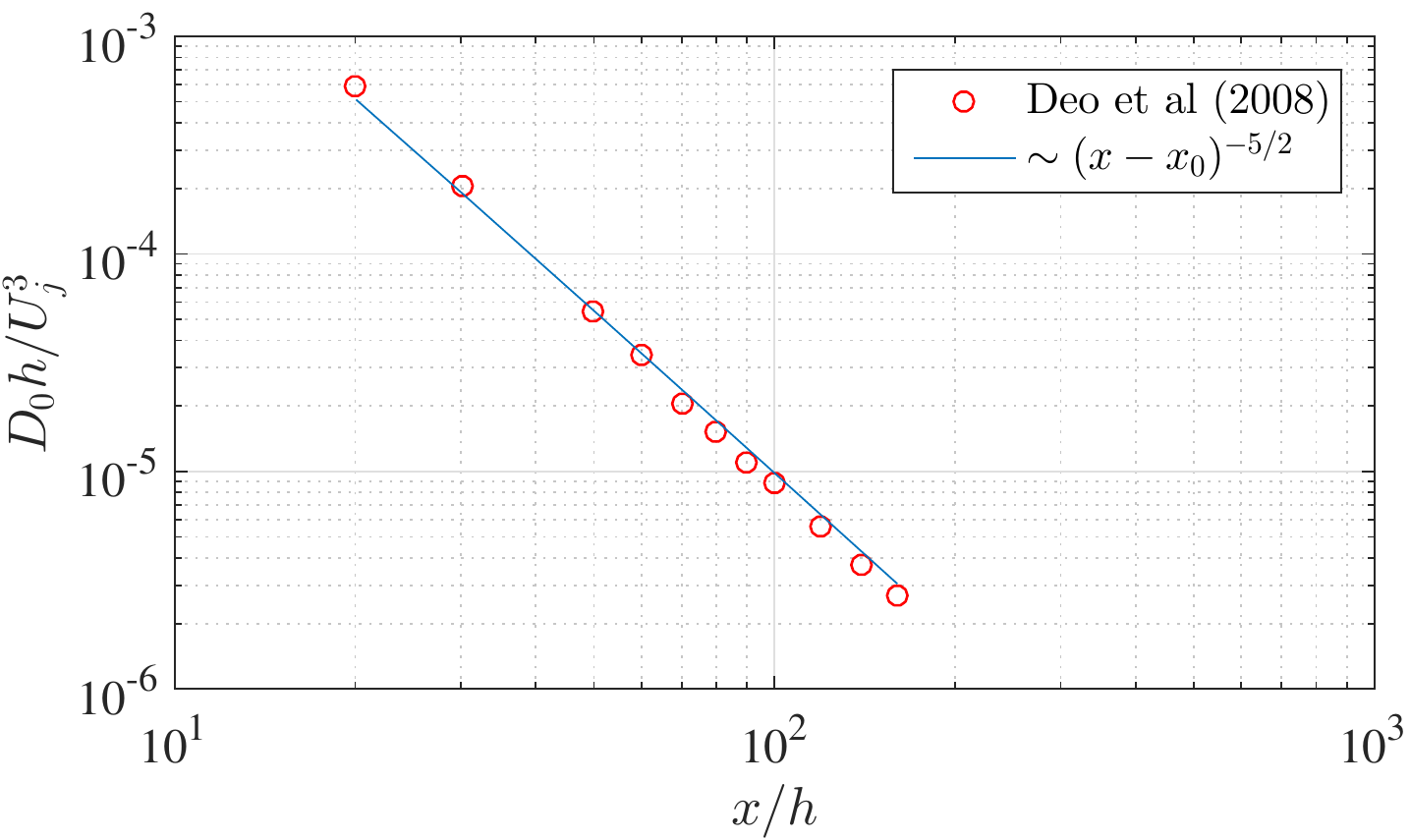}}
\subfloat[][]
{\includegraphics[width = 0.5\columnwidth]{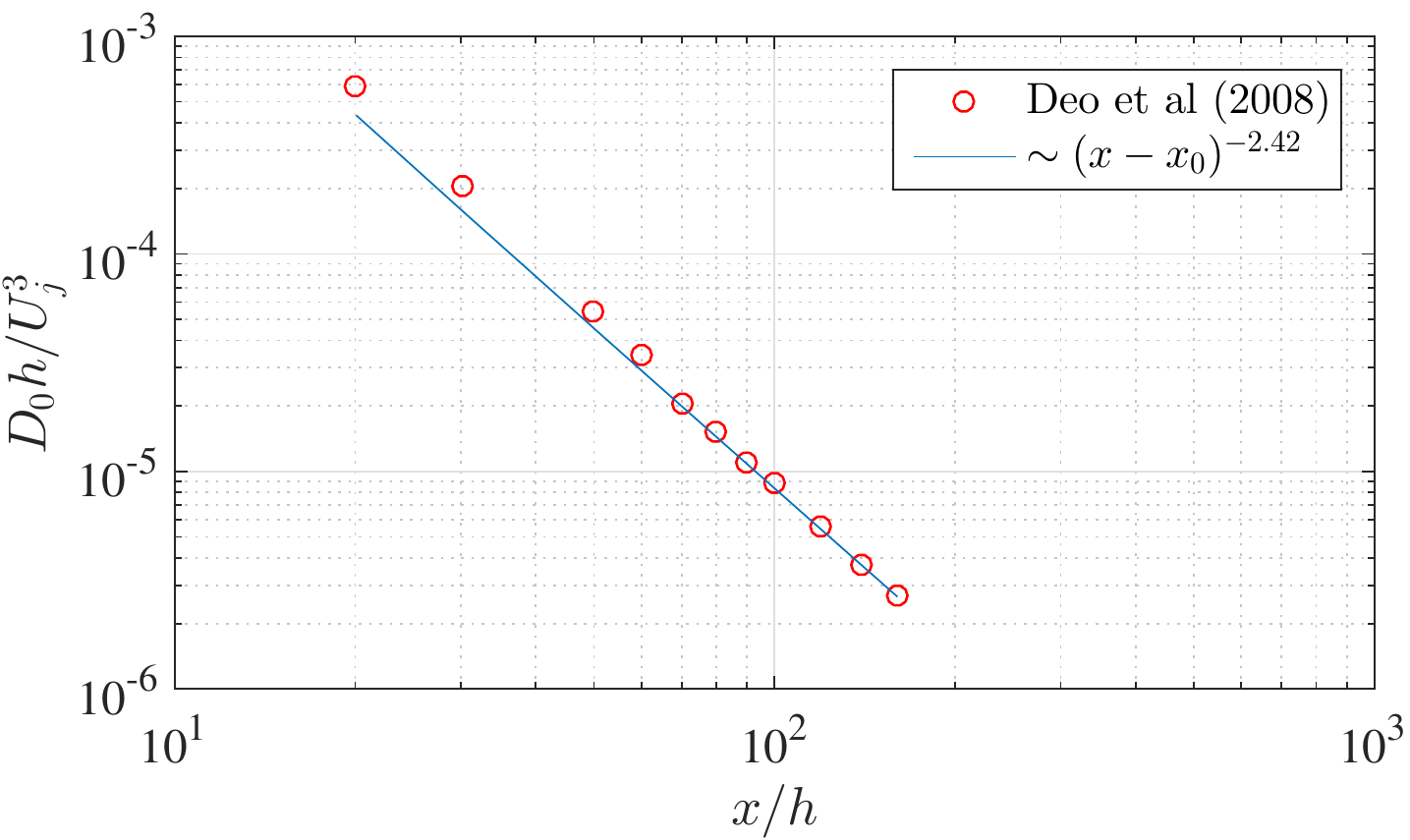}}\\
\caption{Centreline turbulence dissipation rate $D_0$ normalized with
  the inlet speed $U_J$ and the nozzle width $h$ versus $x/h$.  (a)
  Compared to eq. (\ref{eq:DissExp1}) with $\gamma =5/2$ ($m=0$) and
  $x_{0} = {x_{0,A} + x_{0,E}\over 2} \approx -0.65h$ which differs
  from $x_{0,A} \approx 0.65h$ and $x_{0,E} \approx -1.95h$; (b)
  Compared to eq. (\ref{eq:DissExp1}) with $\gamma =2.42$ ($m=0.2353$) and
  $x_{0} \approx 0.5h$ which is about the same as $x_{0,A}$ and
  $x_{0,E}$.}
\label{fig:DeoDiss}
\end{figure}

\subsection{\cite{deoetal2008}} \label{sec:deo}

A fundamental condition to be respected for the validity of the
turbulent planar jet theory in section \ref{sec:theory} is that the
virtual origin used in equations
(\ref{eq:scalingEQ1})-(\ref{eq:scalingEQ2})-(\ref{eq:DissExp1}) must
be unique. In the case of \cite{deoetal2008}, we can only test the
streamwise distance dependencies of eqns. (\ref{eq:scalingEQ1}) and
(\ref{eq:DissExp1}), and this up to $x/h=160$ which is the location of
their furthermost measurements. For values of $m$ ranging between
$m=0$ and $m=1$ we set the corresponding exponents $a$ given by eq.
(\ref{eq:scalingEQ3}) and find the virtual origin $x_{0,A}$ which
returns the best fit of the data to eq. (\ref{eq:scalingEQ1}) in the range
$10\le x/h \le 160$ (reasonable different choices of the lower bound
of this range do not modify the results appreciably). In this way we
obtain a value of $x_{0,A}$ for each $a$ which we plot in figure
\ref{fig:DeoVO}. We apply the same procedure to the dissipation data
provided by \citealp{deoetal2008} and obtain different virtual origins
$x_{0,E}$ which return a best fit to eq. (\ref{eq:DissExp1}) for different
values of $\gamma$ corresponding to values of $m$ between $m=0$ and
$m=1$. In figure \ref{fig:DeoVO} we plot $x_{0,E}$ versus $a$, given
that $\gamma$ is a function of $a$ and $m$ via eq. (\ref{eq:DissExp}) and
that $a$ and $m$ are related by eq. (\ref{eq:scalingEQ3}). The virtual
origin $x_{0}$ must be such that $x_{0} = x_{0,A} = x_{0,E}$ and the
only exponent $a$ where this happens is $a=0.42$ which corresponds to
$m=0.2353$ (see eq. (\ref{eq:scalingEQ3})). These values of $a$ and
$m$ are different from the classical ones, $a=0.5$ and $m=0$, but they
are also different from the non-equilibrium exponents $a=1/3$ and
$m=1$.

In figure \ref{fig:Deou0}(a) we plot $u_{0}/U_{J}$ versus $x/h$ with
the classical fit $(x-x_{0})^{-0.5}$ and in figure
\ref{fig:DeoDiss}(a) we plot $D_{0} h/U_{J}^{3}$ versus $x/h$ with the
classical fit $(x-x_{0})^{-5/2}$ ($\gamma = 5/2$ for $m=0$). We have
chosen the same $x_0$ for these two fits, half way between $x_{0,A}
\approx 0.65h$ and $x_{0,E} \approx -1.95h$ which are the values for
$a=0.5$ in figure \ref{fig:DeoVO}. We compare these fits with those in
figures \ref{fig:Deou0}(b) and \ref{fig:DeoDiss}(b) where we plot the
same data but fitted, respectively, with $(x-x_{0})^{-0.42}$ and
$(x-x_{0})^{-2.42}$ where $x_{0} \approx x_{0,A} \approx x_{0,E}
\approx 0.5h$, as this is the case where a single virtual origin does
exist.

It may be argued that the fits in figures \ref{fig:Deou0} and
\ref{fig:DeoDiss} are slightly better for the classical exponents
$a=0.5$ and $\gamma = 2.5$, but the fits with $a=0.42$ and $\gamma =
2.42$ are not bad either and they are obtained with a consistently
optimal virtual origin whereas those for $a=0.5$ and $\gamma = 2.5$
are not. If the only theoretical option was $a=0.5$ and $\gamma = 2.5$
one might have been able to conclude that the data of
\cite{deoetal2008} fit this option well and perhaps overlook
the appreciable divergence between $x_{0,A}$ and $x_{0,E}$. However,
now that there are more options available, it becomes more difficult
to overlook this difference and conclude.

As already mentioned, the turbulence dissipation scaling eq.
(\ref{eq:neqScal}) is a key pillar underpinning eqns.
(\ref{eq:scalingEQ1})-(\ref{eq:scalingEQ2})-(\ref{eq:DissExp1}).
However, the data which would be necessary to directly test the
validity of eq. (\ref{eq:neqScal}) are not in
\cite{deoetal2008}. Furthermore, \cite{dairay2015} show that eq.
(\ref{eq:neqScal}) appears with $m\approx 1$ in axisymmetric turbulent
wakes only when the Reynolds number is large enough.
\cite{obligado2016} also demonstrated that it is much more difficult
to distinguish between different values of $m$ by fitting streamwise
mean flow profile data than by directly fitting the dissipation
scaling (eq. \ref{eq:neqScal}) in the case of axisymmetric turbulent
wakes. It is therefore important to analyse turbulent planar jet data
with inlet Reynolds numbers much higher than those of
\cite{deoetal2008} where $Re_{G} = 7000$; and it is also important
that these data are complete enough to permit checks of eqns.
(\ref{eq:scalingEQ1})-(\ref{eq:scalingEQ2})-(\ref{eq:neqScal}) and (\ref{eq:DissExp1}). Such data can be found in \cite{Antonia1980}.

\begin{figure} 
\centering
{\includegraphics[width=0.7\columnwidth]{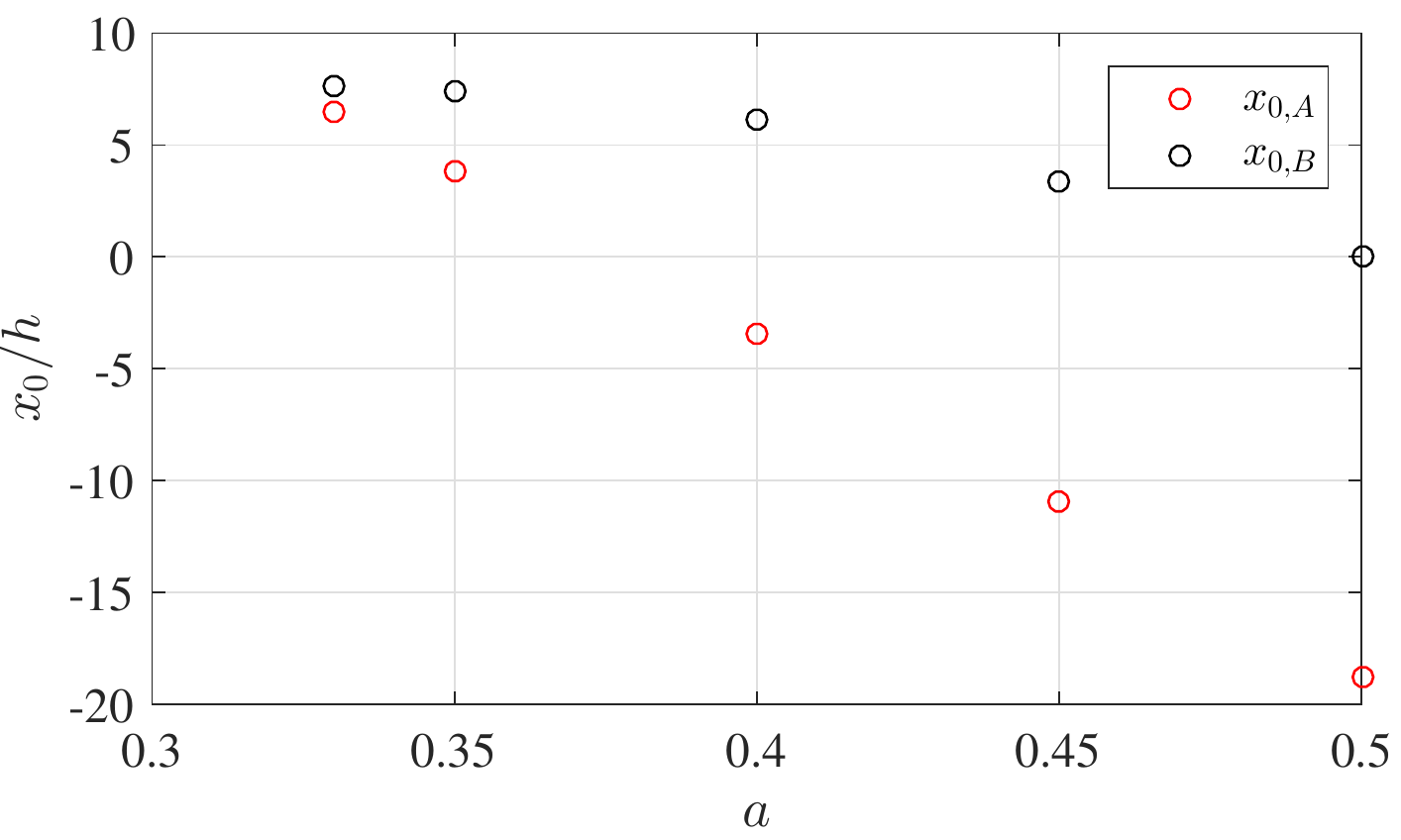}}
\caption{Virtual origins $x_{0,A}$ and $x_{0,B}$ obtained from the
  data of \cite{Antonia1980} ($Re_{G}=42800$) by applying different
  exponents to the power laws eq. (\ref{eq:scalingEQ1}) and
eq.  (\ref{eq:scalingEQ2}) respectively, with $0.33\le a \le 0.5$
  (i.e. $0 \le m \le 1$ from eq. (\ref{eq:scalingEQ3})). The optimal
  exponent is the one where $x_{0} = x_{0,A} = x_{0,E}$, i.e.  $a=1/3$
  corresponding to $m=1$.}
\label{fig:AntoniaVO}
\end{figure}
\begin{figure} 
\centering
\subfloat[][]
{\includegraphics[width = 0.5\columnwidth]{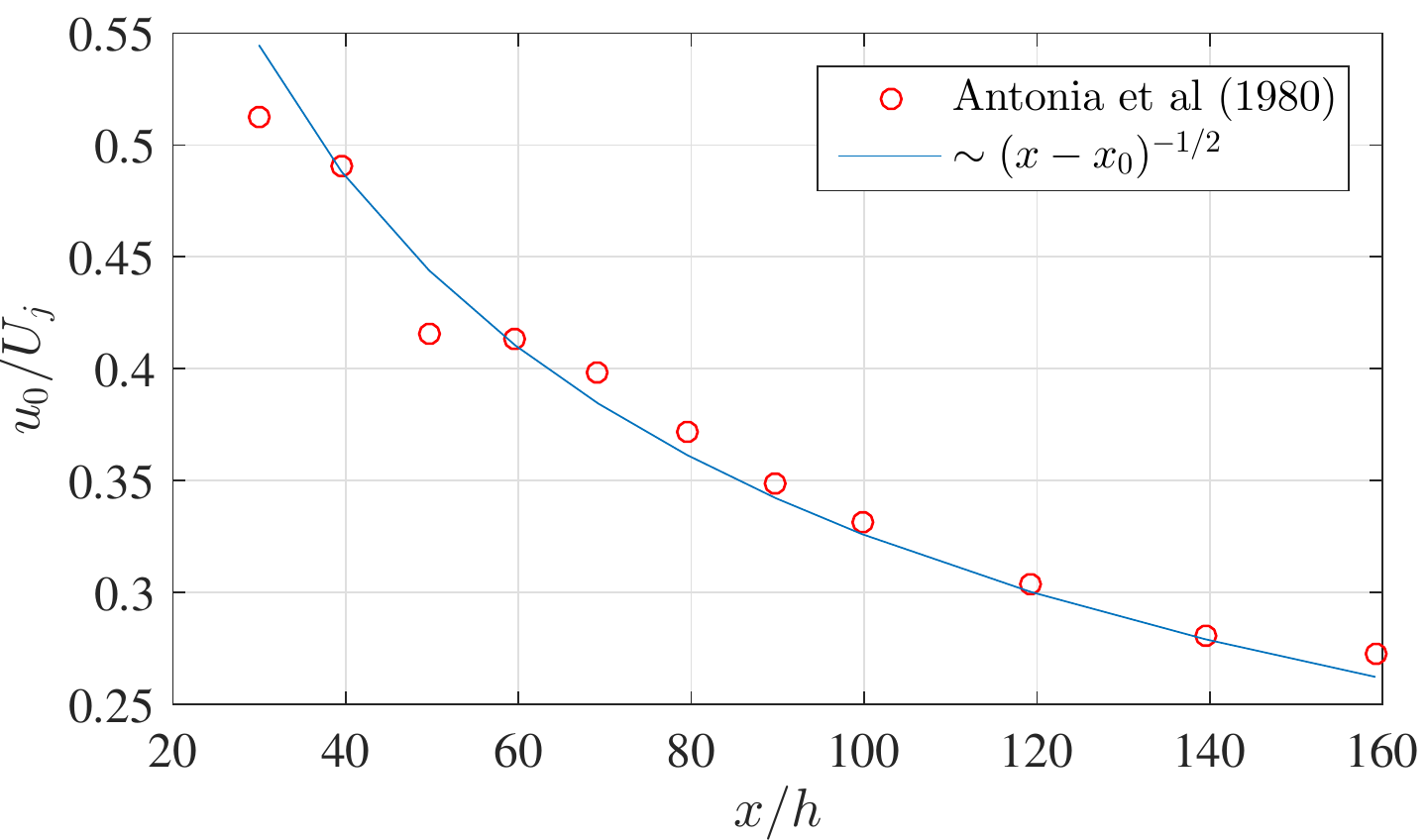}}
\subfloat[][]
{\includegraphics[width = 0.5\columnwidth]{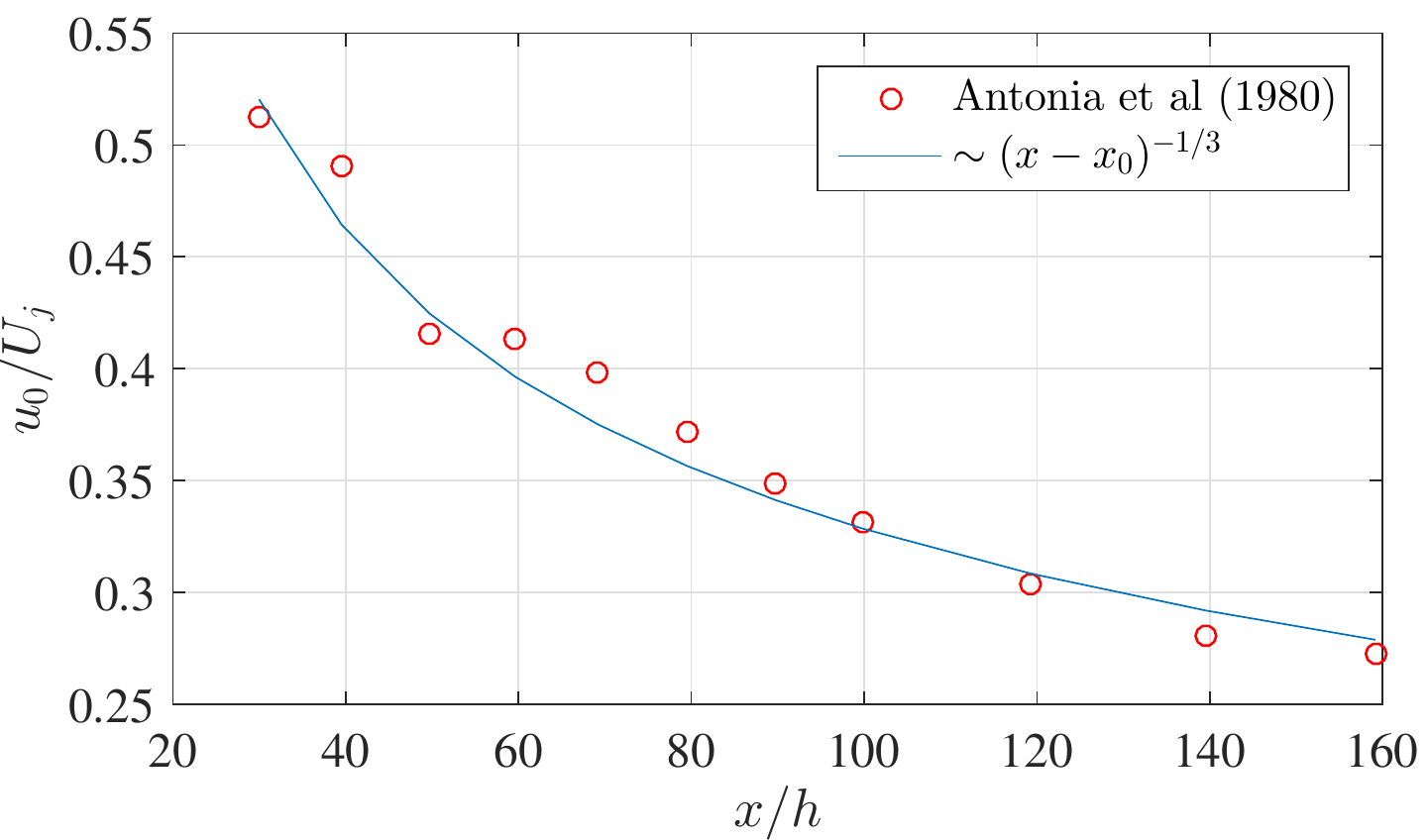}}\\
\caption{Normalised centreline streamwise mean flow velocity
  $u_{0}/U_{J}$ versus $x/h$. (a) Compared to eq. (\ref{eq:scalingEQ1})
  with $a=1/2$ ($m=0$) and $x_{0} = {x_{0,A} + x_{0,B}\over 2} \approx
  -9.4h$ which differs from $x_{0,A} \approx -19h$ and $x_{0,B}
  \approx 0$; (b) Compared to eq. (\ref{eq:scalingEQ1}) with $a=1/3$
  ($m=1$) and $x_{0} \approx 7h$ which is about the same as $x_{0,A}$
  and $x_{0,B}$.}
\label{fig:Antoniau0}
\end{figure}
\begin{figure} 
\centering \subfloat[][] {\includegraphics[width =
    0.5\columnwidth]{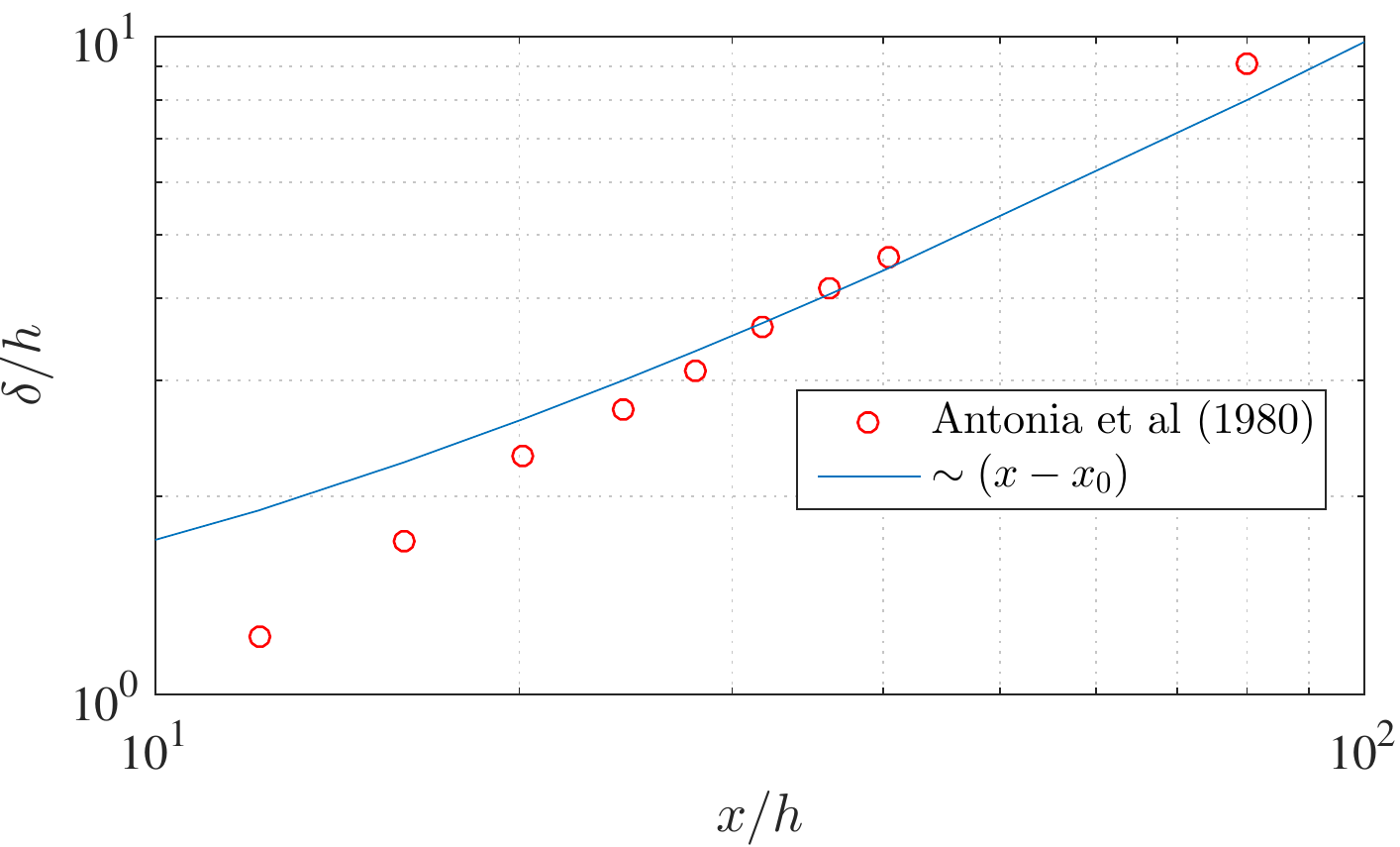}}
\subfloat[][] {\includegraphics[width =
    0.5\columnwidth]{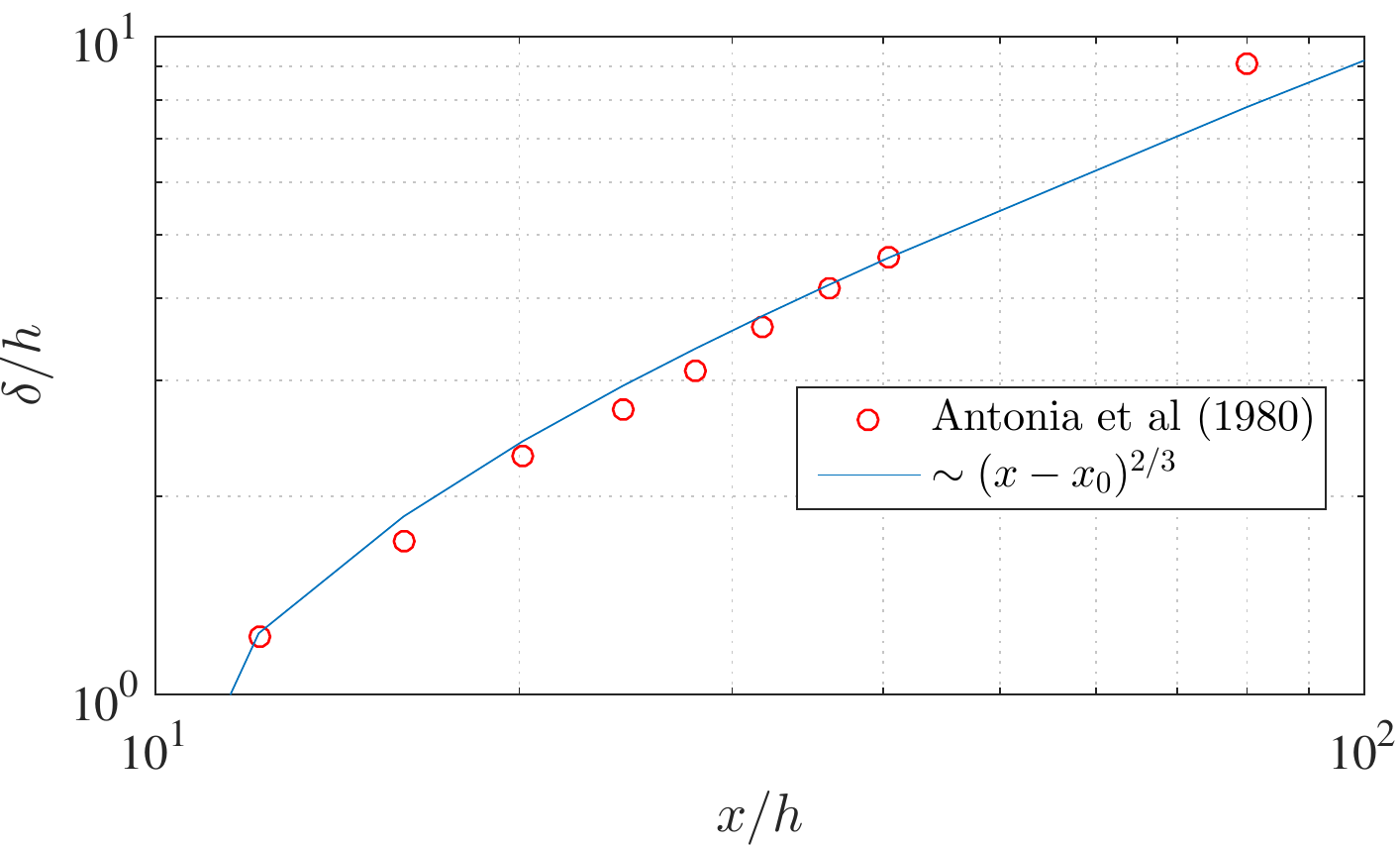}}\\
\caption{Normalised jet width $\delta/h$ versus $x/h$. (a) Compared to eq. 
  (\ref{eq:scalingEQ2}) with $2a=1$ ($m=0$) and $x_{0} = {x_{0,A} +
    x_{0,B}\over 2} \approx -9.4h$ which differs from  both $x_{0,A} \approx
  -19h$ and $x_{0,B} \approx 0$; (b) Compared to eq. (\ref{eq:scalingEQ2})
  with $2a=2/3$ ($m=1$) and $x_{0} \approx 7h$ which is about the same
  as $x_{0,A}$ and $x_{0,B}$.}
\label{fig:Antoniadelta}
\end{figure}
\begin{figure} 
\centering
\subfloat[][]
{\includegraphics[width = 0.5\columnwidth]{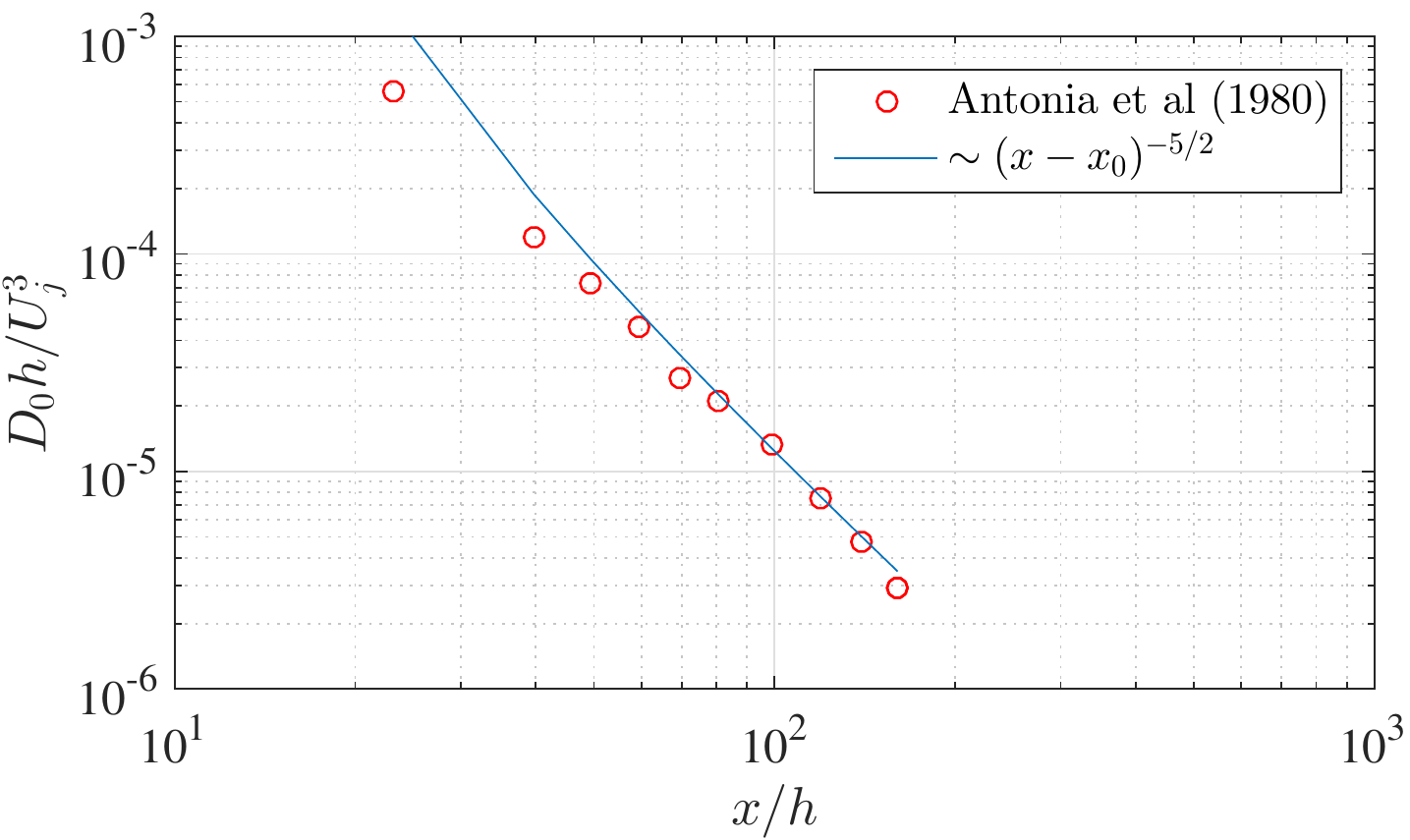}}
\subfloat[][]
{\includegraphics[width = 0.5\columnwidth]{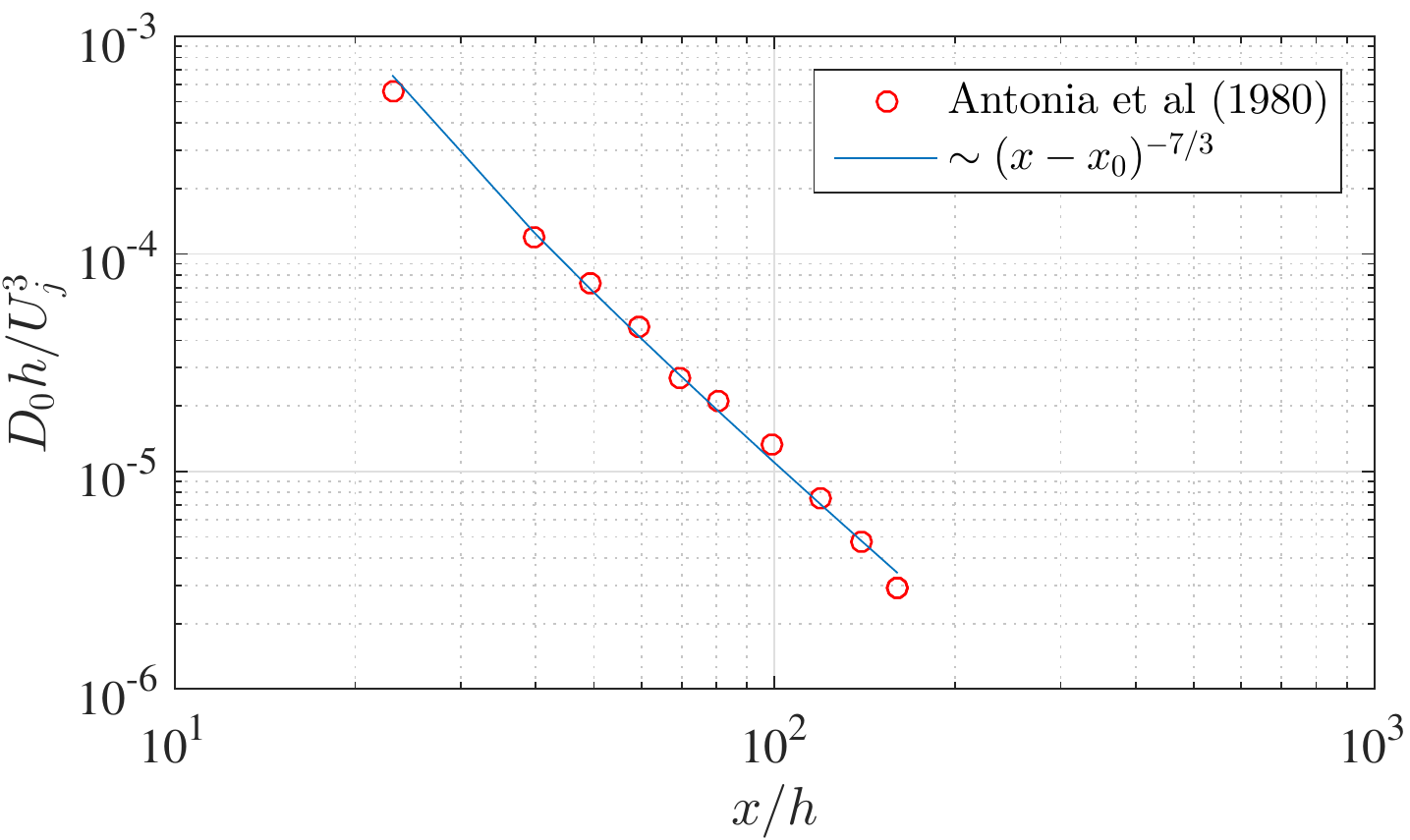}}\\
\caption{Centreline turbulence dissipation rate $D_0$ normalized with
  the inlet speed $U_J$ and the nozzle width $h$ versus $x/h$. (a)
  Compared to eq. (\ref{eq:DissExp1}) with $\gamma =5/2$ ($m=0$) and
  $x_{0} = -9.5h$; (b) Compared to eq. (\ref{eq:DissExp1}) with $\gamma
  =7/3$ ($m=1$) and $x_{0} \approx 7h$.}
\label{fig:Antoniadiss}
\end{figure}

\subsection{\cite{Antonia1980}} \label{sec:anton}

\cite{Antonia1980} report centreline data for $u_0$ in a turbulent
planar jet with inlet/global Reynolds number $Re_{G}=42800$ from $x/h
= 30$ to $x/h = 160$, and also data for $\delta$ from $x/h = 12$ to
$x/h = 100$. In figure \ref{fig:AntoniaVO} we plot the virtual origins
$x_{0,A}$ and $x_{0,B}$ which return the best respective fits of these
data to eqns. (\ref{eq:scalingEQ1}) and (\ref{eq:scalingEQ2}) for values of
$a$ ranging between $1/3$ ($m=1$) and $1/2$ ($m=0$). The only exponent
$a$ where $x_{0,A} \approx x_{0,B}$ is $a\approx 1/3$.  This is
therefore the only exponent $a$ for which the data of
\cite{Antonia1980} can fit both eqns. (\ref{eq:scalingEQ1}) and
(\ref{eq:scalingEQ2}) in a way that respects the momentum flux
conservation eq. (\ref{eq:momFlux}).

It is noticeable that the optimal virtual origin $x_{0,B}$ varies much
less with $a$ than $x_{0,A}$. This is because the exponent $a$ in the
power law (eq. \ref{eq:scalingEQ1}) is smaller than the exponent $2a$ in
the power law (eq. \ref{eq:scalingEQ2}). The exponent $\gamma$ in
eq. (\ref{eq:DissExp1}) is even larger and varies between $5/2$ and $7/3$
in the range $0\le m \le 1$ where $a$ varies from $1/2$ to $1/3$. It
is therefore no surprise that the virtual origin $x_{0,E}$ which
optimises the fit of eq. (\ref{eq:DissExp1}) to the centreline dissipation
data of \cite{Antonia1980} (ten data points from $x/h=20$ to
$x/h=160$) turns out to be about the same for all values of $a$
between $1/3$ and $1/2$ and is in fact quite close to $x_{0,E}/h=5$ on
average. The quality of this centreline dissipation fit does not vary
significantly if $x_{0,E}/h$ is made to vary between 3 and 7 for any
value of $a$ between $1/3$ and $1/2$. The virtual origin $x_{0}/h=7$
which optimises both power law fits (eqns. \ref{eq:scalingEQ1}) and
(\ref{eq:scalingEQ2}) in the case $a=1/3$ (i.e. $m=1$) is also
effectively optimal for the fit of eq. (\ref{eq:DissExp1}) with $\gamma =
7/3$ (i.e. $a=1/3$, $m=1$) to the centreline dissipation data of
\cite{Antonia1980}. No other exponent $a$, or equivalently $m$, can
achieve an optimally good fit of the data of \cite{Antonia1980} to all
three power laws (eqns. \ref{eq:scalingEQ1}), (\ref{eq:scalingEQ2}) and
(\ref{eq:DissExp1}) with one single virtual origin $x_{0} = x_{0,A} =
x_{0,B} = x_{0,E}$.

In figures \ref{fig:Antoniau0}, \ref{fig:Antoniadelta} and
\ref{fig:Antoniadiss} we plot streamwise profiles of $u_{0}/U_{J}$,
$\delta/h$ and $D_{0}h/U_{J}^{3}$ respectively using the data of
\cite{Antonia1980}. The left plots show fits of these data to the
classical power laws which correspond to $m=0$, all with the same
virtual origin $x_0$ as must of course be the case. However, given the
wide difference between $x_{0,A}$ and $x_{0,B}$ (see figure
\ref{fig:AntoniaVO}) for $m=0$, i.e. $a=0.5$, the single virtual
origin $x_0$ in all figures \ref{fig:Antoniau0}(a),
\ref{fig:Antoniadelta}(a) and \ref{fig:Antoniadiss}(a) has been chosen
to be midway between $x_{0,A}$ and $x_{0,B}$. The fits in the right
plots \ref{fig:Antoniau0}(b), \ref{fig:Antoniadelta}(b) and
\ref{fig:Antoniadiss}(b) are to the non-equilibrium power laws which
correspond to $m=1$. In this case the virtual origin is unambiguous
and naturally the same for all the plots as this is the only case
where the data of \cite{Antonia1980} are best fitted with one same
virtual origin for all three quantities $u_0$, $\delta$ and $D_0$.
The equilibrium ($m=0$) fit of $u_0$ is arguably a little better than
the non-equilibrium fit ($m=1$) of $u_0$, but the non-equilibrium fits
of $\delta (x)$ and $D_{0}(x)$ are both clearly superior to the
equilibrium fits of these two quantities. All in all, the data of
\cite{Antonia1980} seem to favour the non-equilibrium power-law
dependencies on streamwise distance and we now use further data from
their paper for a direct check of the non-equilibrium dissipation
scaling which underpins the non-equilibrium fits in figures
\ref{fig:Antoniau0}, \ref{fig:Antoniadelta} and \ref{fig:Antoniadiss}.
  
\begin{figure}
\centering
\includegraphics[width=0.5\columnwidth]{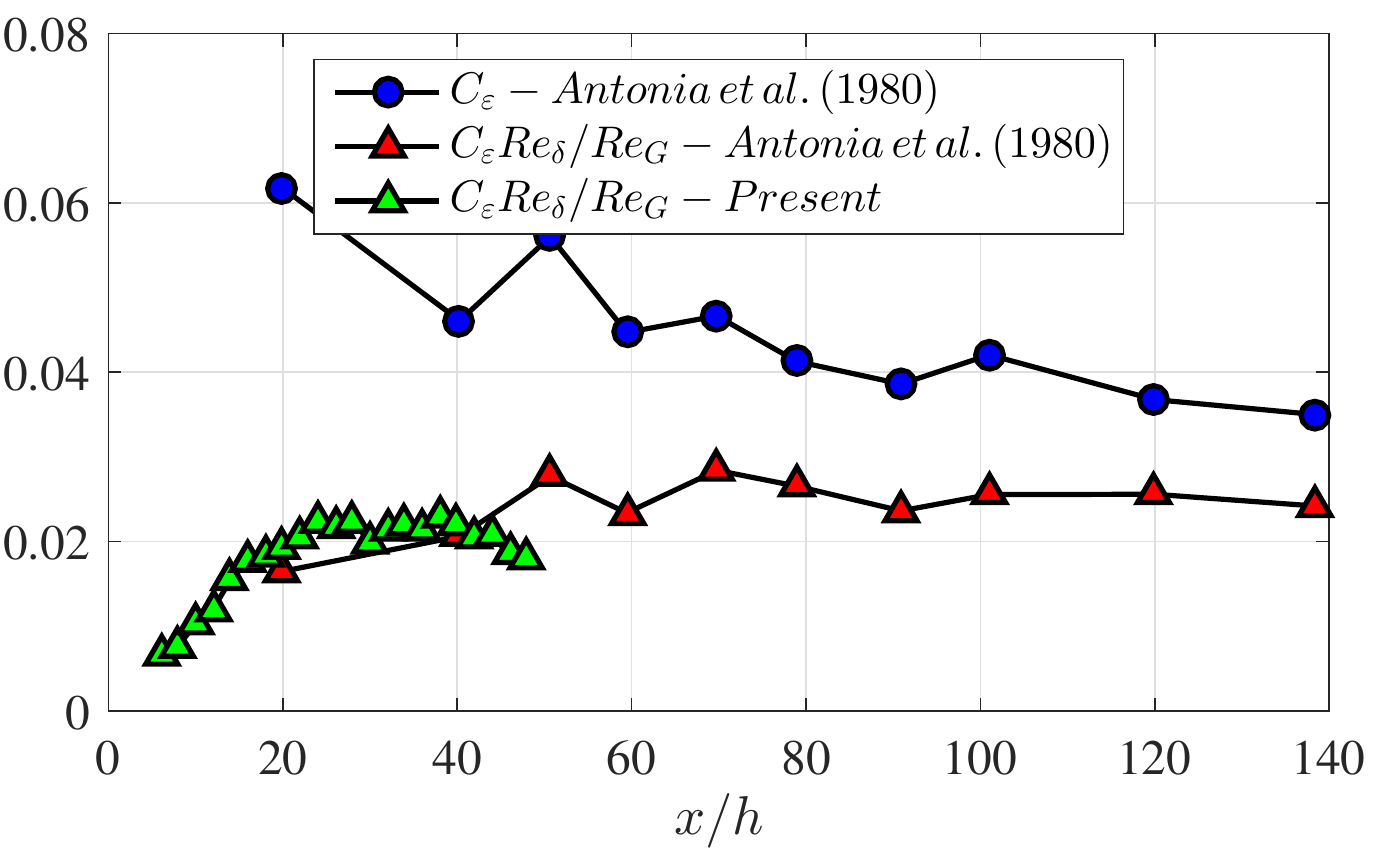}
\caption{Centreline dissipation coefficient $C_{\varepsilon}\equiv
  2D_{0} \delta/(3u'^3)$ (blue circles) and $C_{\varepsilon} \cdot
  Re_{\delta}/Re_{G}$ (red triangles), both from \cite{Antonia1980}
  where $Re_{G}=42800$ (red triangles)). The green triangles are
  $C_{\varepsilon} \cdot Re_{\delta}/Re_{G}$ from our present data
  ($Re_{G}=20000$, see section \ref{sec:expset}) where
  $C_{\varepsilon}$ is also calculated from $C_{\varepsilon} \equiv
  2D_{0} \delta/(3u'^3)$ on the centreline. The quantities in this
  figure are plotted as functions of normalised streamwise distance
  $x/h$. }
\label{fig:Antonia}
\end{figure}

\cite{Antonia1980} also provided centreline streamwise profile data in
the range $20\le x/h\le 140$ for the rms turbulent velocity $u'$
normalised by $u_0$, i.e.  $u'/u_0$, and for $D_{0}\delta/u_{0}^{3}$.
It is therefore possible to obtain, from their data, the turbulence
dissipation coefficient $C_{\varepsilon}\equiv 2D_{0} \delta/(3u'^3)$
which we plot in figure \ref{fig:Antonia} against $x/h$. This figure
shows that $C_{\varepsilon}$ decreases with increasing $x/h$ in the range
$20\le x/h\le 140$. 
However, $C_{\varepsilon} Re_{\delta}$ appears to remain constant in
the range $40\le x/h\le 140$ (figure \ref{fig:Antonia}), which is
consistent with the non-equilibrium exponent $m=1$ in
eq. (\ref{eq:neqScal}).
The largest difference between two values of $C_{\varepsilon}
Re_{\delta}$ in this range is 15\% of the mean (over the same range)
of $C_{\varepsilon} Re_{\delta}$ whereas it is 46\% for
$C_{\varepsilon}$. In the range $50\le x/h\le 140$ these percentages are
even more convincing as they are 3.5\% and 46\% respectively. The data
of \cite{Antonia1980} support $C_{\varepsilon} Re_{\delta} = Const$
(i.e. $m=1$) rather than $C_{\varepsilon} =Const$ (i.e. $m=0$) quite
clearly in the range $50\le x/h\le 140$ and perhaps even $40\le x/h\le
140$.

There are of course two potential caveats in this conclusion both of
which result from the fact that the measurements of \cite{Antonia1980}
were taken with single hot wire anemometry. Strictly speaking,
$C_{\varepsilon}$ should be defined as $C_{\varepsilon}\equiv D_{0}
\delta/K_{0}^{3/2}$ rather than $C_{\varepsilon}\equiv 2D_{0}
\delta/(3u'^3)$ and all fluctuating velocity gradients should be
accessed for a measurement of $\varepsilon$ which does not rely on
assumptions. \cite{Antonia1980} used the isotropic approximation of
$\varepsilon$ which is accessible with single hot wire measurements and
thereofre relied on the assumption of small-scale isotropy. These
issues are addressed in the following section and in section
\ref{sec:dissipation}.

In the following section we describe our turbulent planar jet
experiment and validate it against previously published data. We take
Hot Wire Anemometry (HWA) measurements with both single and cross
wires and investigate the validity of the assumptions and predictions
of the theory described in section \ref{sec:theory}. Our measurements
do not extend beyond $x/h = 54$, but we do measure and report in
sections \ref{sec:dissipation}, \ref{sec:selfsimilarity} and
\ref{sec:scalings} profiles of $U$, $V$, $R_{xy}$, $K$ and
$\varepsilon$. Our global/inlet Reynolds number is $Re_G=20000$ and
therefore nearly three times larger than $Re_{G} = 7000$ in
\cite{deoetal2008}. It is also about half the value of $Re_G$ in
\cite{Antonia1980}, and the factor 2 between our $Re_G=20000$ and the
$Re_{G} = 42800$ of \cite{Antonia1980} can help us assess the
universal non-equilibrium expectation that the constant in
$C_{\varepsilon} \cdot Re_{\delta} = Const$ is in fact proportional to
$Re_{G}$, in full agreement with $m=1$.

\section{Experimental apparatus and measurements} \label{sec:expset}
\begin{figure}
\centering
\subfloat [][]
{\includegraphics[width=0.5\columnwidth]{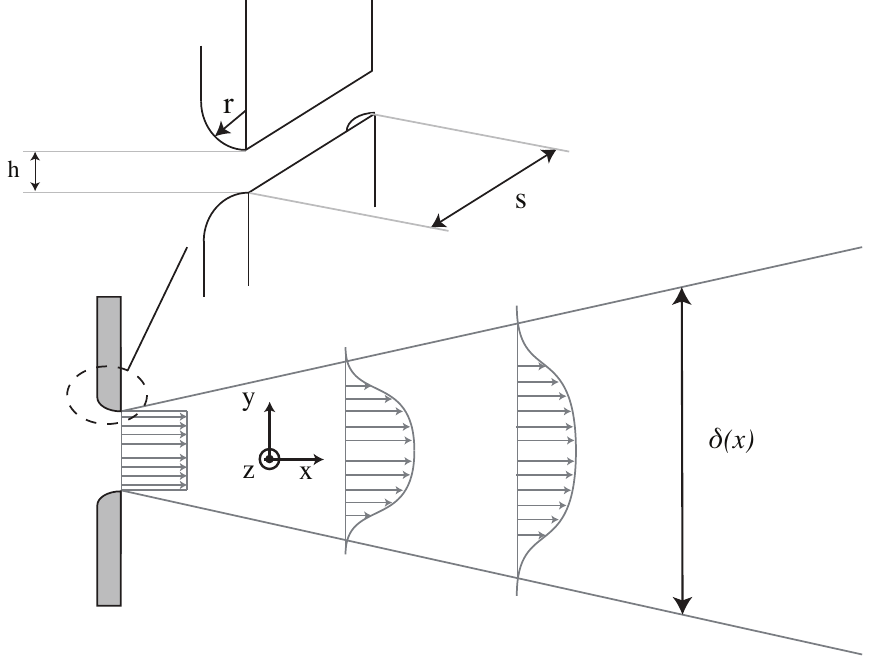}}
\subfloat[][]
{\includegraphics[width=0.5\columnwidth]{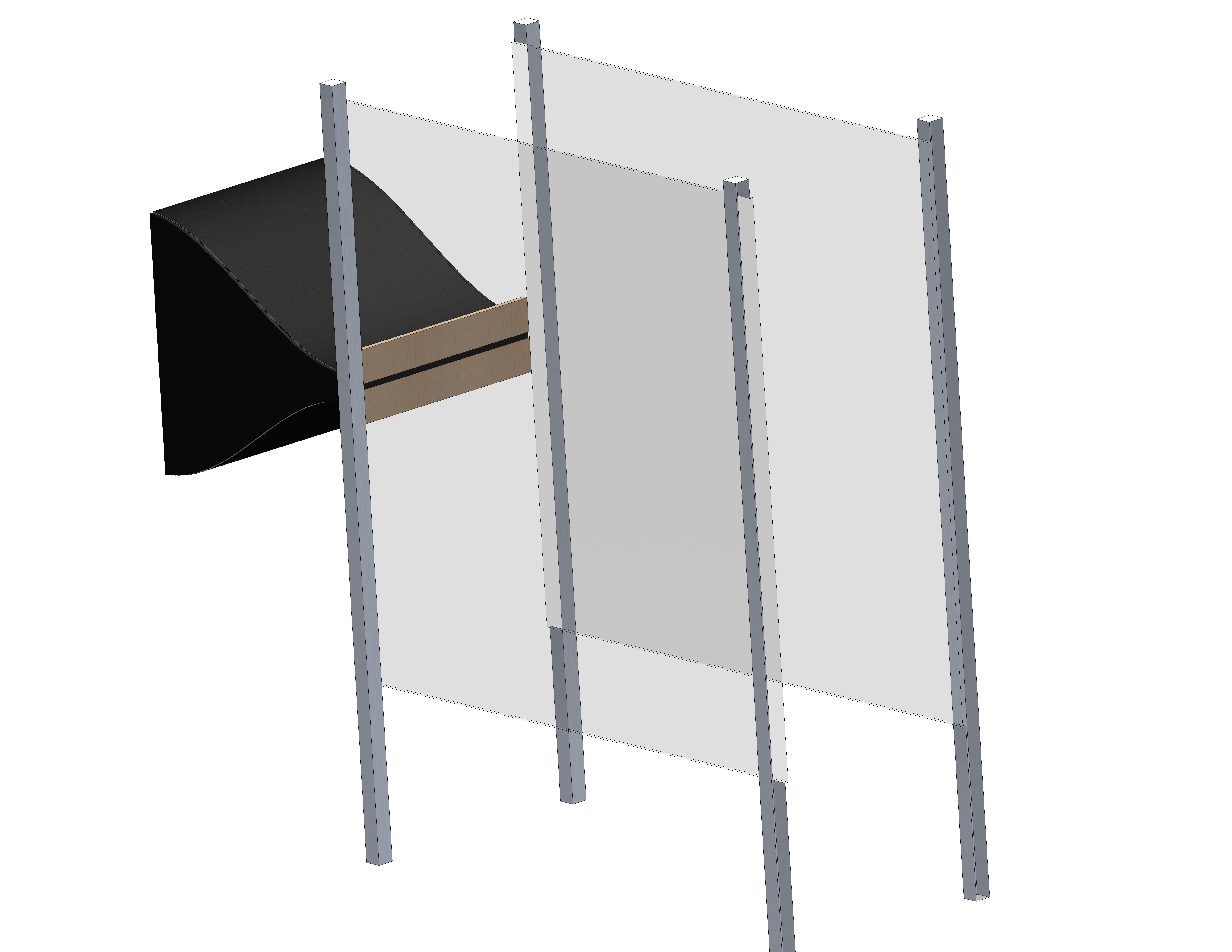}}\\
  \caption{a) Schematic representation of the planar jet flow, with
    detail of the fillet radius $r$; $\delta(x)$ is representative of
    the jet width at each streamwise location $x$.  b) Isometric
    representation of the experimental apparatus employed to generate
    the planar jet.}
\label{fig:planjet}
\end{figure}

Our planar jet flow is generated using a centrifugal blower which
collects air from the environment and then forces
it into a plenum chamber. In order to reduce the inflow turbulence
intensity level and remove any bias due to the feeding circuit, the
air passes through two sets of flow straighteners before entering a
convergent duct (having area ratio equal to about 8). At the end of
the duct there is a letterbox slit with aspect ratio $s/h=31$
and $h=15mm$ (see Figure \ref{fig:planjet}a).  Figure
\ref{fig:planjet}a includes a schematic of the contoured inlet: in
order to produce a top hat velocity profile at the jet exit ($x=0$),
the two longest sides of the slit are filleted with a radius $r=2h$
(see Figure \ref{fig:planjet}b), following the careful recommendation
by \cite{deo}. The jet exhausts into ambient air and is confined in
the spanwise direction by two perspex walls (see Figure
\ref{fig:planjet}b) of size $100h \times 100h$ placed in $x-y$
planes. The aspect ratio $s/h =31$ is sufficiently large to ensure
that the flow can be considered planar as documented in the published
literature (e.g. \citealp{gutmarkwygnanski1976,
  gardon&akfirat1966}). Furthermore, the effect of the boundary layer
which develops on the bounding perspex walls is estimated to
  affect less than 3\% of the overall spanwise extent $s$ at 100 $h$
from the jet exit section. The jet rig is located in a room much
larger in all directions than the jet width $\delta$ at $x=100h$, so
that the effects of the ceiling, floor and room walls on the
  entrainment and development of the jet flow are reduced to a
minimum.
The inlet velocity $U_J=20m/s$ is set and stabilized using a PID
feedback controller which takes as input the thermo-fluid-dynamic
conditions of the flow measured by a thermocouple and a Pitot
tube. The thermocouple measures the temperature of the working fluid
about 5 $cm$ upstream of the letterbox slit in the convergent part of
the nozzle, whilst the Pitot tube is located such that the pressure
measurements are carried out within the potential core of the jet
flow. These data are acquired using a Furness Control micromanometer
FCO510, then manipulated by the in house PID controller which
outputs the voltage to be supplied to the blower's driver in order to
achieve the desired flow speed.

The velocity signal is measured using both one- and two-component hot
wires (herein referred to as SW and XW respectively) driven by a
Dantec Streamline constant temperature anemometer (CTA). Considering
the large dynamic range that characterizes the planar jet flow, we
operate both the SW and the XW with an overheat ratio of 1.2. Both the
SW and the XW are etched in house; the sensing length of the wire is
$\approx$1 $mm$, whilst the wire diameter is $5 \mu m $. For the XW,
the separation between the two wires is about 1$mm$. Data are sampled
at a frequency of 50 $KHz$ using a 16-bit National Instruments NI-6341
(USB) data acquisition card. Each SW measurement lasts for 60$s$,
which was estimated to be a sufficiently long time for convergence of
the turbulent statistics studied here. This was checked by taking
longer time SW measurements, up to two minutes at the furthermost
investigated location (i.e. $x/h=50$) where the integral length-scale
is the largest and checking the convergence of the longitudinal
integral length-scale $L_{uu}$. (The number of integral scales within
a 60$s$ sampling period is about 30000 at $x/h=50$ and is of course
higher at locations closer to the nozzle exit.) The acquisition time
for XW measurements was increased to 120$s$ in all cases because they
involve the cross-stream velocity which is of the order of 2-3\% of
the streamwise velocity.

Cross-stream profiles were acquired with the SW probe from $x/h=0$ to
$x/h=50$ with a 2$h$ spacing. These measurements were taken for
 inlet Reynolds number $Re_G=U_J h/\nu=20000$.
We ascertained that the jet is indeed planar by also taking
measurements at $z= \pm 10h$ and verifying that there are no
statistical differences between the three sampled values of $z$.

Cross-stream profiles were also taken with the XW probe in order to
measure both the cross-stream mean velocity component($V$) and the
relevant component $R_{xy}$ of the Reynolds stress tensor, for the
same inlet Reynolds number. Cross-stream profiles were measured at 10
different streamwise locations ranging from 14$h$ to 54$h$. The probe
displacement through the flow field is ensured by a high precision
traverse system controlled by a in-house driving system. We verified
that no significant differences exist between SW and XW measurements
of same statistics.

The SW calibration is carried out at the beginning and end of each run
and is obtained by fitting data acquired at seven different inlet
speeds (ranging from 0 to 100\% of $U_J$) with a 4-th order polynomial
curve. For the XW, a similar procedure is applied with the additional
introduction of 9 angles of the probe with respect to the streamwise
direction (in the $x - y$ plane), ranging from -30$^{\circ}$ to
30$^{\circ}$. This range of angles was chosen on the basis of previous
planar jet investigations, e.g. \cite{Browne1984}, and by checks done
with a wider range of angles (i.e. $\pm 35^{\circ}$).  Experimental
runs showing differences between calibrations at beginning and end of
the run larger than 1\% are discarded and repeated. Particular care
was also taken with respect to temperature drift during runs: in all
cases, there were no excursions larger than 0.3$K$ between start and
end of run.

\begin{figure} 
\centering
\subfloat[][]
{\includegraphics[scale=0.5]{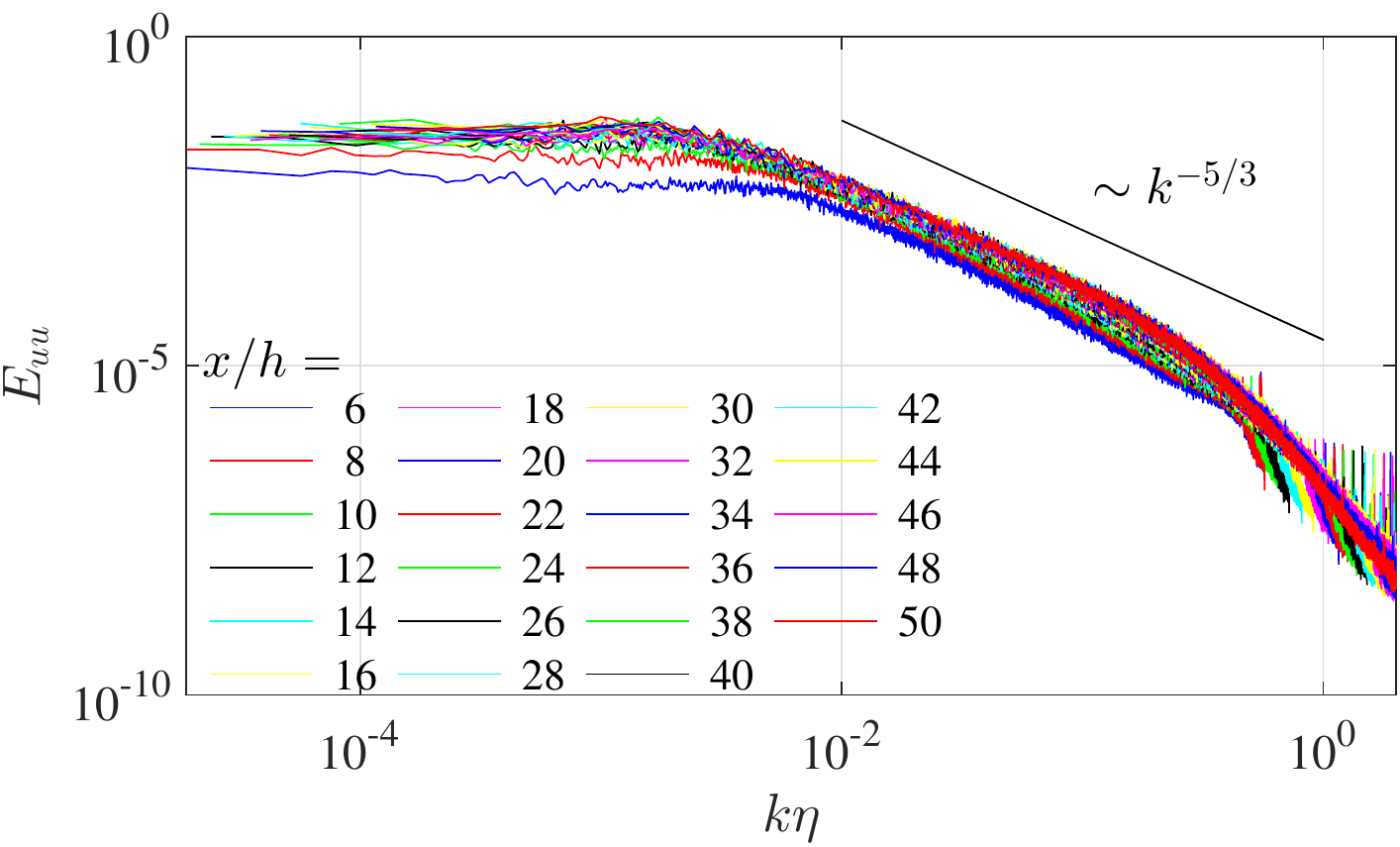}}\\
\subfloat[][]
{\includegraphics[scale=0.5]{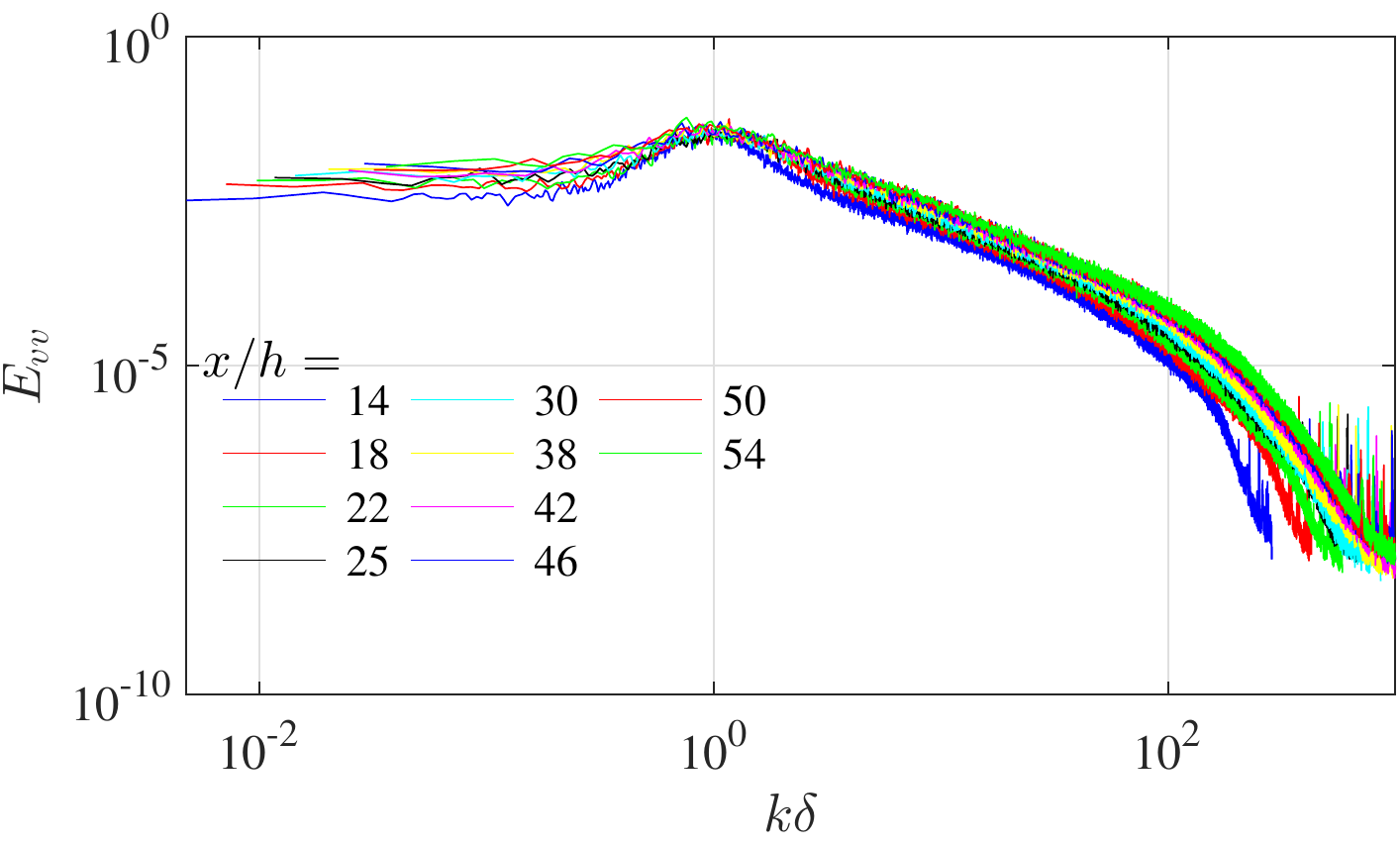}}
\caption{a) Streamwise fluctuating velocity spectra measured along the
  jet centreline using SW data plotted against the normalised
  longitudinal wavenumber $k\eta$. b) Cross-stream fluctuating
  velocity spectra measured along the jet centreline using XW data
  plotted against the normalised longitudinal wavenumber $k\delta$.}
\label{fig:spectra}
\end{figure}

The velocity spectra $E_{uu}$ (of the streamwise fluctuating velocity)
and $E_{vv}$ (of the cross-stream fluctuating velocity) provide
information about large scale and small scale resolution of our
measurements as well as presence of coherent/periodic structures.
Figure \ref{fig:spectra}a is a plot of $E_{uu}$ on the jet centreline
region $6 \leq x/h \leq 50$. Data are plotted against the longitudinal
wavenumber on the basis of the Taylor hypothesis ($k= 2 \pi f /u_0$
where $f$ stands for frequency) multiplied by the Kolmogorov
lengthscale $\eta \equiv (\nu^{3}/\varepsilon)^{1/4}$. The temporal
resolution of the wire is not enough to resolve the dissipative scales
immediately past the potential core. However, in the region of major
interest for the present study, namely $x/h>18$ as established in the
following sections, the small scales are resolved sufficiently
well. The large scales are also well resolved given the small
wavenumber plateau in Figure \ref{fig:spectra}(a). As the streamwise
distance increases, the increasing value of $E_{uu}$ at low
wavenumbers is due to the increase of the longitudinal integral
length-scale $L_{uu}$ given that $L_{uu} = E_{uu}(0) u_0/(4 u'^2)$
\citep{tennekes&lumley}. Similar observations and comments can be made
for the cross-stream spectrum $E_{vv}$.

We plot the lateral velocity spectra $E_{vv}$ calculated along the jet
centreline against $k\delta$ rather than $k\eta$ in Figure
\ref{fig:spectra}b to bring out the fact that the peak in this
spectrum scales with $\delta (x)$. A peak can clearly be spotted at
$k\delta \approx 1$, corresponding to $f\delta/u_0\approx 0.16$ (in agreement with \citealp{deoetal2008}), at all investigated centreline
streamwise distances. These peaks must be associated with jet coherent
structures.

\begin{figure} 
\centering
\subfloat[][]
{\includegraphics[scale=0.5]{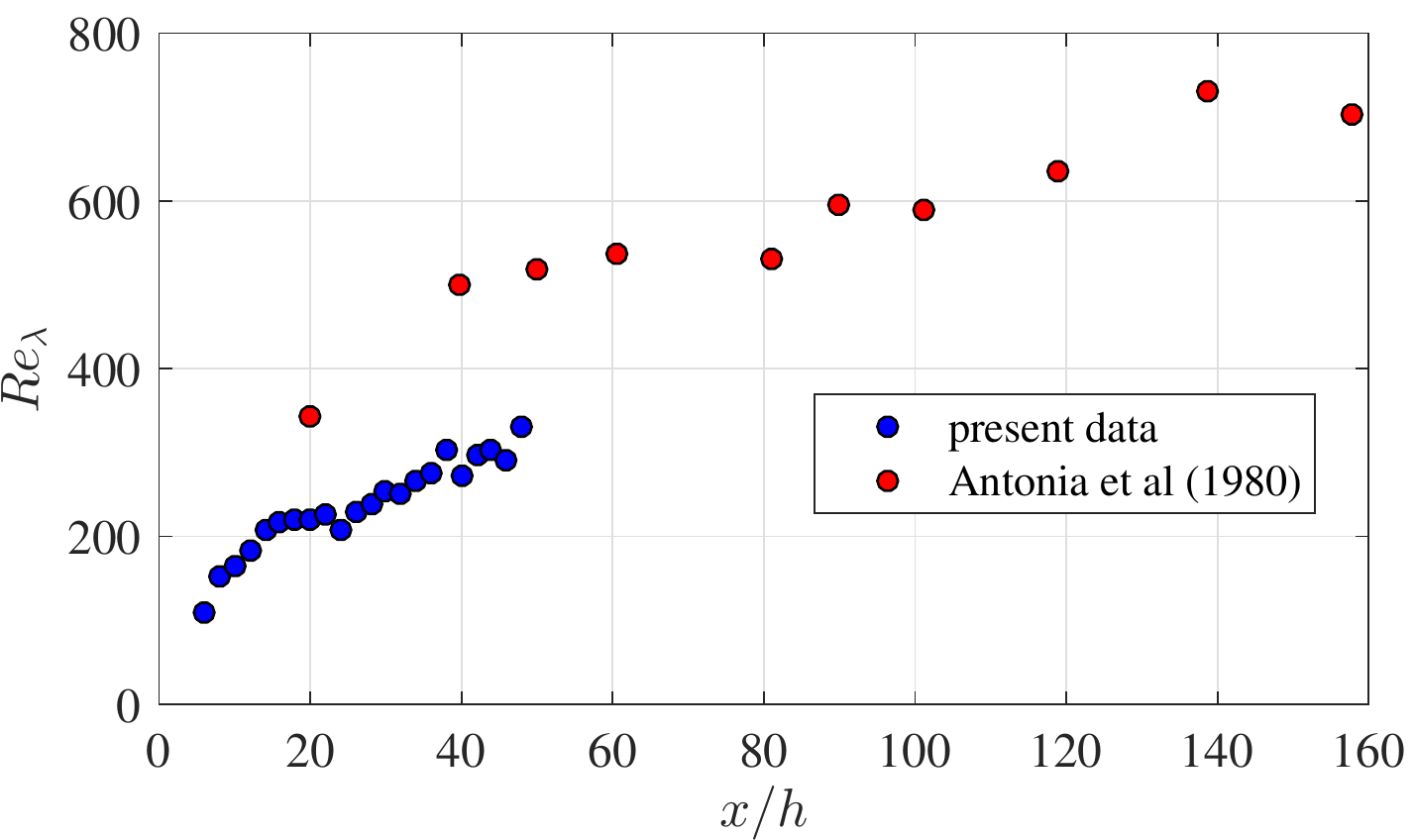}}
\caption{Local Reynolds number ($Re_{\lambda}$) along the jet
  centreline as a function of the streamwise distance
  $x/h$. $Re_G=20000$ for the present data. $Re_G=42800$ for the data
  of \cite{Antonia1980}.}
\label{fig:Relambda}
\end{figure}

The estimate of the turbulent dissipation rate $\varepsilon$ is
obtained from its isotropic surrogate, i.e. $\varepsilon_{ISO}=15 \nu
\overline{(\partial u/\partial x)^2}$, by integrating the one
dimensional spectrum $E_{uu}$ following
\begin{equation}
\overline{(\partial u/\partial x)^2} = \int_0^{\infty} k^2 E_{uu} dk 
\label{eq:2ndDer}
\end{equation}
where $u$ is the streamwise turbulent fluctuating velocity.  We also follow
\cite{Antonia1980} and chose to estimate $\varepsilon$ from
$\varepsilon_{iso}$ rather than from XW data because of the better
resolution of the SW data. This choice is supported by the DNS results
of \cite{stanley} at Reynolds number $Re_{G}=3000$, which show that
along the centreline there is only a 3\% difference between
$\varepsilon$ from $\varepsilon_{iso}$, and that this difference
slightly rises at the location of the jet shear layer to no more than
10\%. The DNS calculations of \cite{stanley} were limited to a
streamwise distance $x/h=12$ and it is therefore reasonable to expect
the correspondence between $\varepsilon$ from $\varepsilon_{iso}$ to
improve at higher $Re_G$ and higher values of $x/h$ given that the
local Reynolds number and the Kolmogorov length-scale increase with
downstream distance (e.g. \citealp{gutmarkwygnanski1976}). Hence, the
DNS of \cite{stanley} support our centreline dissipation measurements
and those of \cite{Antonia1980} which were obtained from SW data by using $\varepsilon = \varepsilon_{iso}$ to infer $\varepsilon$. We use
our SW dissipation measurements in section \ref{sec:dissipation} to
establish the turbulent dissipation scalings. In figure
\ref{fig:Relambda} we plot $Re_{\lambda} \equiv {u'\lambda\over \nu}$
as a function of $x/h$, where the Taylor length $\lambda$ is obtained
from $\varepsilon_{iso} = 15\nu u'^{2}/\lambda^{2}$. Note that
$Re_{\lambda}$ is larger than about 200 and increases with $x/h$ in
the range $10\le x/h\le 50$ for our data. The DNS of
\cite{goto&vassilicos2015} and \cite{goto&vassilicos2016} have shown
that non-equilibrium dissipation scalings such as eq. (\ref{eq:neqScal})
with $m = 1$ are well-defined for values of the Taylor length Reynolds
numbers $Re_{\lambda}$ larger than about 100 to 200.

The other use that we make of our dissipation measurements is to
demonstrate self-similarity of dissipation cross-stream profiles. In
section \ref{sec:selfsimilarity} we obtain such profiles for both
$\varepsilon_{iso}$ and $\varepsilon_{XW}=\nu (3\overline{(\partial
  u/\partial x)^2}+6\overline{(\partial v/\partial x)^2}) $ (where $v$
is the cross-stream turbulent flutuating velocity) and provide support
for self-similarity of both.

For dissipation calculations, a 4-th order Butterworth filter was
applied to the signals with cut-off frequency such that $k_{max} \,
\eta \approx 1.3$ where $k_{max}$ is the maximum longitudinal
wavenumber.

\subsection{Comparison with previously published data}

We now compare our data for the centreline mean flow velocity $u_0$
and the jet width $\delta$ with data in the published literature.

\begin{figure} 
\centering
{\includegraphics[width=.7\columnwidth]{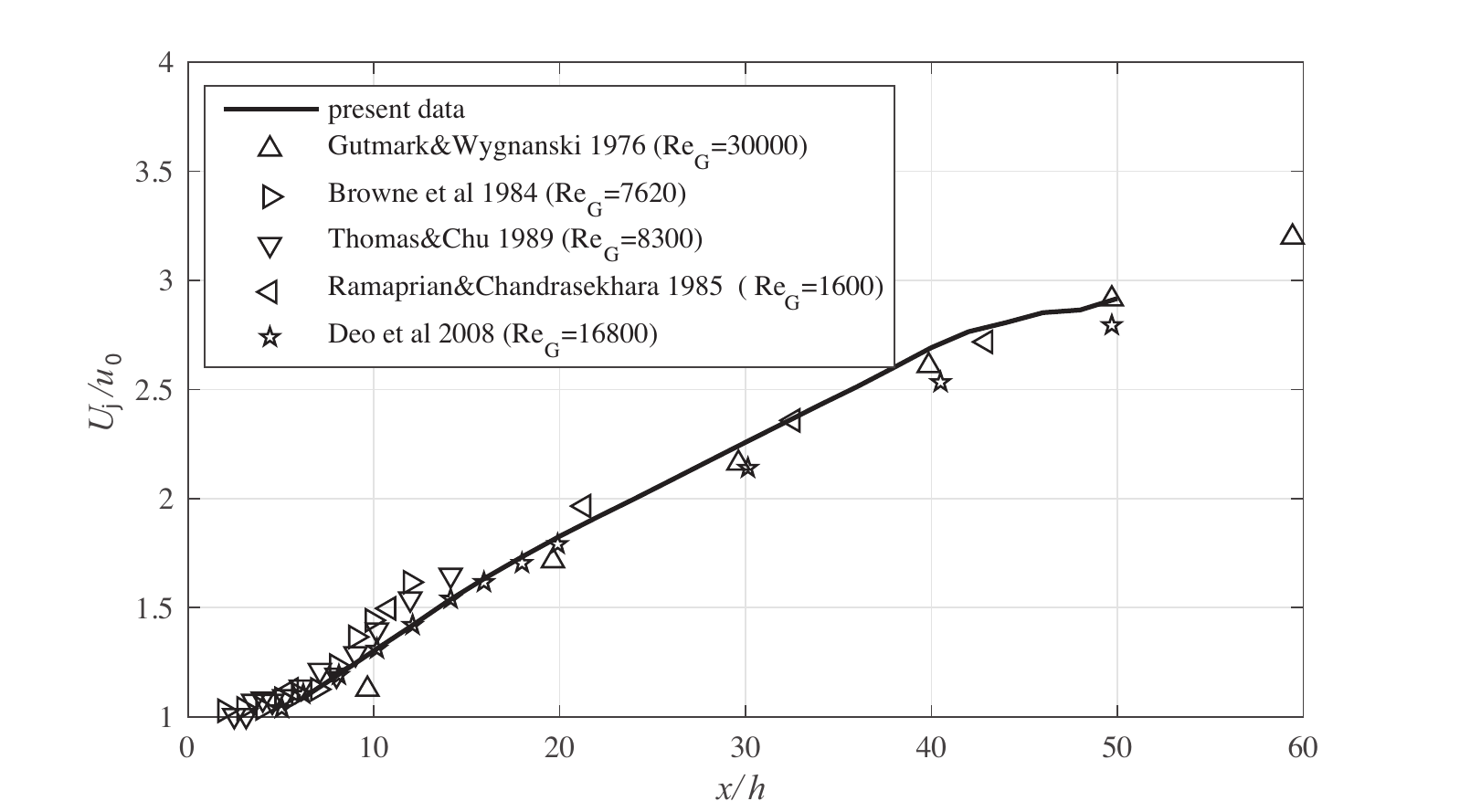}}
\caption{Centreline mean flow velocity $u_0$ as a function of the
  streamwise distance. Data are normalized using the inlet speed $U_J$
  and are plotted as $U_{J}/u_0$ versus $x/h$. The continuous line is
  representative of the current dataset and symbols refer to previous
  investigations. Our data show a rather good agreement with
  \cite{deoetal2008}, \cite{gutmarkwygnanski1976} and
  \cite{RamaprianChandra} between $x/h$ about 10 and $x/h =50$ (except
  for a data point by \cite{gutmarkwygnanski1976} near $x/h = 10$
  which is 5\% off). Some discrepancies can be detected at distances
  shorter than $x/h \approx 15$, which are not too far from the
  potential core, the length of which, as discussed by
  \cite{deoetal2008}, is a decreasing function of inlet Reynolds
  number; some small differences (smaller than 5\%) can be detected
  with the data of \cite{ThomasandChu1989} and \cite{Browne1984},
  which are characterized by a significantly smaller $Re_G$ and are in
  the region $x/h\le 15$.}
\label{fig:uouj}
\end{figure}
\begin{figure} 
\centering
{\includegraphics[width=0.7\columnwidth]{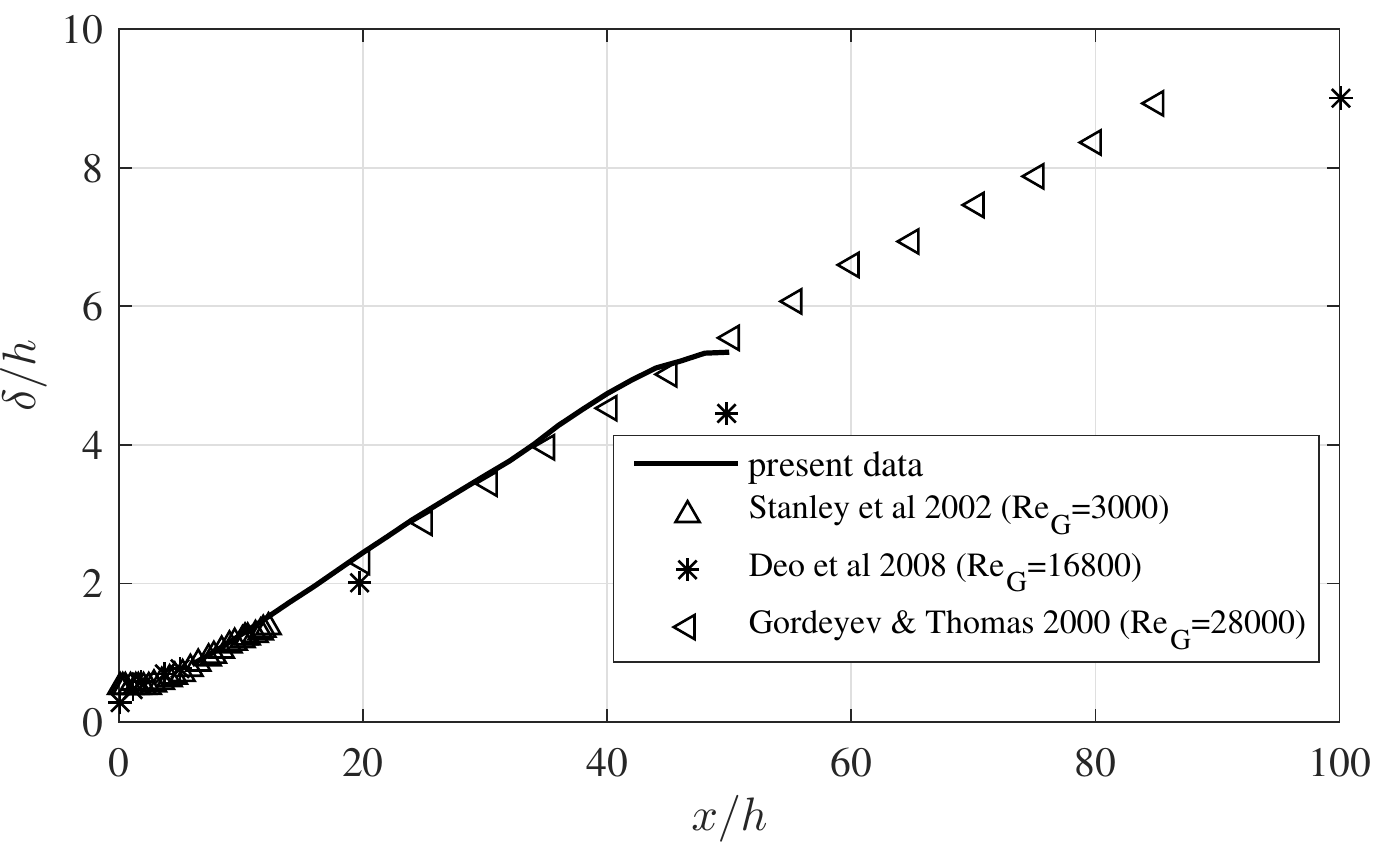}}
\caption{Comparison of the jet width $\delta (x)$ defined in eq.
  (\ref{eq:delta}) between present measurements and previously
  published data. Our data compare very well with those of
  \cite{gordeyev&thomas} and \cite{stanley}. A larger discrepancy
  between our data and the data of \cite{gordeyev&thomas} on the one
  hand and the data of \cite{deoetal2008} on the other is detected at
  $x/h=50$.}
\label{fig:delta_comp}
\end{figure}

Figure \ref{fig:uouj} shows the mean velocity $u_0$ measured along the
jet centreline normalized with the inlet speed and plotted as
$U_J/u_0$ versus normalised distance from inlet, $x/h$. Data from the
present experiment (continuous line) are compared to different
experiments with inlet Reynolds number ranging from values as low as
$Re_G=1600$ to $Re_G=30000$ (see legend of Figure \ref{fig:uouj}). Our
data compare very well with \cite{deoetal2008},
\cite{gutmarkwygnanski1976} and \cite{RamaprianChandra} in the range
$10<x/h<50$, with differences smaller than 3\%. At shorter distances,
some discrepancies can be detected but these positions are quite close
to the potential core, which as discussed at length by
\cite{deoetal2008} depends significantly on inlet conditions,
including inlet Reynolds number.  Also, some discrepancies, smaller
than 5\%, can be detected with the results of \cite{ThomasandChu1989}
and \cite{Browne1984} in the region $x/h\le 15$.

Figure \ref{fig:delta_comp} shows a comparison of our measured values
of the jet width with those obtained in previous investigations: an
extremely good matching can be ascertained through the whole domain,
particularly with the experimental data of \cite{gordeyev&thomas} and
the numerical simulation of \cite{stanley}. Some differences are
detected with the data of \cite{deoetal2008}, but a thorough
comparison with their data cannot be carried out given the small
number of streamwise locations where they reported jet width
measurements (four locations in the range $0 < x/h < 100$).

All in all, particularly in the region of greatest interest for the
present investigation (i.e. $20\leq x/h\leq 50$), the overall
behaviours of our centreline mean flow velocity and jet width data do
agree quite well with previously published literature.

\section{Turbulence dissipation scaling}  \label{sec:dissipation}

\begin{figure}
\centering \subfloat[][]
{\includegraphics[scale=0.45]{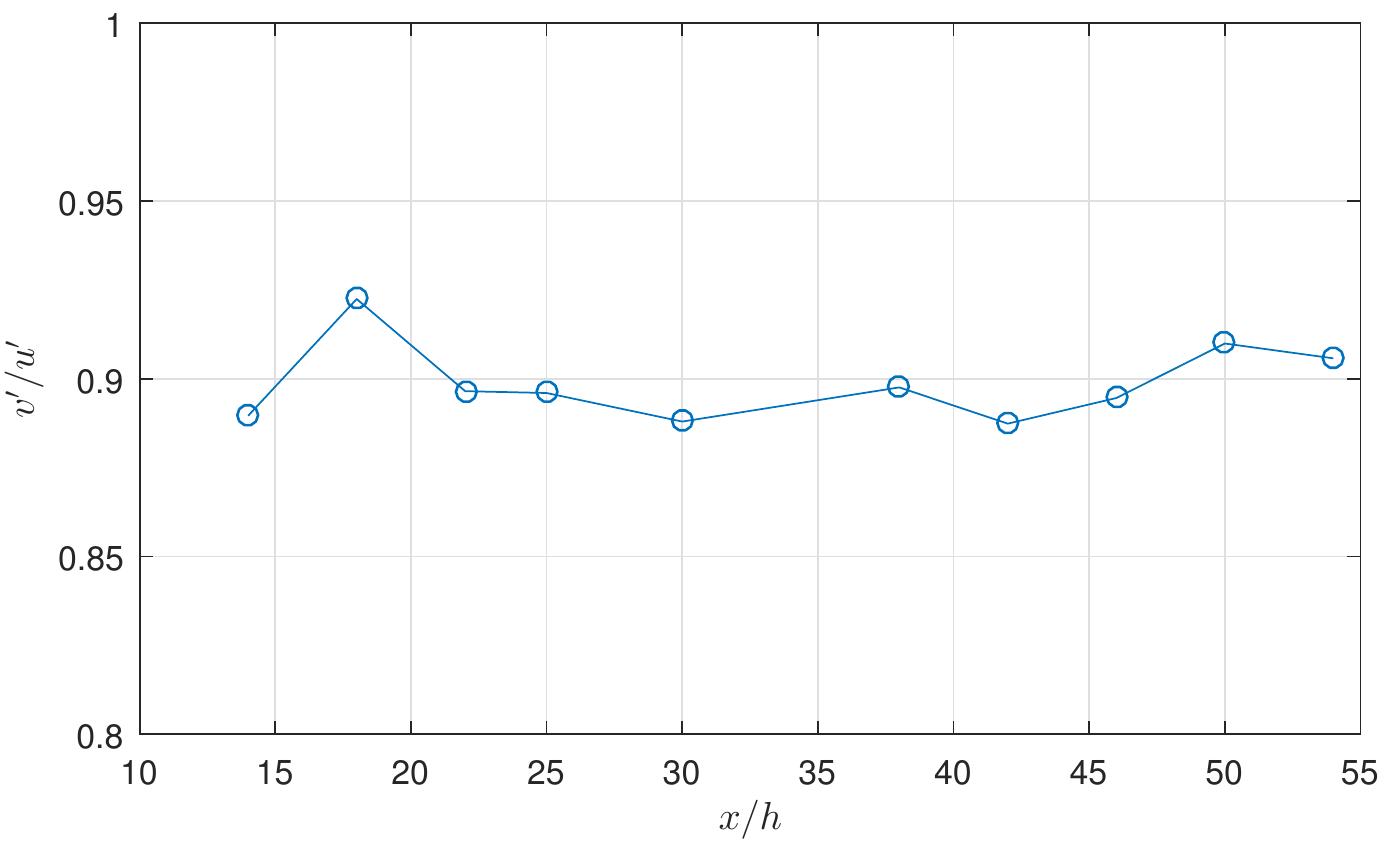}}
\subfloat[][]
{\includegraphics[scale=0.45]{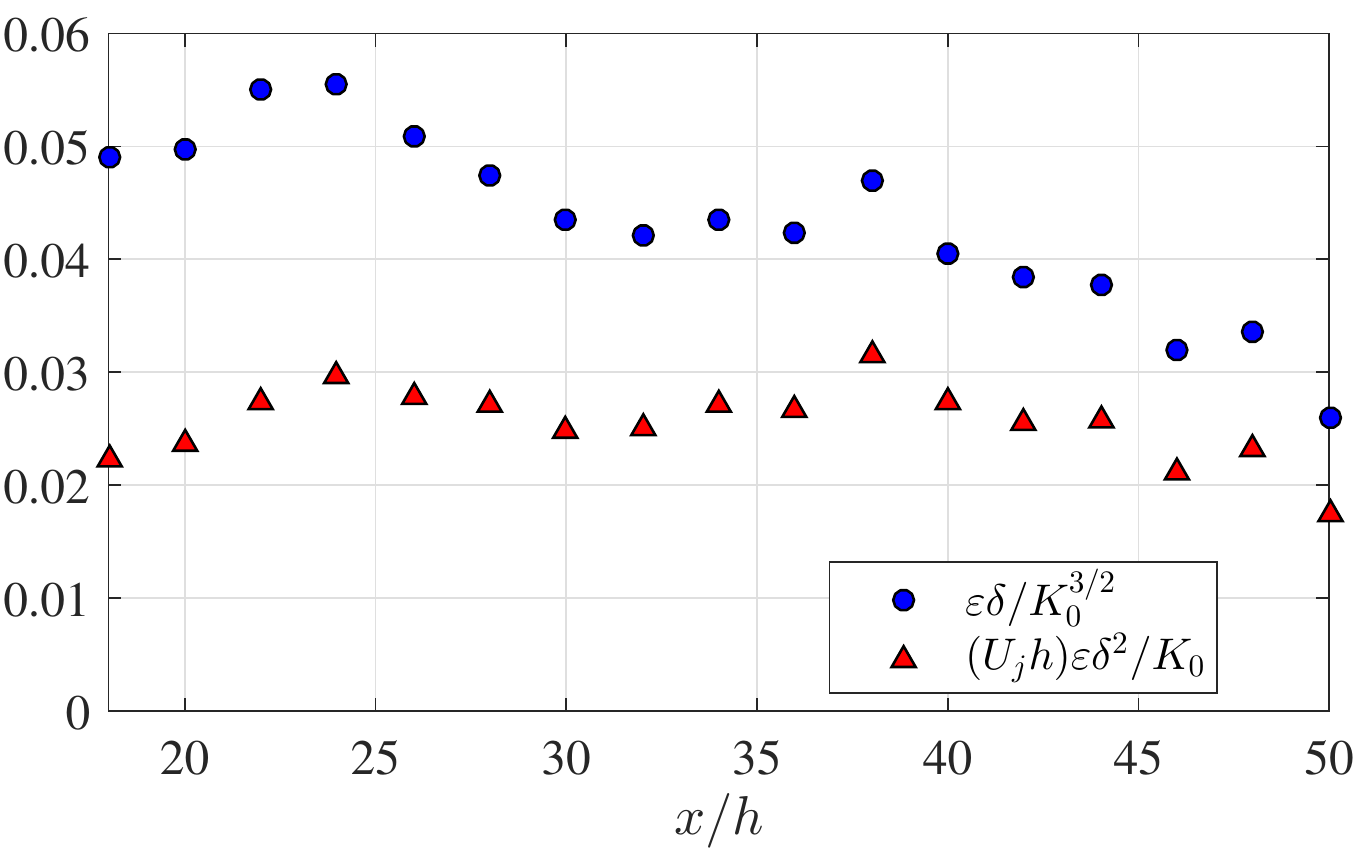}}
  \caption{(a) $v'/u'$ versus $x/h$ on the centreline. (b) Centreline
    dissipation coefficient $\epsilon_{iso}\delta/K_{0}^{3/2}$ (blue
    circles) and $U_{J}h\epsilon_{iso}\delta^{2}/K_{0}$ (red
    triangles) versus streamwise distance $x/h$ on the centreline.
    The inlet Reynolds number is $Re_G=20000$ and $K_0$ is estimated
    from the centreline $u'^2$ and $v'^2$ by using $K_{0}={1\over
      2}(u'^2+2v'^2)$.}
\label{fig:Ceps}
\end{figure}

The global/inlet Reynolds number $Re_{G}$ differs by a factor higher
than 2 between our data and the data of
\cite{Antonia1980}. Nevertheless, figure \ref{fig:Antonia} shows that
our data for $C_{\varepsilon} Re_{\delta}$ collapse quite closely with
those of \cite{Antonia1980} if $C_{\varepsilon} Re_{\delta}$ is
divided by $Re_{G}$ and is plotted as $C_{\varepsilon}
Re_{\delta}/Re_{G}$. This is what one would expect from eq.
(\ref{eq:neqScal}) with $m=1$ if the centreline $u'^{2}$ scales as
$K_0$. It is known that the ratio of the two transverse rms velocities
$v'/w'$ is about constant with $x/h$ in most free turbulent shear
flows \citep{Townsend}, and it has been confirmed for turbulent planar
jets that $v'^{2}/w'^{2}$ is indeed constant and in fact very close to
unity on the centreline (\citealp{bashir}, \citealp{gutmarkwygnanski1976}). With our XW measurements we accessed both $u'^{2}$ and
$v'^{2}$, and in figure \ref{fig:Ceps}(a) we plot $v'/u'$ versus $x/h$
on the centreline. The ratio $v'/u'$ remains about constant around the
value $v'/u' \approx 0.9$ in the range $14 \le x/h \le 54$. It is
therefore reasonable to expect the centreline $u'^{2}$ to
approximately scale as $K_0$ in the jet experiment of
\cite{Antonia1980} and the support for $C_{\varepsilon} \sim
Re_{\delta}^{-1}$ in Figure \ref{fig:Antonia} provided by their data
to actually be support for eq. (\ref{eq:neqScal}) with $m=1$ in the range
$40\le x/h\le 140$.

We now use our own data to test eq. (\ref{eq:neqScal}). We estimate the
centreline dissipation by calculating $\varepsilon_{iso}$ on the
centreline and the centreline turbulent kinetic energy as $K_{0} =
{1\over 2}(u'^{2} + 2v'^{2})$ given that we can assume $v'^{2}=w'^{2}$
on the centreline. We plot the results of these centreline
calculations in figure \ref{fig:Ceps}(b) as
$U_{J}h\varepsilon_{iso}\delta^{2}/K_{0}$ and as
$\varepsilon_{iso}\delta/K_{0}^{3/2}$, which would be constant in $x/h$
if eq. (\ref{eq:neqScal}) were to hold with $m=0$. It is quite clear from
figure \ref{fig:Ceps}(b) that $\varepsilon_{iso}\delta/K_{0}^{3/2}$ is,
overall, a decreasing function of $x/h$ and therefore not a constant
in the range $20\le x/h\le 50$. As can be seen in figure
\ref{fig:Antonia}, $(U_{J}h)\varepsilon_{iso}\delta^{2}/K_{0}$ cannot be
expected to be close to a constant in the very near field $x/h \le
20$, but figure \ref{fig:Ceps}(b) shows that it is definitely constant
in the range $20\le x/h\le 50$. These results support eq.
(\ref{eq:neqScal}) with $m=1$, i.e. the non-equilibrium dissipation
scalings. On the basis of our data and the data of \cite{Antonia1980}
we conclude that eq. (\ref{eq:neqScal}) with $m=1$ is well supported in
the region $20\le x/h\le 140$. Further studies will be needed in the
future to explore dissipation scalings in the region beyond $x/h =
140$.

\begin{figure} 
\centering
\includegraphics[scale=0.5]{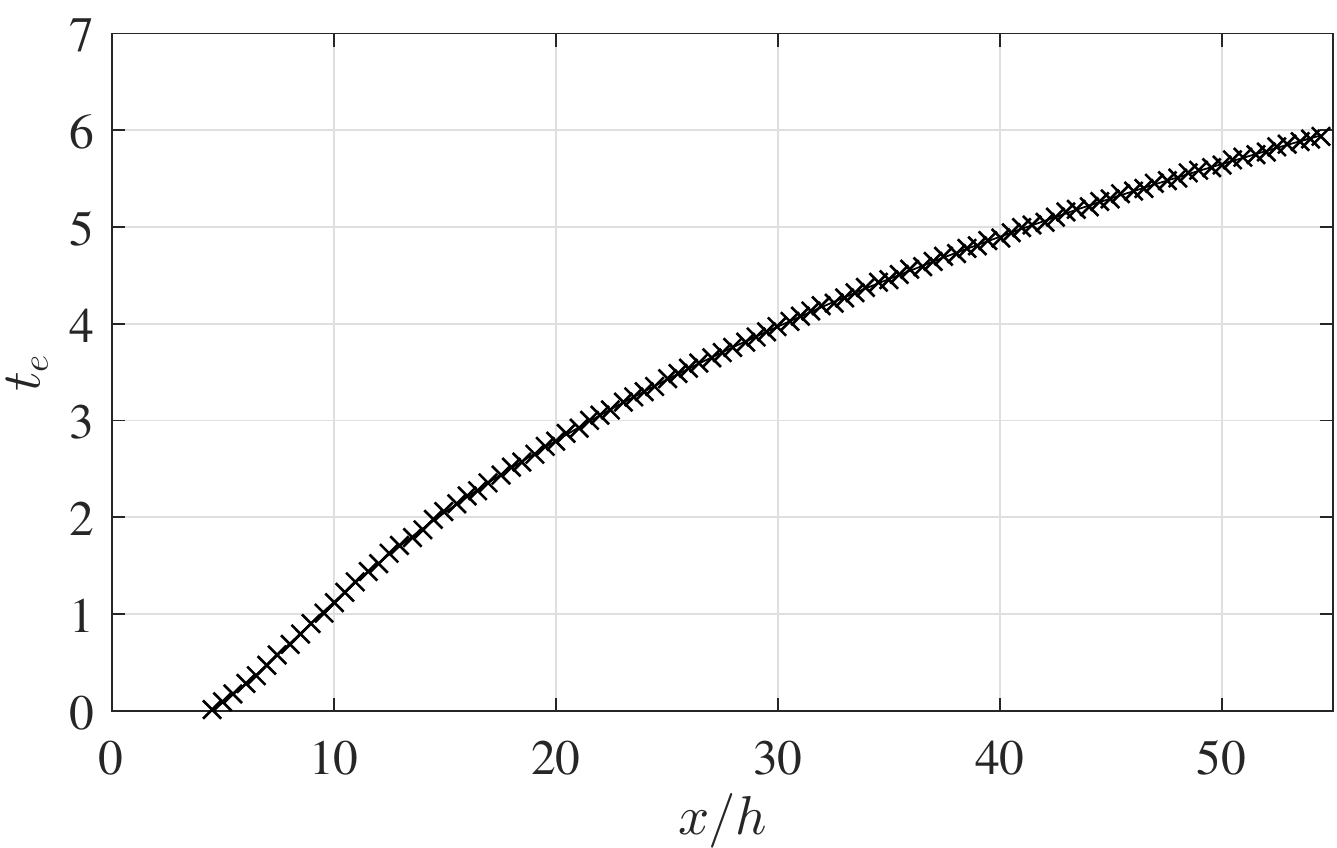}
\caption{Eddy turnover time $t_e$ as a function of the normalised
  streamwise distance $x/h$ on the centreline. The integral in
  eq. (\ref{eq:et}) has been calculated by setting $x_{1} = 5h$. The
  inlet Reynolds number is $Re_G=20000$.}
\label{fig:eddytt}
\end{figure}

Given that the turbulence dissipation is determined by the turbulence
cascade \citep{vassilicos2015, goto&vassilicos2016} which occurs over
about one turnover time, it is worth checking that the streamwise
range where eq. (\ref{eq:neqScal}) holds with $m=1$ is long enough for the
turbulence to evolve by at least a few turnover times over this
range. We therefore estimate the number of turnover times, $t_e$, on
the centreline as follows:
\begin{equation}
  t_{e} (x)= \frac{1}{u_0} \int_{x_1}^{x} \, \frac{u'}{L_{uu}} dx
 \label{eq:et}
\end{equation}
where the integral is computed along the centreline and $x_1$ is an
arbitrary starting point. Figure \ref{fig:eddytt}, where we plot
$t_{e}$ versus $x/h$, shows that the distance between $x=20h$ and
$x=54h$ corresponds to about three turnover times for the present paper's data.

\section{Self-similarity}  \label{sec:selfsimilarity}
\begin{figure} 
\centering
\subfloat[][]
{\includegraphics[width = 0.5\columnwidth]{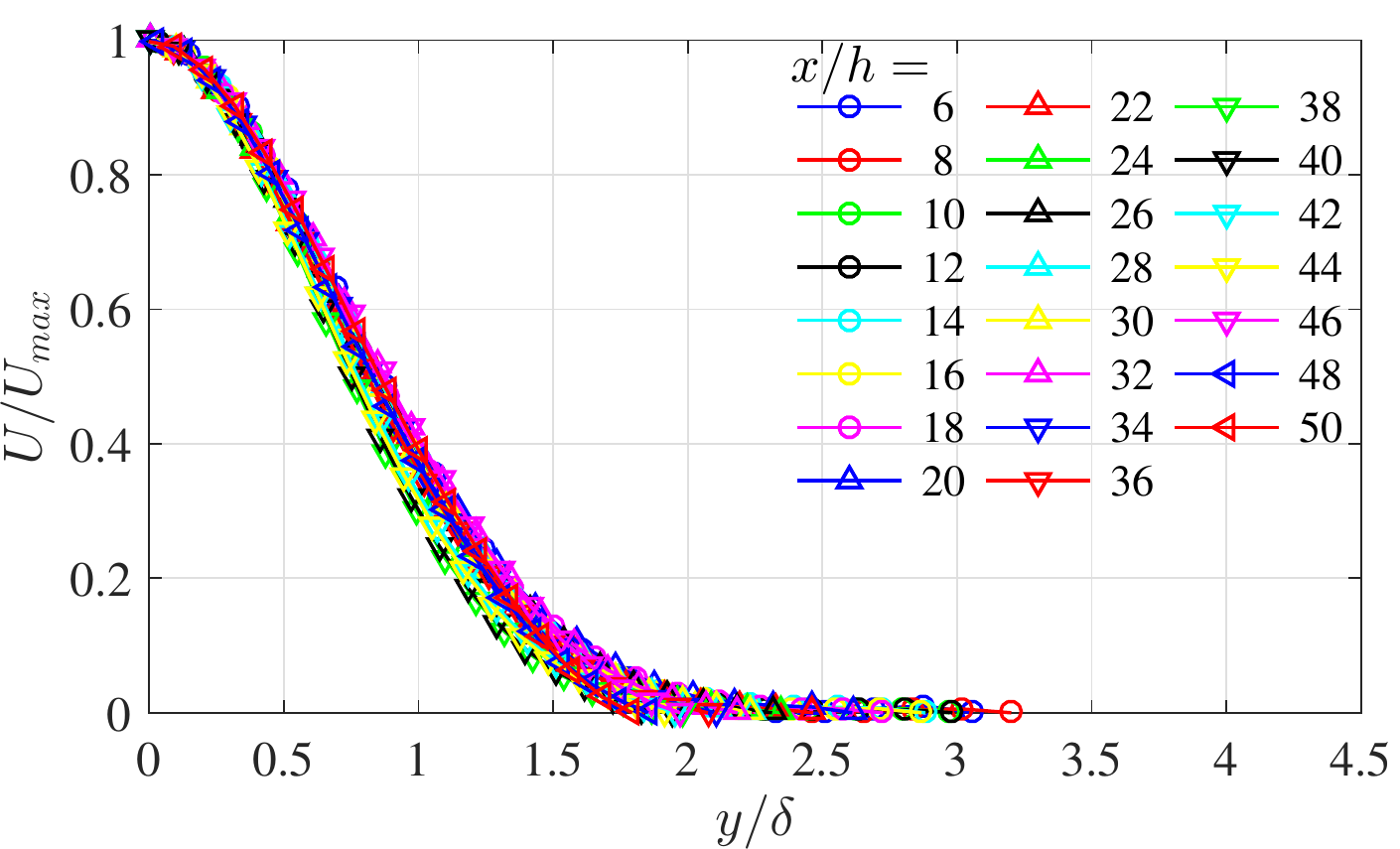}}
\subfloat[][]
{\includegraphics[width = 0.5\columnwidth]{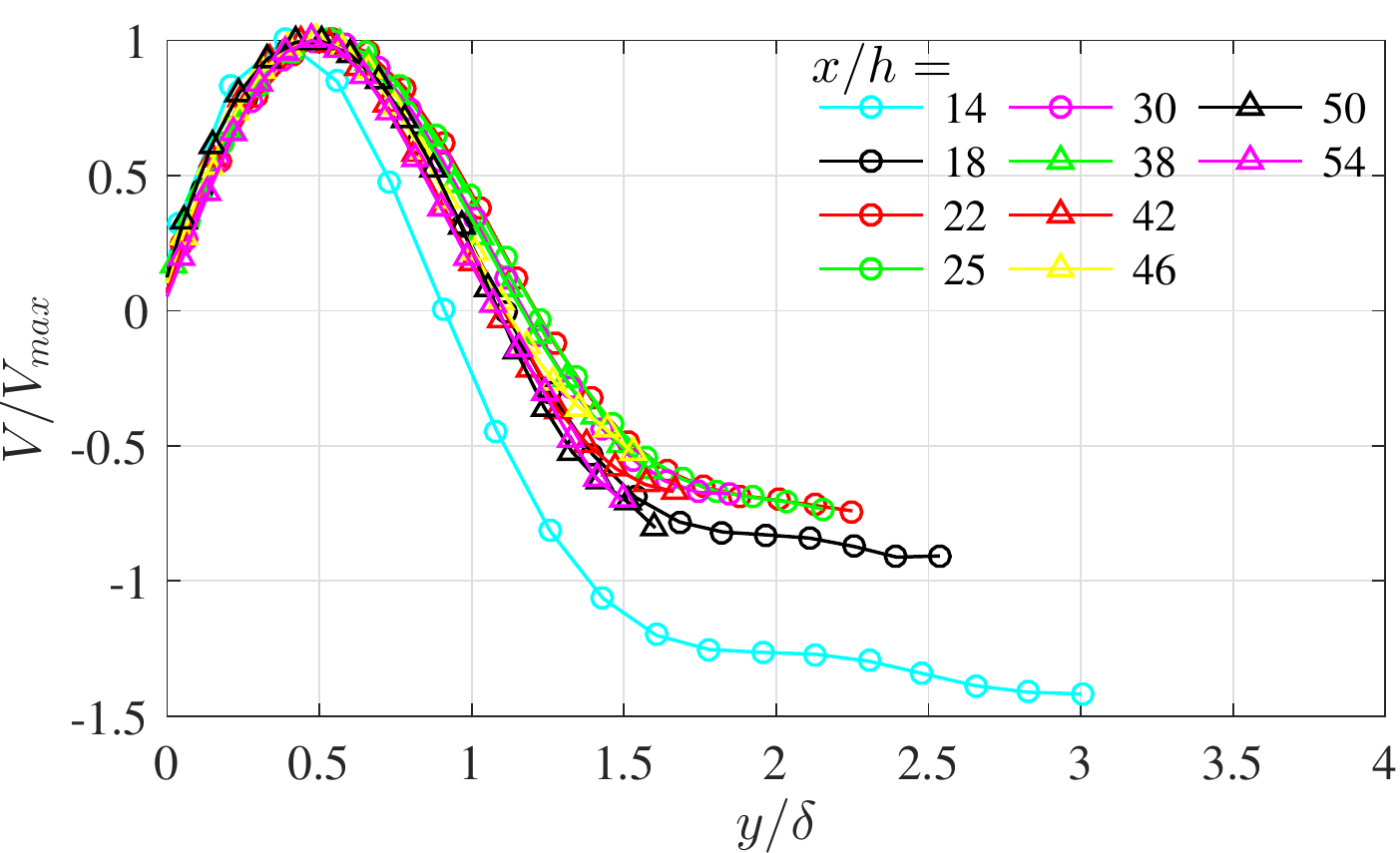}}\\
\subfloat[][]
{\includegraphics[width=0.5\columnwidth]{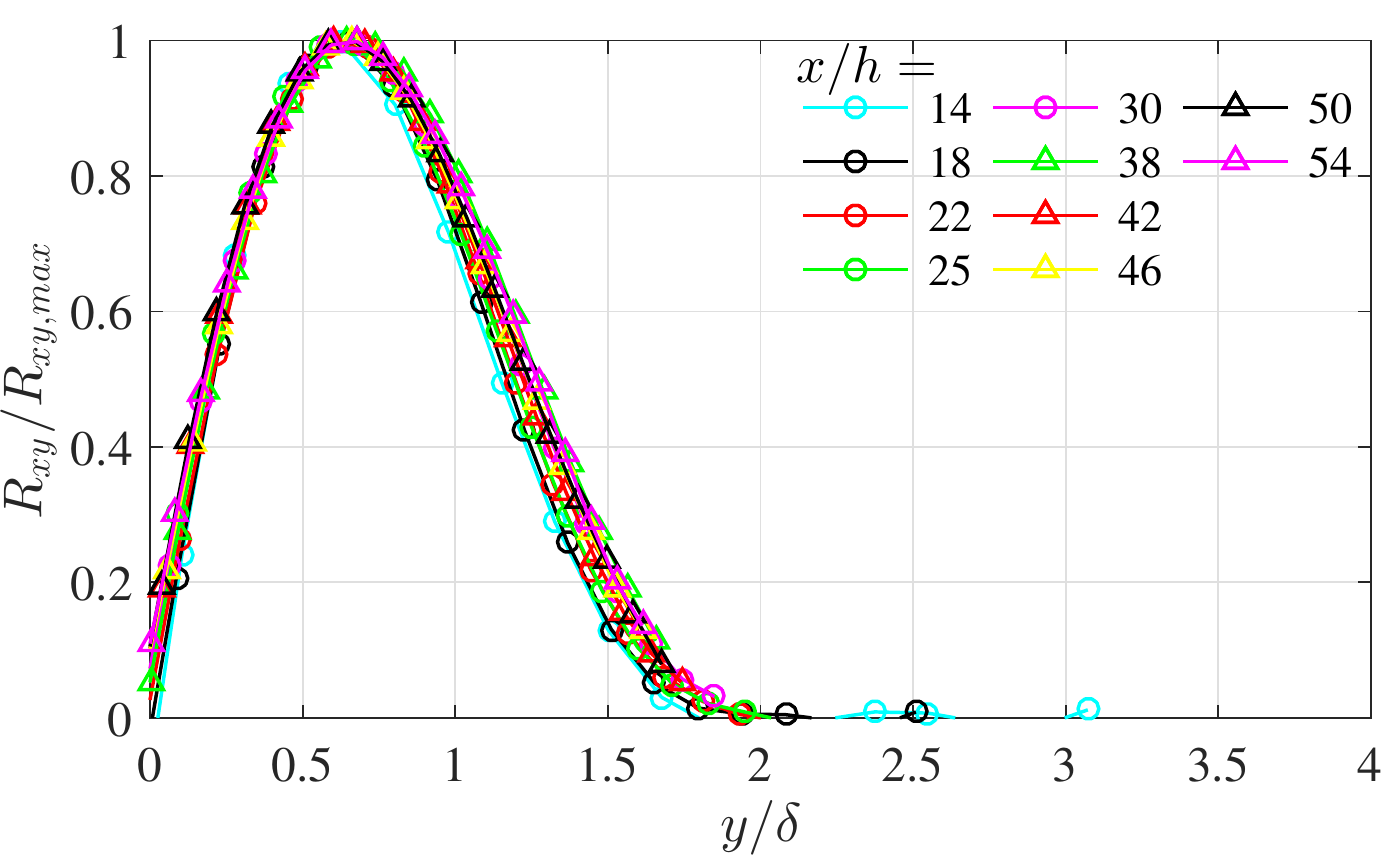}}
\subfloat[][]
{\includegraphics[width=0.5\columnwidth]{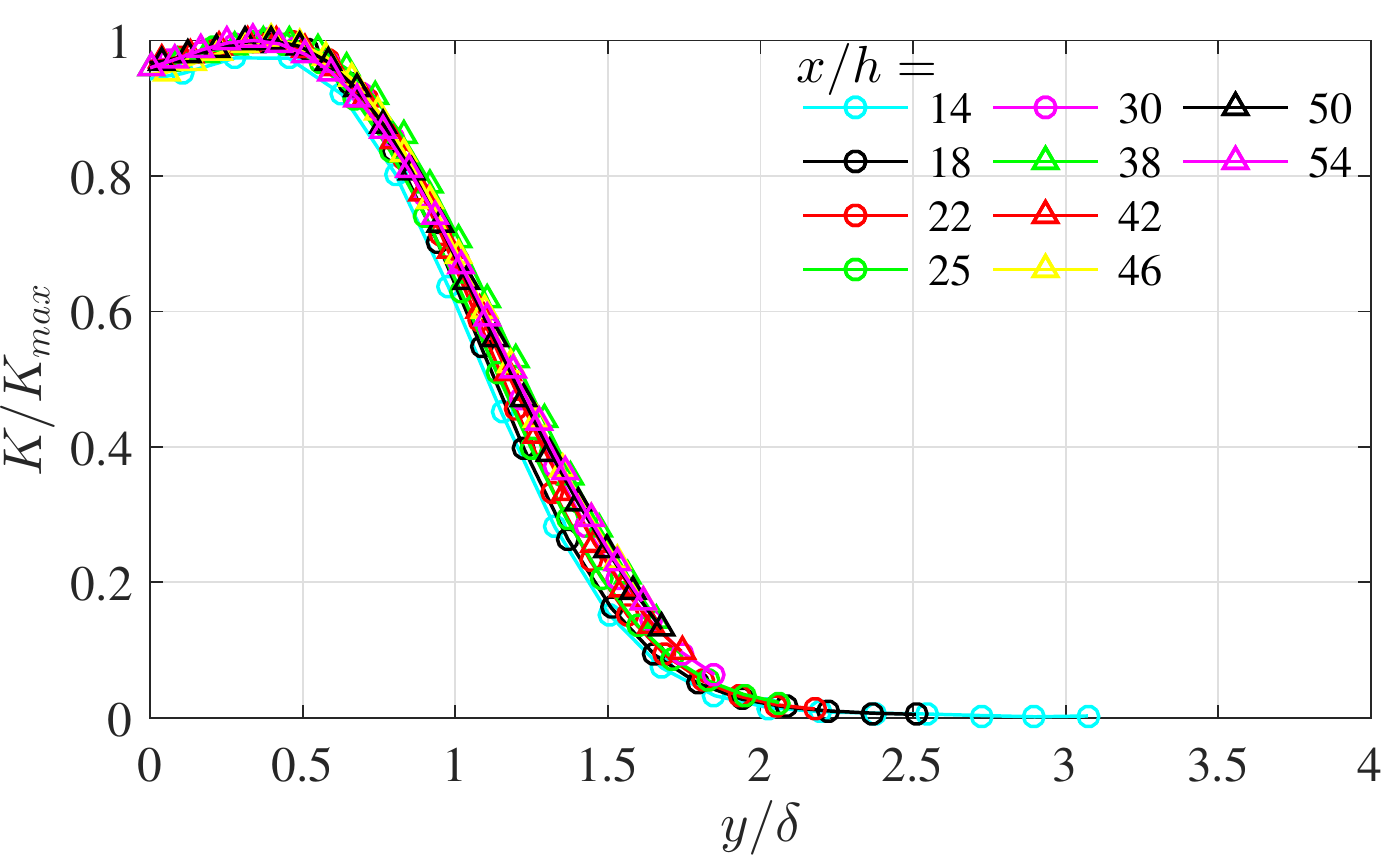}}\\
\caption{Self-similarity of (a) $U$, (b) $V$, (c) Reynolds shear
  stress and (d) turbulent kinetic energy. The data are normalised by
  their maximum values and plotted as functions of normalised
  cross-stream coordinate $y/\delta (x)$. Streamwise mean flow data
  (a) are shown starting from $x$=6$h$ with 2$h$ spacing. Data in
  (b), (c) and (d) are acquired using the XW in the range
  $x$=14$h$-54$h$. The inlet Reynolds number is $Re_G=20000$. }
\label{fig:sim}
\end{figure}

\begin{figure}
\centering 
\subfloat[][]
{\includegraphics[scale=0.45]{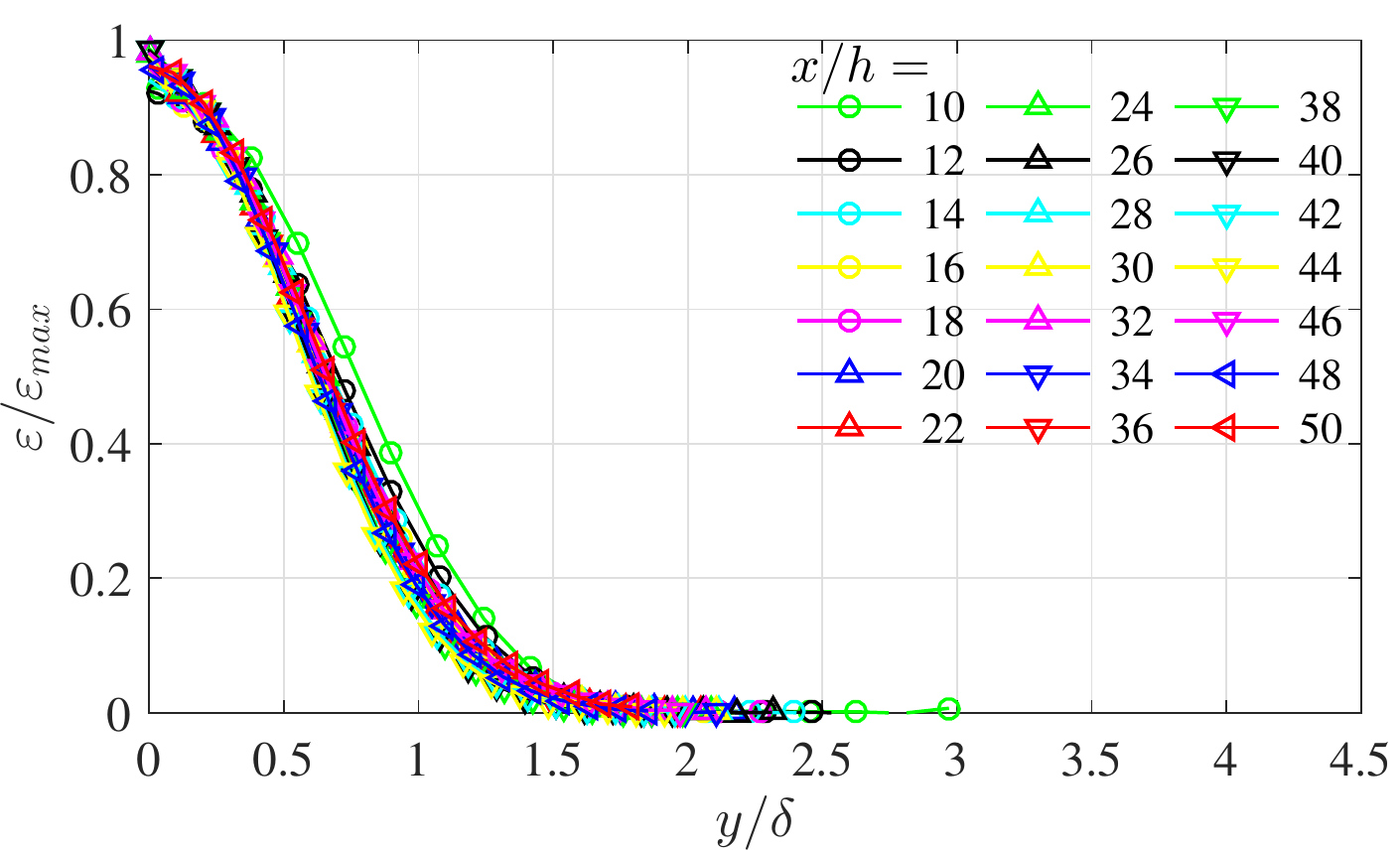}}
\subfloat[][]
{\includegraphics[scale=0.45]{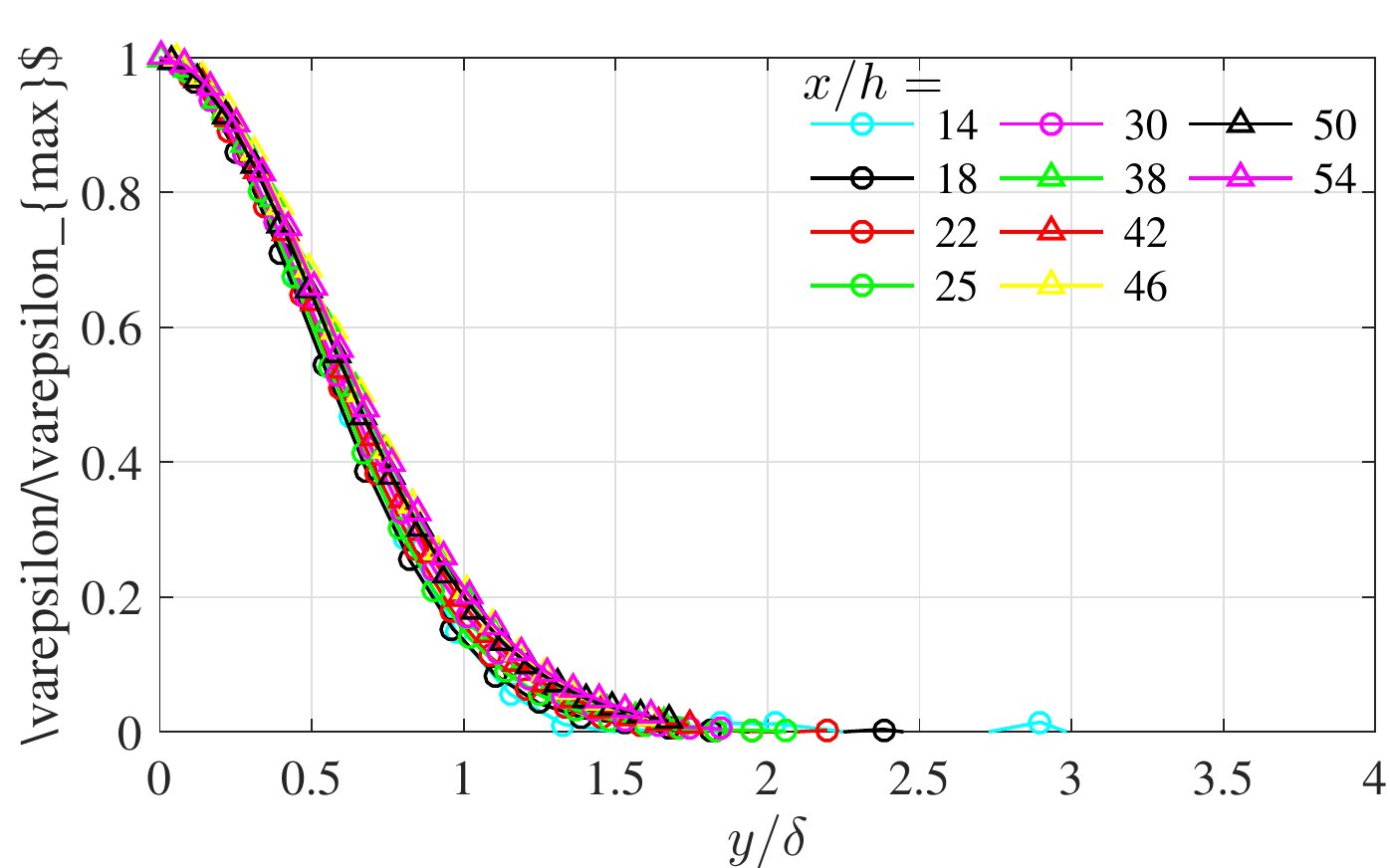}}\\
  \caption{Turbulent dissipation $\varepsilon$ normalised by its
    maximum value versus $y/\delta (x)$. (a) Obtained from SW data as
    $\varepsilon = \varepsilon_{ISO}$. (b) Obtained from XW data as
    $\varepsilon = \varepsilon_{XW}=\nu (3\overline{(\partial
      u/\partial x)^2}+6\overline{(\partial v/\partial x)^2})$.  The
    inlet Reynolds number is $Re_G=20000$.}
\label{fig:eps}
\end{figure}

Besides the turbulence dissipation scaling (eq. \ref{eq:neqScal}), the
other assumptions of the theory in section \ref{sec:theory} which are
accessible by our experiment are the self-similarity of the $U$
profile, which implies that the profiles of $V$ and $R_{xy}$ are also
self-similar, the self-similarities of the profiles of $K$ and
$\varepsilon$, and $K_{0}\sim R_{0}$ which is only needed if $m\not =
1$. In figures \ref{fig:sim} and \ref{fig:eps} we plot profiles of
these five quantities against the similarity coordinate $y/\delta$
normalised by the respective maximum values at each streamwise
distance $x$ ($V$ is actually normalized by the innermost
  maximum value, without loss of generality as long as the profile is
  self-similar). The turbulent kinetic energy in \ref{fig:sim}(d) is
estimated as $K={1\over 2}(u'^{2}+2v'^{2})$. The turbulence
dissipation is estimated as $\varepsilon_{iso}$ in figure
\ref{fig:eps}(a) and as $\varepsilon_{XW}$ in figure \ref{fig:eps}(b).
The results support self-similarity of these profiles from $x=18h$ to
the furthermost streamwise distance of our measurements, i.e. $x=54h$:
$\varepsilon_{iso}$ and $\varepsilon_{XW}$ appear equally self-similar in
this range.

In summary, our data support the self-similarities of $U$, $V$,
$R_{xy}$, $K$ and $\varepsilon$ as well as the dissipation scaling
(eq. \ref{eq:neqScal}) with $m=1$ in the range $20\le x/h\le 50$ (perhaps
even $18\le x/h\le 50$). If we assume that the self-similarities of
the $U$, $V$, $R_{xy}$, $K$ and $\varepsilon$ profiles extend further downstream,
then the validity of the non-equilibrium self-similar theory of high
Reynolds number turbulent planar jets may reach till at least $x/h
=140$ given the results of sections \ref{sec:anton} and
\ref{sec:dissipation}.

\begin{figure}
\centering
{\includegraphics[scale=0.45]{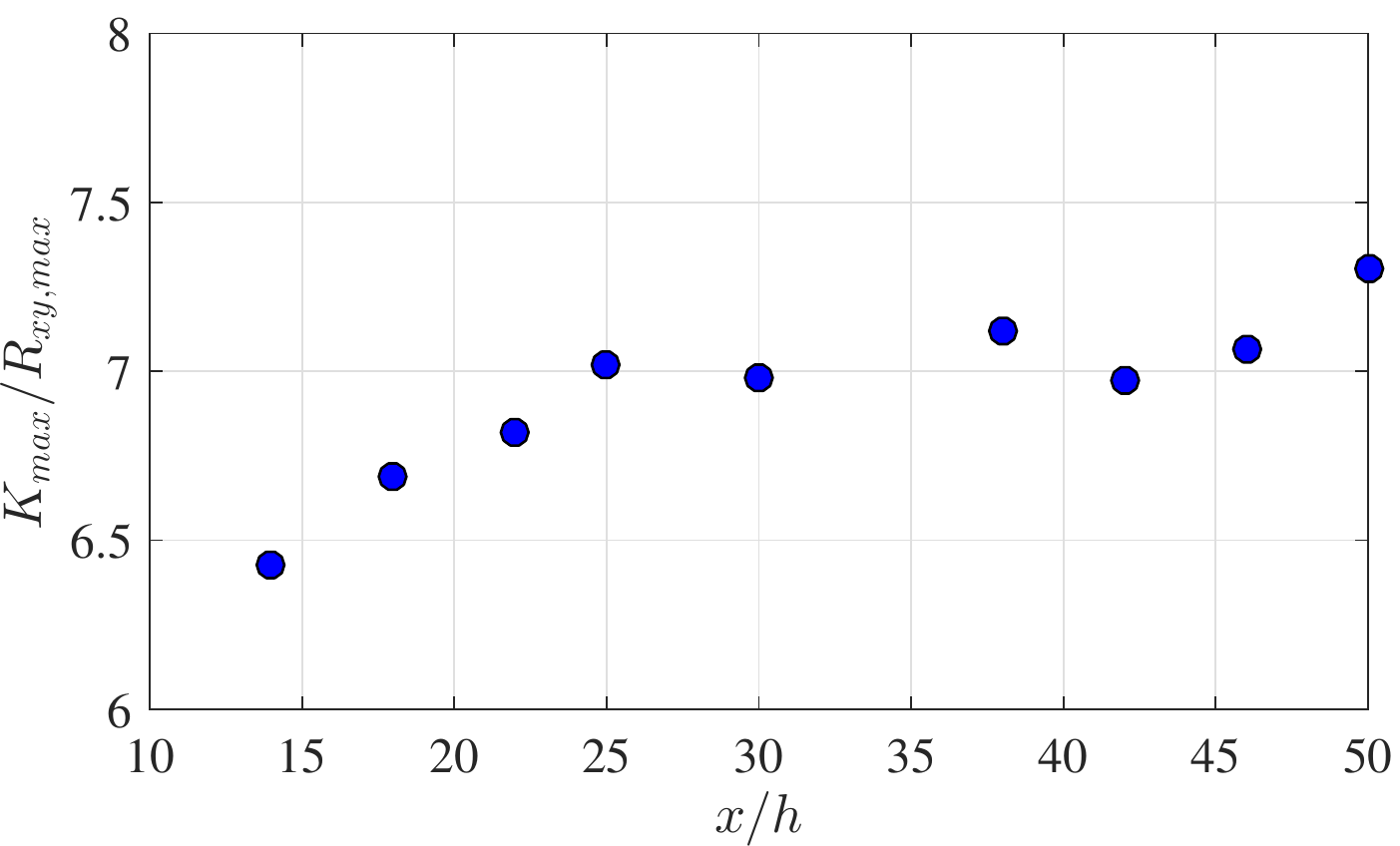}}
  \caption{Maximum of $K$ divided by the maximum of $R_{xy}$ (both
    maxima over all values of $y$ at a given $x$). This ratio
    $K_{max}/R_{xy,max}$ is plotted as a function of the streamwise
    distance $x/h$. Data obtained at $Re_G=20,000$.}
\label{fig:K0R0}
\end{figure}

As a final comment for this section, the theory of section
\ref{sec:theory} is inconclusive if $m\not = 1$, in which case eqns.
(\ref{eq:scalingEQ1})-(\ref{eq:scalingEQ2})-(\ref{eq:scalingEQ3}) are
obtained by making the additional assumption $K_{0} \sim R_{0}$.  In
Figure \ref{fig:K0R0} the ratio of the maximum value of $K$ to the
maximum value of the Reynolds shear stress $R_{xy}$
suggests that this extra condition is satisfied for $x/h \ge 25$.

\section{Scalings}  \label{sec:scalings}
\begin{figure}
\centering \subfloat[][]
           {\includegraphics[scale=0.45]{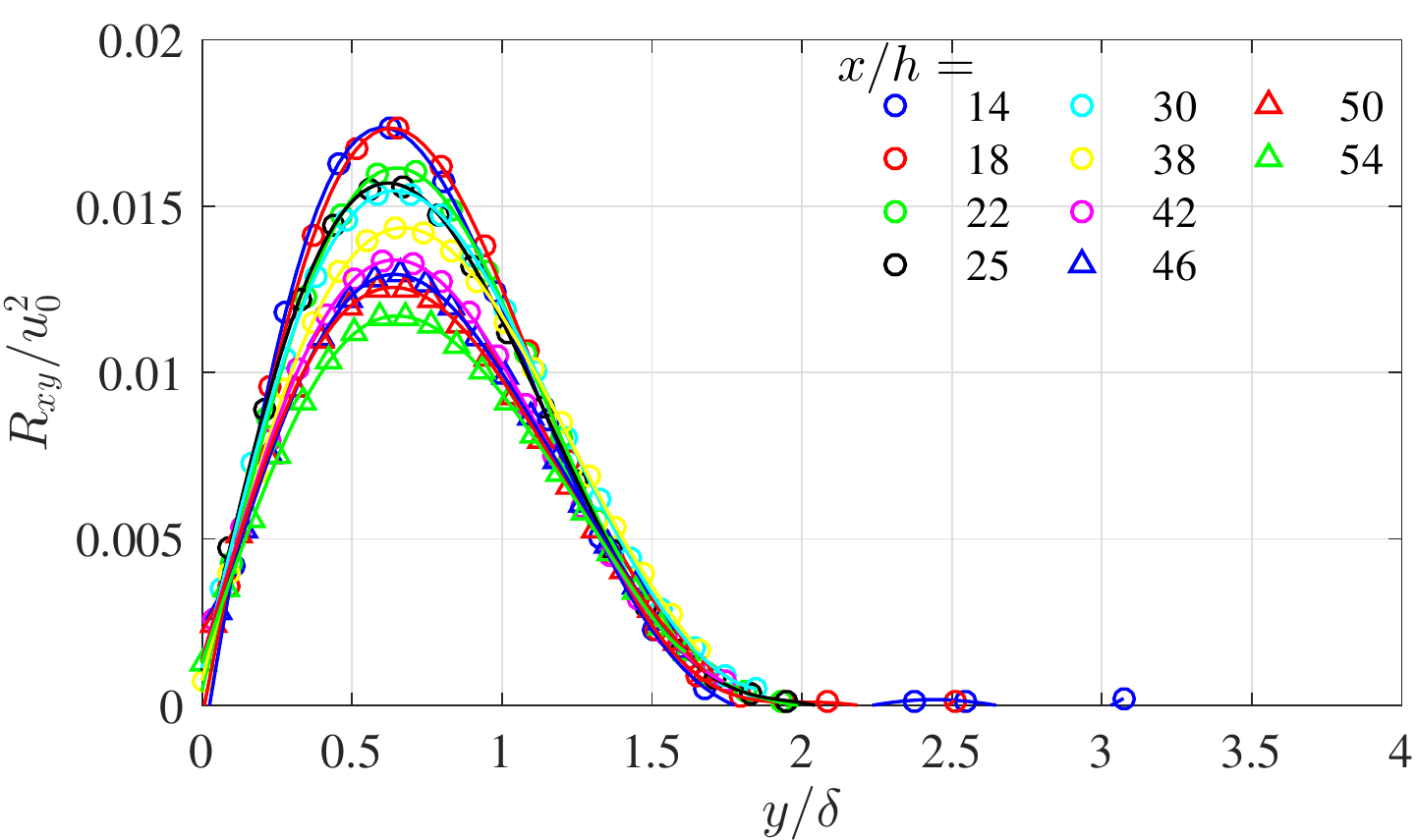}}
           \subfloat[][]
                    {\includegraphics[scale=0.45]{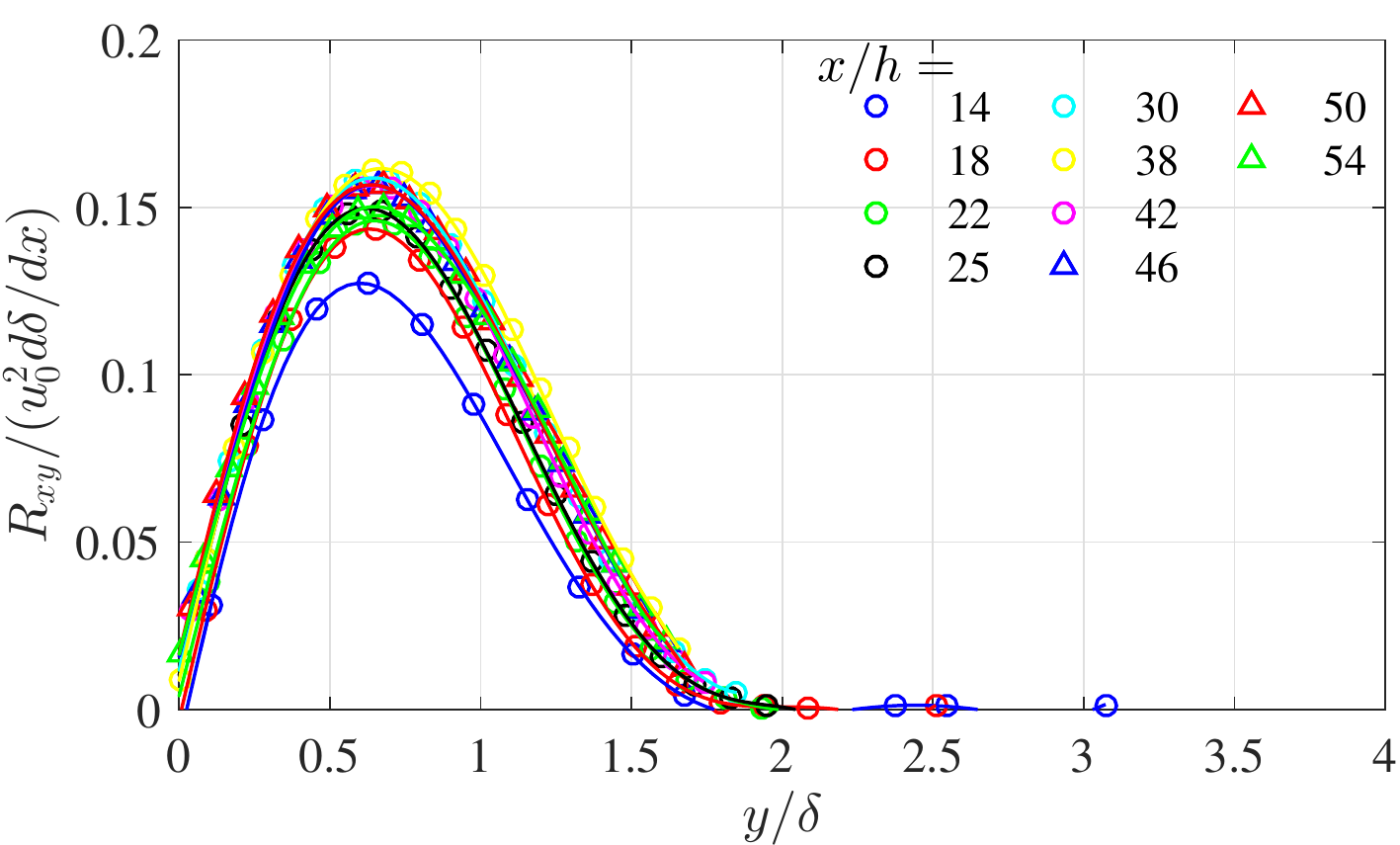}}\\
  \caption{Reynolds shear stress $R_{xy}$ normalized by (a)
    $u_{0}^{2}$ and (b) by $u_0^2 \frac{d\delta}{dx}$ as in the George
    scaling $R_0\sim u_0^2 \frac{d\delta}{dx}$. Values of
    $\frac{d\delta}{dx}$ are obtained by fitting our $\delta (x)$ data
    with a power law and then differentiating the resulting fit. The
    profiles are plotted against $y/\delta(x)$. Data acquired for
    inlet Reynolds number $Re_G=20000$. }
\label{fig:ReynStress}
\end{figure}

Having found experimental support for the self-similarity of $U$ and
the self-similar behaviours of $R_{xy}$ and $V$ that it implies, we
now turn our attention to the George scaling $R_{0} \sim u_{0}^{2}
d\delta /dx$ \citep{george1989}. This scaling is also a consequence of
the self-similarity of $U$ and it differs in general from the scaling
$R_{0} \sim u_{0}^{2}$ that one finds in various textbooks
(e.g. \citealp{tennekes&lumley}). The theory of section
\ref{sec:theory} makes it clear, however, that one particular instance
where $d\delta/dx$ is constant and these two $R_0$ scalings are the
same is when $K_{0}$ is proportional to $R_{0}$ and the centreline
dissipation scales as $D_{0} \sim (Re_{G}/Re_{\delta})^{m}
K_{0}^{3/2}/\delta$ with $m=0$. In other words, if the collapse of
$R_{xy}/u_{0}^{2}$ profiles differs from the collapse of
$R_{xy}/(u_{0}^{2}d\delta/dx)$ profiles (both versus $y/\delta$) and
if $K_{0}\sim R_{0}$, then $m \neq 0$. Figure \ref{fig:ReynStress}
shows clearly that $R_0$ does not scale with $u_{0}^{2}$, and also
provides support for $R_{0} \sim u_{0}^{2} d\delta /dx$ in the range
$18 \le x/h \le 54$. Given figure \ref{fig:K0R0} which suggests that
$K_{0}\sim R_{0}$ holds for $x/h \ge 25$, this is additional support
for $m \neq 0$, in agreement with our conclusion concerning
$\varepsilon$ in section \ref{sec:dissipation}.

\begin{figure}
\centering \subfloat[][]
           {\includegraphics[scale=0.45]{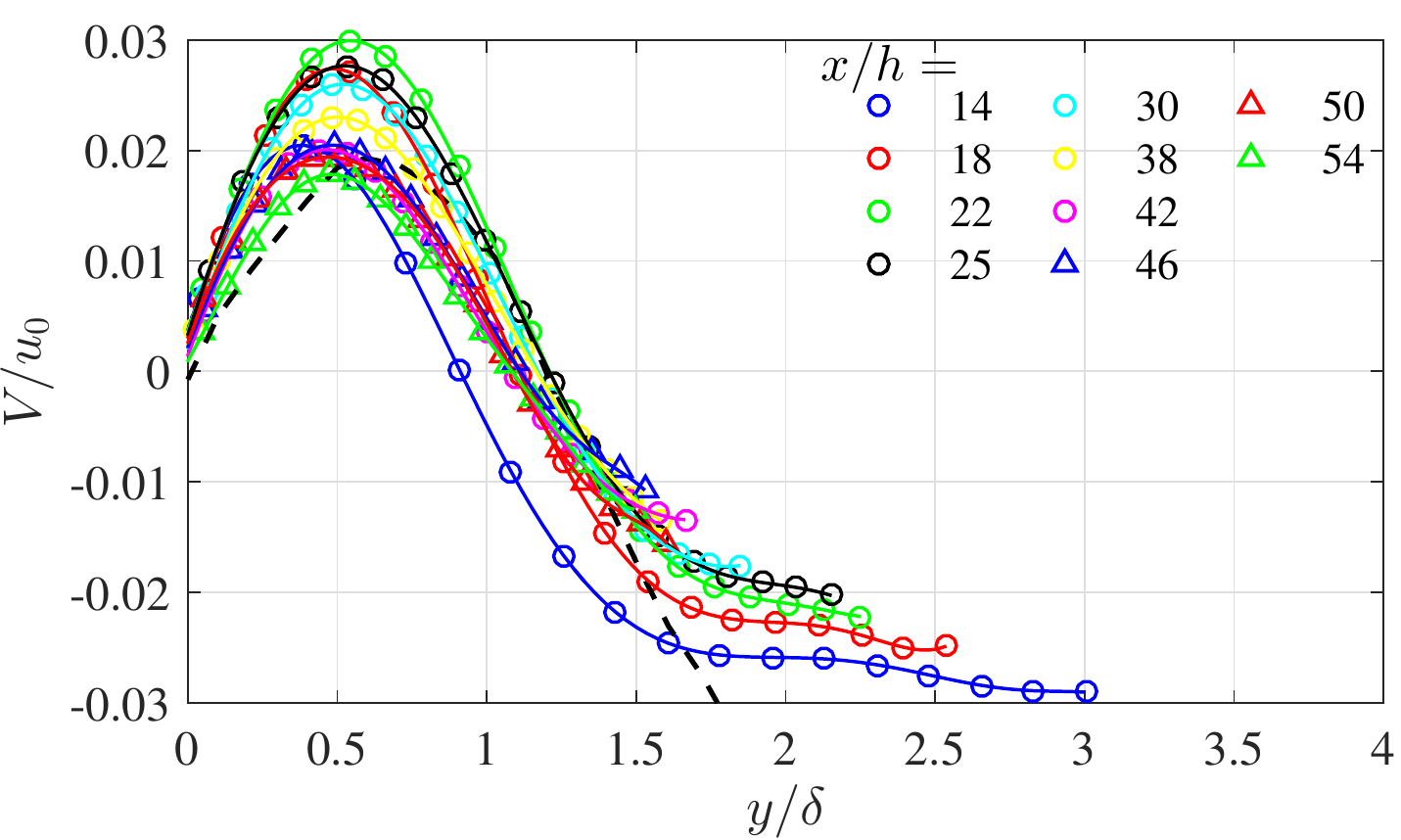}}
           \subfloat[][]
                    {\includegraphics[scale=0.45]{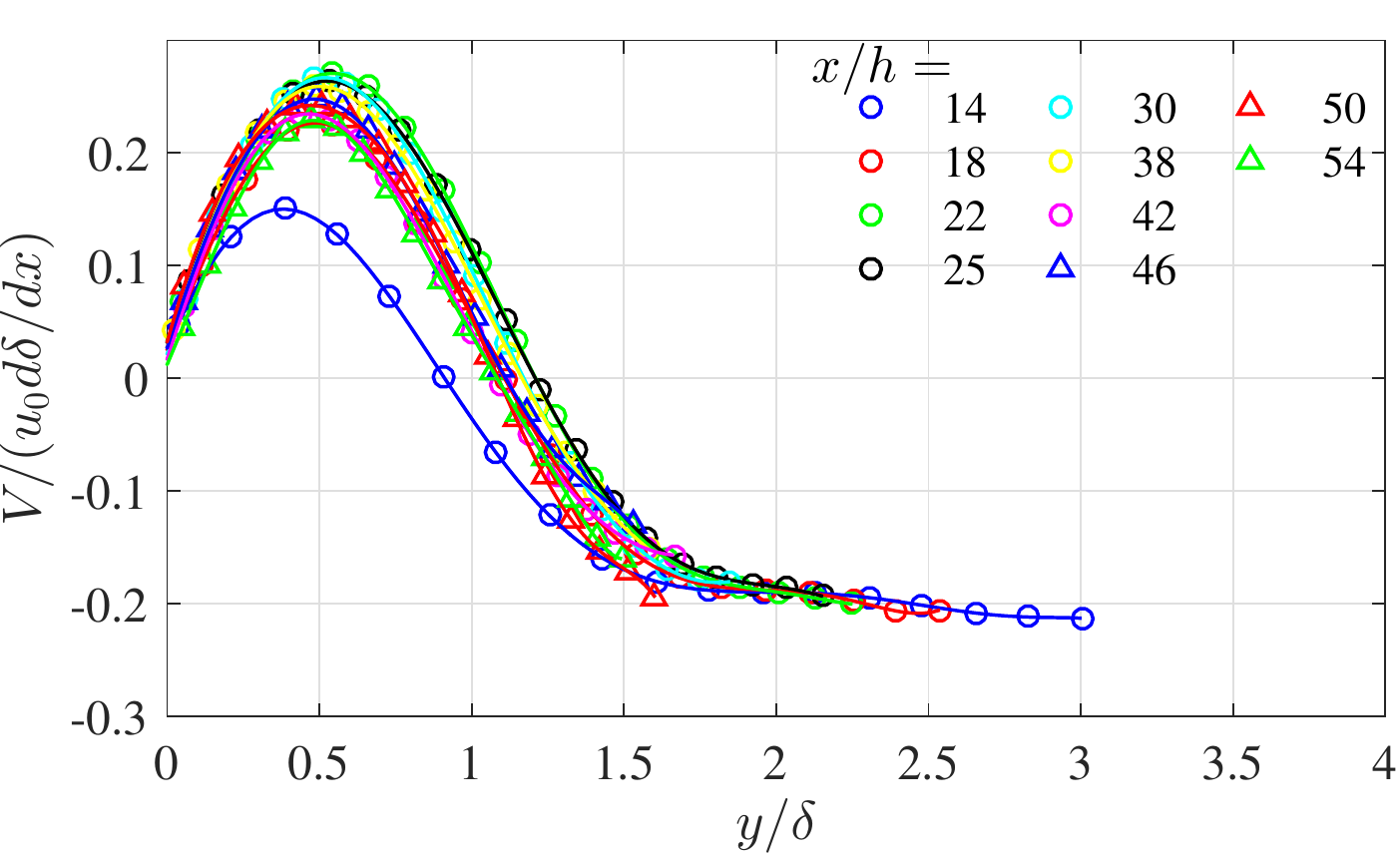}}\\
  \caption{Mean cross-stream velocity $V$ normalized by (a) $u_{0}$
    and (b) $u_0 \frac{d\delta}{dx}$. Values of $\frac{d\delta}{dx}$
    are obtained by fitting our $\delta (x)$ data with a power law and
    then differentiating the resulting fit. The profiles are plotted
    against $y/\delta(x)$. Data acquired for inlet Reynolds number
    $Re_G=20000$. In (a) the dashed line is representative of the mean
    cross-stream velocity profile from the numerical simulation of
    \cite{stanley} at $x/h=11$ and $Re_G=3000$.}
\label{fig:CrossVel}
\end{figure}

Another important implication of the theory involves the mean
cross-stream velocity $V(x,y)$ which is self-similar with $v_0(x) \sim
u_0(x) d\delta/dx$, see eq. (\ref{eq:Vform}) and
eq. (\ref{eq:v0u0del}), if $U(x,y)$ is self-similar. As already
mentioned, $d\delta/dx$ is different depending on the value of $m$:
$d\delta/dx = const$ for $m=0$ but $d\delta/dx \sim (x-x_0)^{-1/3}$
for $m=1$. In figure \ref{fig:CrossVel}(a) we plot the mean
cross-stream velocity profiles scaled according to the classical
dissipation formula corresponding to $m=0$ and in figure
\ref{fig:CrossVel}(b) we plot the same profiles but scaled according
to the non-equilibrium dissipation formula corresponding to $m=1$. The
data collapse significantly better in figure \ref{fig:CrossVel}(b)
than in figure \ref{fig:CrossVel}(a) in the range $18 \leq x/h \leq
54$.  (In \ref{fig:CrossVel}(a) we also plot DNS data from
\citealp{stanley} for comparison, see dashed line.)

The last scalings of the theory in section \ref{sec:theory} to be
checked are those of the self-similar mean flow profiles, namely the
dependencies on $x$ of $u_0$ and $\delta$, which can also help us
assess $v_0(x) \sim u_0(x) d\delta/dx$ more closely. The theoretical
predictions for $u_{0}$ and $\delta$ when the dissipation scaling is
$D_{0} \sim (Re_{G}/Re_{\delta}) K_{0}^{3/2}/\delta$ (i.e. $m=1$ in
eq. (\ref{eq:neqScal}) as evidenced by our data and the data of
\citealp{Antonia1980}) are given by equations
(\ref{eq:scalingEQ1})-(\ref{eq:scalingEQ2}) with $a=1/3$. These
predictions are based on the self-similarity behaviours of $U$,
$\varepsilon$ and $K$ which are supported by the experimental results
in the previous section. We therefore expect our data to be consistent
with eq.  (\ref{eq:scalingEQ1})-\ref{eq:scalingEQ2}) and $a=1/3$.

\begin{figure} 
\centering
    {\includegraphics[width=0.7\columnwidth]{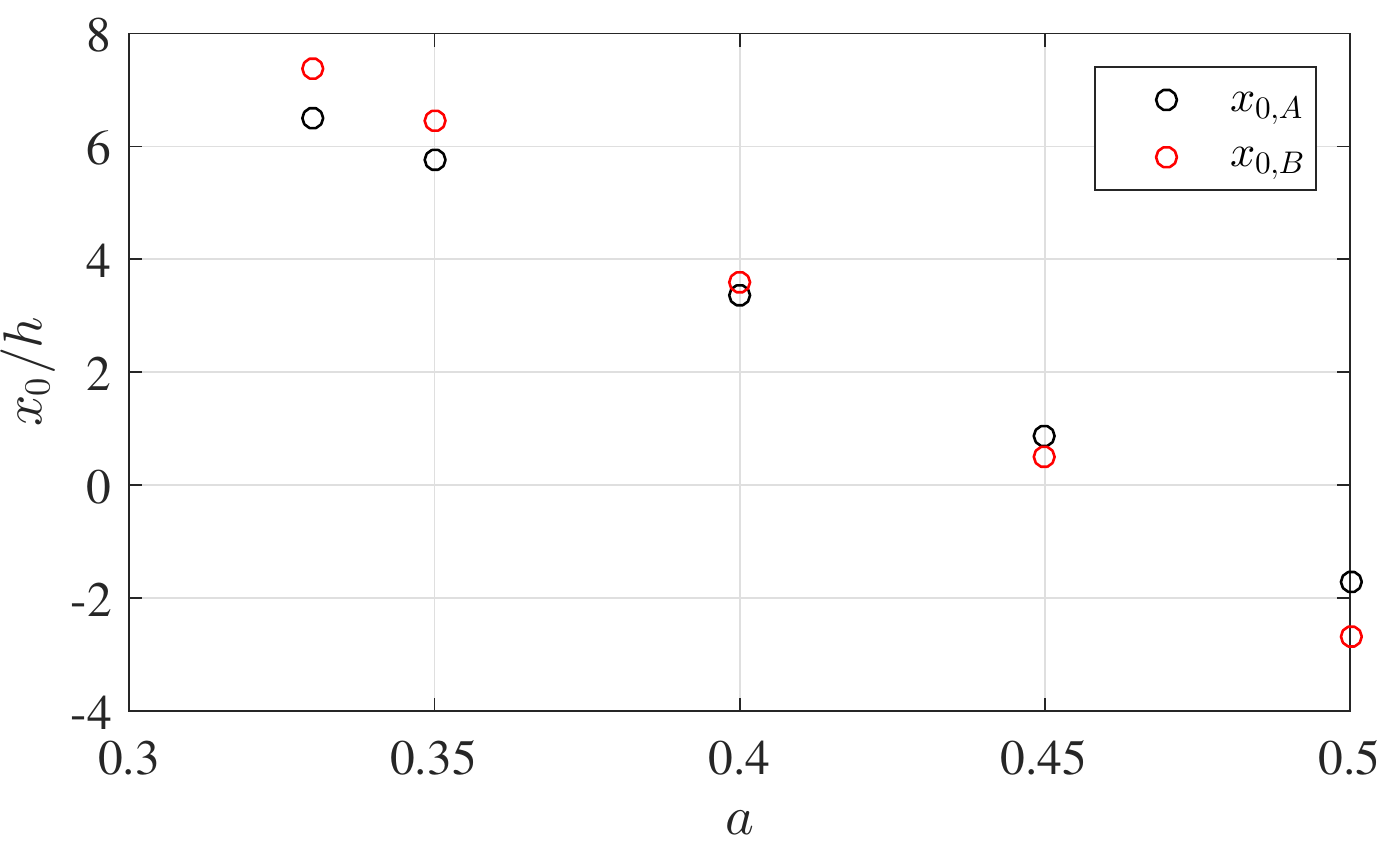}}
\caption{Virtual origins obtained by applying different exponents to
  the power laws eq. (\ref{eq:scalingEQ1}) (for $x_{0,A}$) and
  eq. (\ref{eq:scalingEQ2}) (for $x_{0,B}$) with $1/3 \le a \le 1/2$. From
  our data, acquired for inlet Reynolds number $Re_G=20000$.}
\label{fig:ourV00X}
\end{figure}

For consistency with the method followed in section \ref{sec:origin},
we first seek the virtual origins $x_{0,A}$ and $x_{0,B}$ which,
respectively, best fit our $u_{0}(x)$ and $\delta (x)$ data in the
range $18 \le x/h \le 50$. We limit ourselves to this range because
our SW measurements do not extend downstream of $x/h =50$ and because
the non-equilibrium dissipation scaling $D_{0} \sim
(Re_{G}/Re_{\delta}) K_{0}^{3/2}/\delta$ holds downstream of about
$x/h=18$ or $20$. In figure \ref{fig:ourV00X} we plot the resulting
$x_{0,A}$ and $x_{0,B}$ for different values of the exponent
$a$. Unfortunately, the values of $x_{0,A}$ and $x_{0,B}$ are quite
close to each other for all exponents $a$ in the range $1/3 \le a \le
1/2$ and there is no clear way to chose a value of this exponent on
such a basis. All exponents $a$ in this range can and do return good
fits of our $u_{0}(x)$ and $\delta (x)$ data with a choice of virtual
origins $x_{0,A}$ and $x_{0,B}$ that are quite close to each other in every case.

\begin{figure}
\centering
\subfloat[][]
   {\includegraphics[scale=0.4]{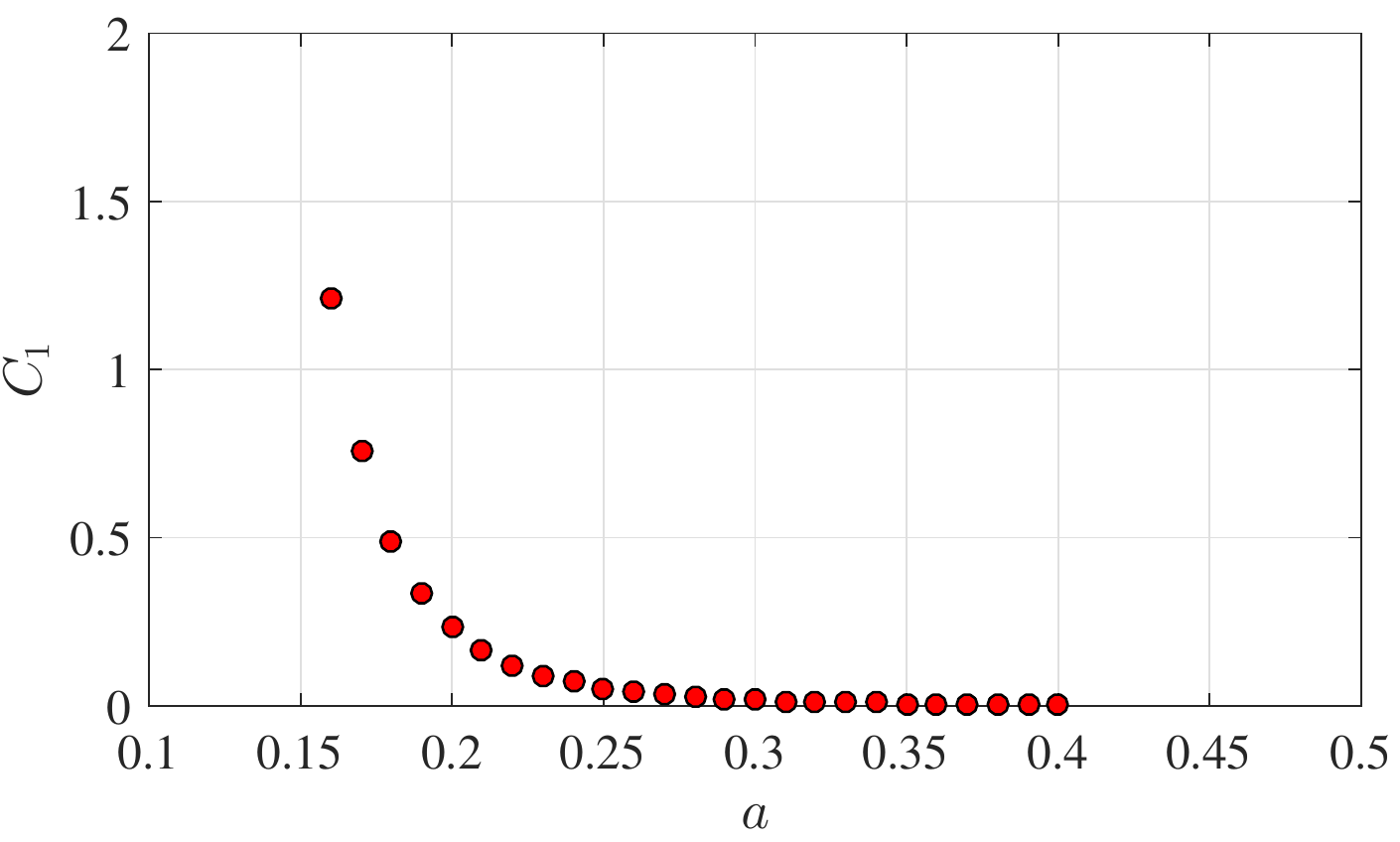}}
\subfloat[][]
   {\includegraphics[scale=0.4]{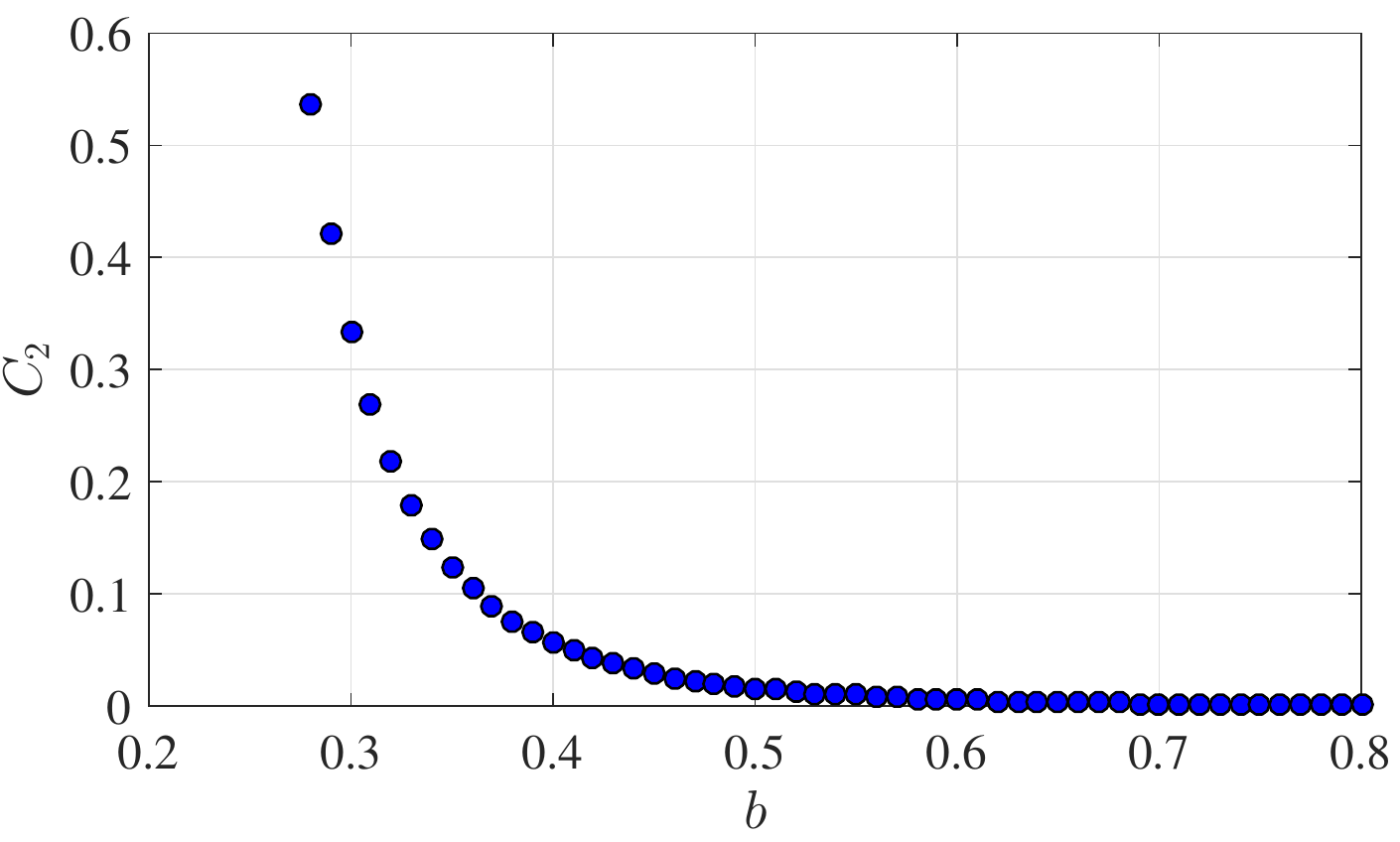}}\\
\caption{$C_1$ and $C_2$ values obtained from linear fits of the
  derivatives $\frac{d}{dx}(u_0(x)/U_j)^{-1/a}$ and
  $\frac{d}{dx}(\delta(x)/h)^{1/b}$. From our data, acquired for inlet
  Reynolds number $Re_G=20000$. }
\label{fig:C1C2}
\end{figure}

\begin{figure}
\centering \subfloat[][]
           {\includegraphics[scale=0.4]{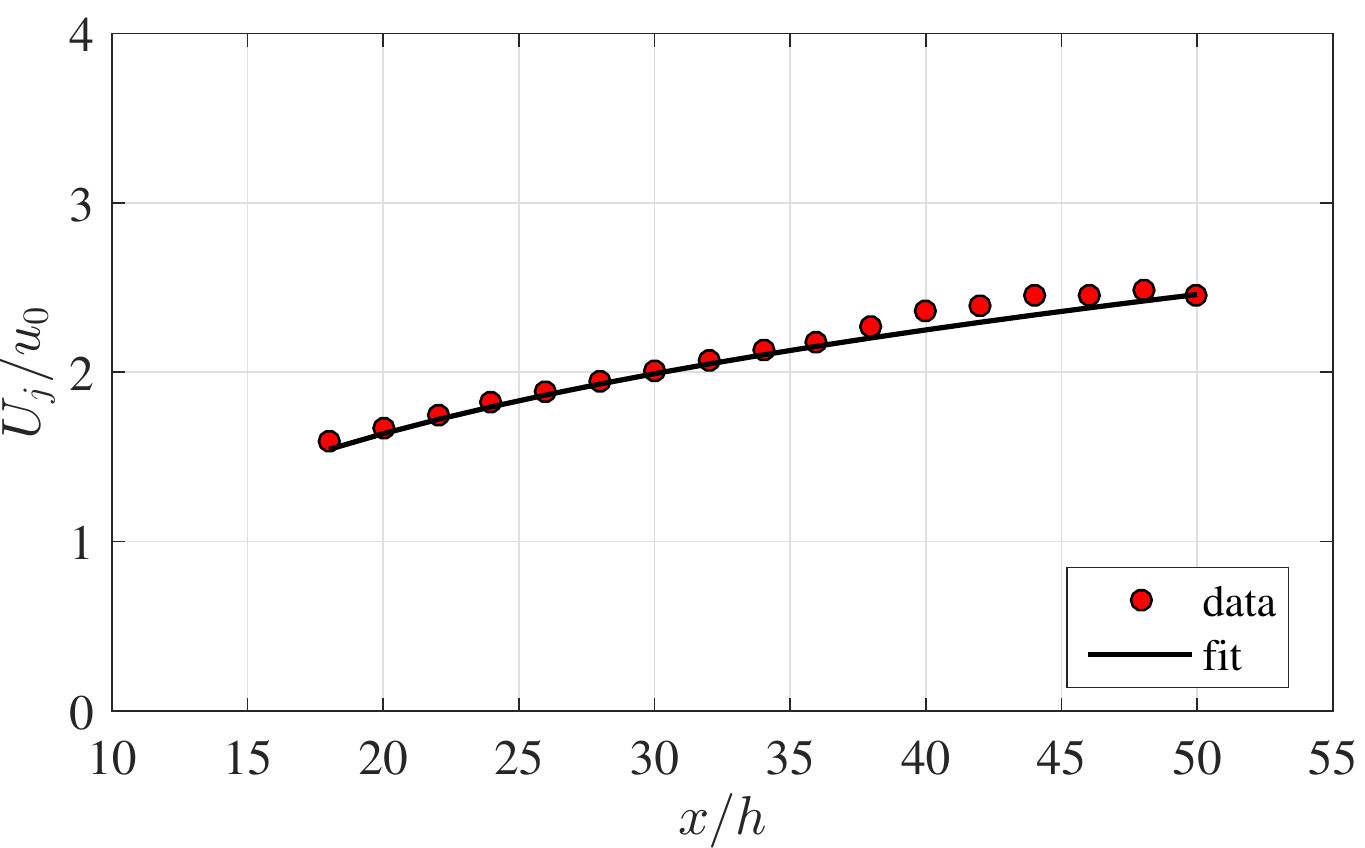}}
           \subfloat[][]
                    {\includegraphics[scale=0.4]{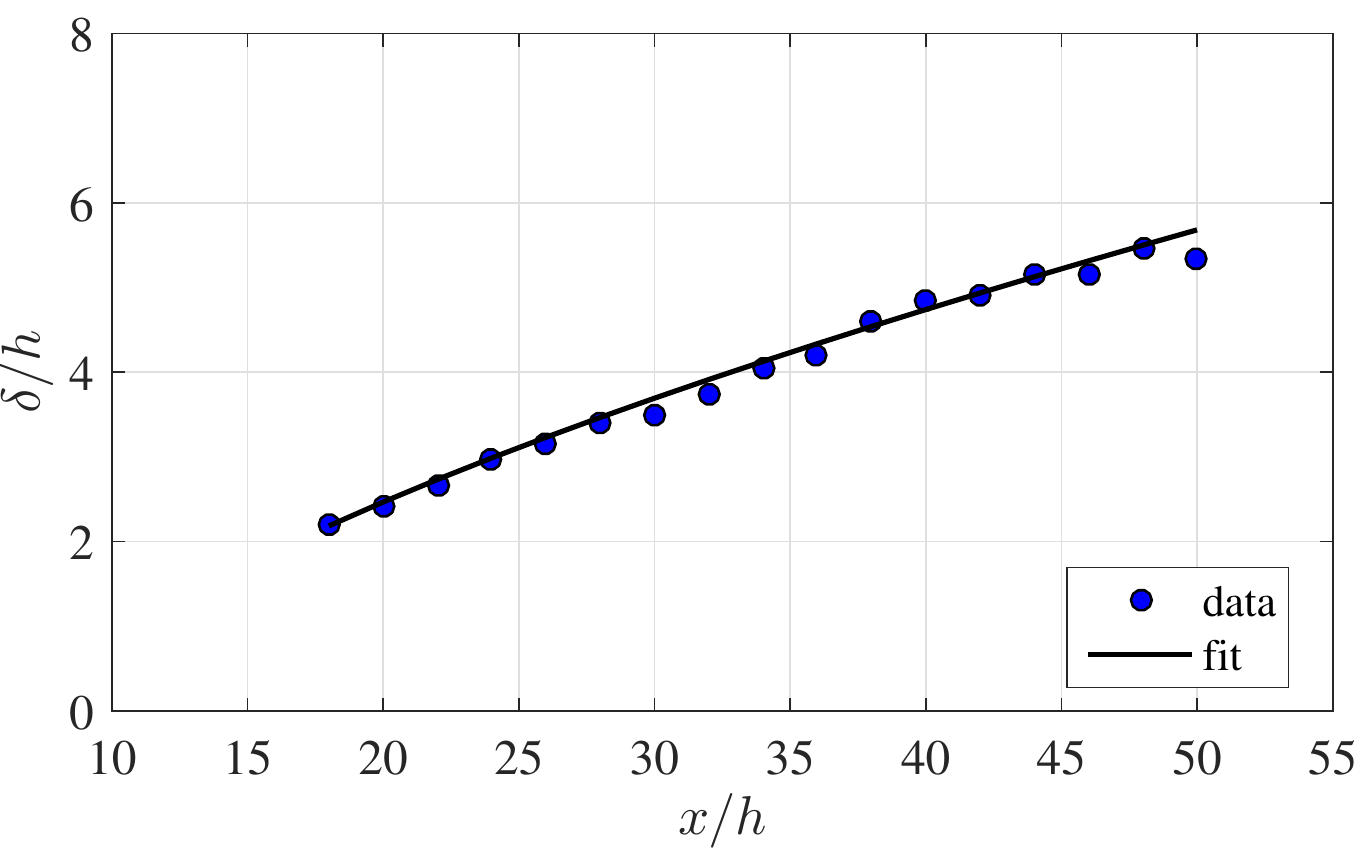}}\\
\caption{Fits of the centreline mean flow velocity $u_0$ and the jet
  width $\delta$ according to the non-equilibrium scaling laws given
  in eqns.  (\ref{eq:scalingNEQ_u_0}-\ref{eq:scalingNEQ_delta}), with fit
  coefficients given in Table \ref{tab:fit}. Data acquired for inlet
  Reynolds number $Re_G=20000$. }
\label{fig:udeltaFit}
\end{figure}

We therefore turn to the approach of \cite{Nedic2013} which is to
estimate the two derivatives $\frac{d}{dx}[(u_0(x)/U_{J})^{-1/a}]$ and
$\frac{d}{dx}[(\delta(x)/h)^{1/b}]$ for a range of values of $a$ and
$b$ and then evaluate the best linear fits of these two derivatives
expressed as $C_1 x/h+A^{-1/a}$ and $C_2 x/h+B^{1/b}$. The constants
$A$ and $B$ are the proportionality constants in $u_0(x)/U_J = A
((x-x_{0,A})/h)^{-a}$ and $\delta(x)/h = B ((x-x_{0,B})/h)^{b}$.
Again, we apply this analysis to the range $18\le x/h \le 50$. In
figure \ref{fig:C1C2} we plot the resulting values of $C_1$ and $C_2$
as functions of $a$ and $b$ respectively. The point of this method is
to chose the exponents $a$ and $b$ for which $C_1$ and $C_2$
vanish. However, it turns out that both $C_1$ and $C_2$ are very close
to 0 for any value of $a$ in the range $1/3\le a \le 1/2$ and any
value of $b$ in the range $2/3\le b \le 1$. The conclusion is
therefore the same: any exponents $a$ and $b=2a$ in the range $1/3\le
a\le 1/2$ can fit our $u_{0}(x)$ and $\delta (x)$ data equally well in
the range $18\le x/h \le 50$. We have checked that all these good fits
can be achieved with values of $x_{0,A}$ and $x_{0,B}$ that are close
to each other.

We stress the point that the exponents $a=1/3$ and $b=2a=2/3$, which
follow from our definite finding that $m=1$, are consistent with our
$u_{0}(x)$ and $\delta (x)$ data. Any other exponents $a$ and $b=2a$
are not in agreement with $m=1$. We therefore set $a=1/3$ and
$b=2a=2/3$ and determine the virtual origins $x_{0,A}$ and $x_{0,B}$
and proportionality constants $A$ and $B$ which provide the best fits
of our $u_0 (x)$ and $\delta (x)$ data.  We plot our data and our
non-equilibrium ($m=1$) fits in Figure \ref{fig:C1C2} and list the
values of $x_{0,A}$, $x_{0,B}$, $A$ and $B$ in Table \ref{tab:fit}. As
expected from figure \ref{fig:ourV00X}, $x_{0,A}$ and $x_{0,B}$ do
turn out to be very close to each other, as required by the theory. We
repeat that one could fit this data equally well with the classical
exponents $a=1/2$ and $b=2a=1$ implied by eq. (\ref{eq:scalingEQ3}) if
$m=0$, including with virtual origins $x_{0,A}$ and $x_{0,B}$ that are
very close to each other. The difference is that $m=0$ is not
supported by our data and by the data of \cite{Antonia1980} whereas
$m=1$ is.

\begin{table}
  \begin{center}
\def~{\hphantom{0}}
  \begin{tabular}{cccccccc}
      $A$  	 & $a$    &   $x_{0,A}$     &  $R^2_u$  &    $B$      & $b$    &  $x_{0,B}$ &$R^2_{\delta}$\\[3pt]
      1.4     &  1/3 	&    7.7$h$	    &   0.991      &    0.48	   &	2/3    &    8$h$ &    0.991
  \end{tabular}
  \caption{Fit coefficients for $u_0(x)/U_{J}= A ((x-x_{0,A})/h)^{-a}$
    and $\delta(x)/h= B ((x-x_{0,B})/h)^{b}$ where $a$ and $b$ have
    been set to $a=1/3$ and $b=2a=2/3$. The values of the
    determination coefficients of the two fits are also reported. From
    our data which were acquired at inlet Reynolds number
    $Re_G=20000$.}
  \label{tab:fit}
  \end{center}
\end{table}

\begin{figure}
\centering
\subfloat[][]
{\includegraphics[scale=0.45]{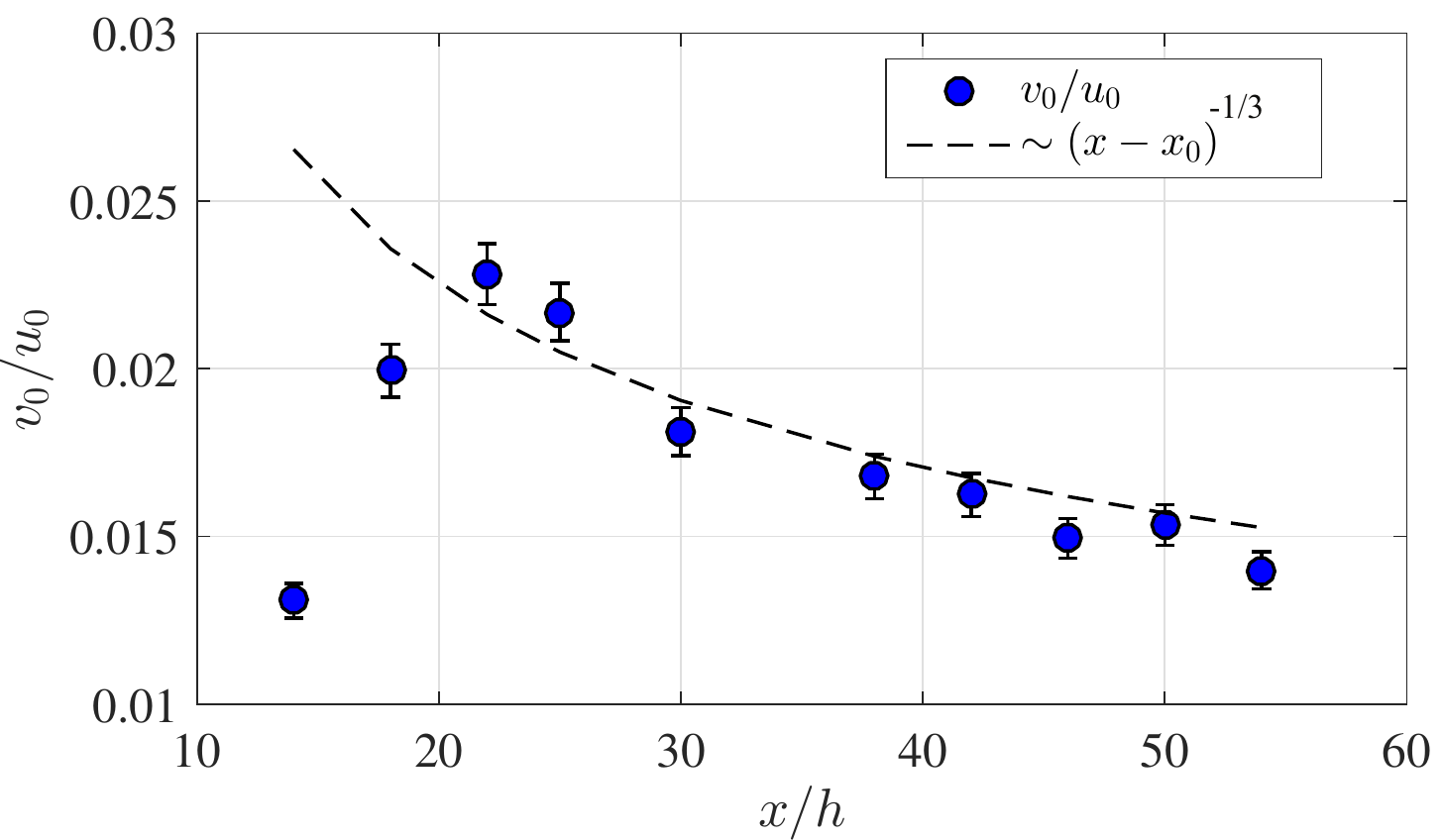}}
\subfloat[][]
{\includegraphics[scale=0.45]{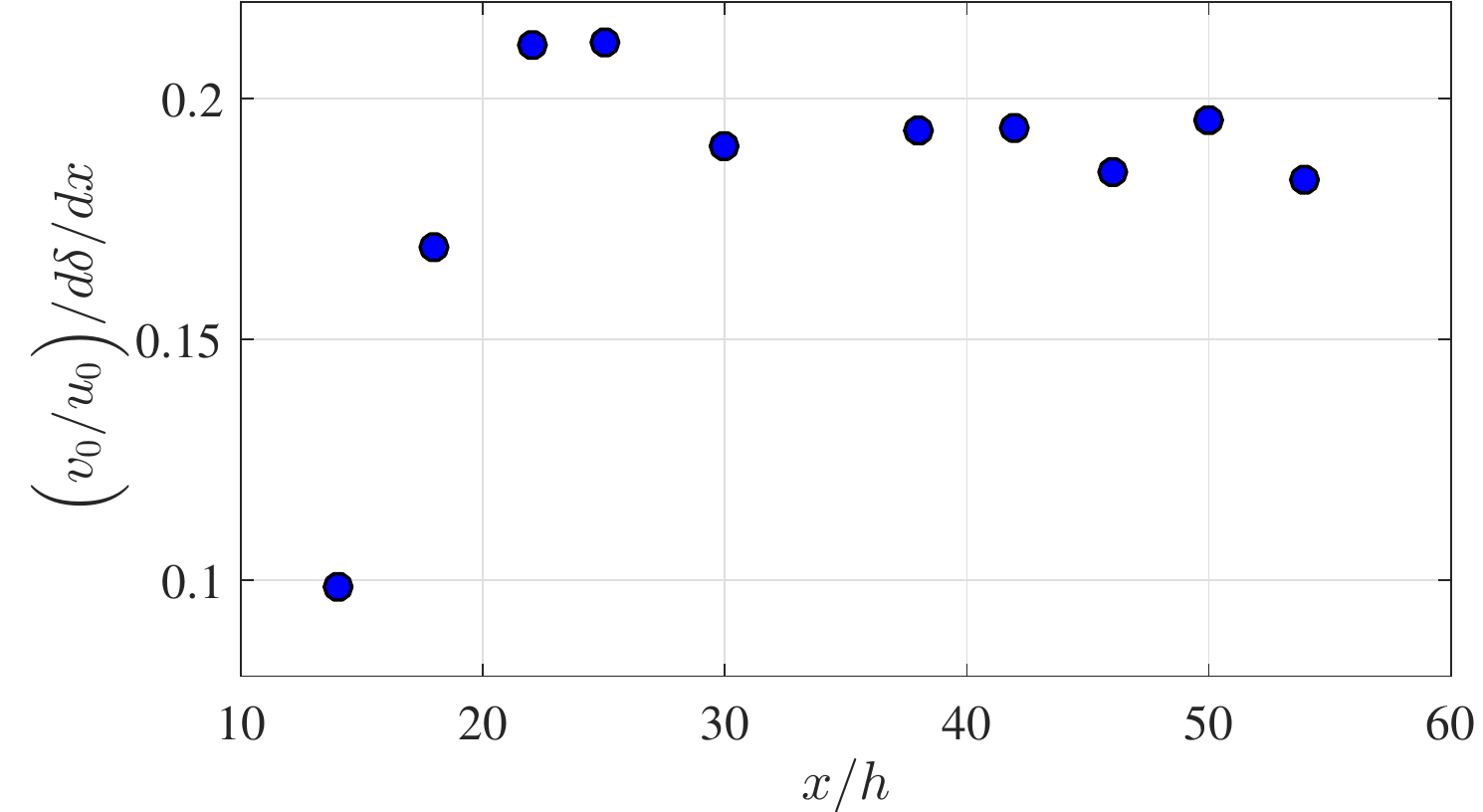}}\\
  \caption{(a) Streamwise profile of the ratio $v_0/u_0\sim
    d{\delta}/dx$. The dashed black line is representative of the
    $\sim (x-x_0)^{-1/3}$ power law dependence of $d\delta/dx$ on
    $x$. The error bars are obtained by repeating the measurements at a given $x/h$ between three and five times; (b) $\frac{v_0/u_0}{d{\delta}/dx}$ (where ${d{\delta}/dx}$ is obtained from the fit of $\delta (x)$ with the parameters listed
    in Table \ref{tab:fit}) plotted as a function of the streamwise
    distance $x/h$. Data obtained at $Re_G=20000$. }
\label{fig:vouo}
\end{figure}

We now turn our attention to $v_{0}/u_{0} \sim d\delta/dx$, see
eq. (\ref{eq:v0u0del}). Given that the self-similar behaviours of the
five profiles studied here are supported by our data in the range $18
\le x/h \le 54$ and that the dissipation scaling given by $m=1$ is in
good agreement with our data for values of $x/h$ larger than about 20,
we expect to find $v_{0}/u_{0} \sim d\delta/dx \sim (x/h -
x_{0}/h)^{2a-1}$ at values of $x/h$ larger than about 20 with
$a=1/3$. In Figure \ref{fig:vouo}(a) we plot $v_{0}/u_{0}$ as a
function of $x/h$ and compare it to $(x/h - x_{0}/h)^{-1/3}$ where
$x_{0}$ is taken from Table \ref{tab:fit} (not fitted anew) as $x_{0}
= (x_{0,A}+x_{0,B})/2$ which is very close to both $x_{0,A}$ and
$x_{0,B}$.  After an initial growth associated with the progressive
build-up of entrainment following the potential core, one can see in
figure \ref{fig:vouo}(a) a clear streamwise decrease of $v_0/u_0$ for
$x/h > 20$ which is evidence that $m \not = 0$. As shown in the plots
(a) and (b) of figure \ref{fig:vouo}, this decreasing trend is in good
agreement with $d\delta/dx \sim (x/h - x_{0}/h)^{-1/3}$ as predicted
by the theory for $m=1$. The proportionality coefficient in
$d\delta/dx \sim (x/h - x_{0}/h)^{-1/3}$ is obtained from figure
\ref{fig:vouo}(b).

We close this section with a comment concerning turbulent viscosity
modelling which assumes $R_{xy}$ to equal $-\nu_{T} dU/dy$. The usual
algebraic model for the turbulent viscosity $\nu_{T}$ is $\nu_{T} \sim
u_{0} \delta$ (see \citealp{pope}, \citealp{davidson2004}) and it
returns the scalings reported in equations
(\ref{eq:scalingEQ1})-(\ref{eq:scalingEQ2})-(\ref{eq:scalingEQ3}) with $m=0$
if the mean velocity profiles are assumed to be self-similar. However,
in the present case where the data support the non-equilibrium
dissipation scaling $D_{0} \sim (Re_{G}/Re_{\delta})
K_{0}^{3/2}/\delta$ (i.e. $m=1$) rather than the equilibrium one
($m=0$) and $v_{0}/u_{0} \sim d\delta/dx \sim (x/h - x_{0}/h)^{-1/3}$
rather than $v_{0}/u_{0} = Const$, the turbulent viscosity needs to be
$\nu_{T} \sim U_{J} h$ to return the scalings of $u_0$ and $\delta$
which are consistent with these non-equilibrium dissipation scalings
and self-similarity. The same holds for axisymmetric turbulent wakes
where $\nu_{T} \sim u_{0} \delta$ needs to be replaced by $\nu_{T}
\sim U_{\infty} L_{B}$ in the presence of the non-equilibrium
dissipation scaling (see \citealp{dairay2015} and
\citealp{obligado2016}), $U_{\infty}$ being the incoming freestream
velocity and $L_B$ the size of the wake-generating body.

\section{Conclusions} \label{sec:concl}
The non-equilibrium dissipation law which has been found in
grid-generated turbulence \citep{vassilicos2015}, axisymmetric
turbulent wakes \citep{obligado2016}, forced and decaying periodic
turbulence \citep{goto&vassilicos2015} and turbulent boundary layers
\citep{Nedic2017} is also present in  turbulent planar jets. It does not
matter if the local Reynolds number decays with downstream distance
like it does in grid-generated turbulence and axisymmetric wakes or
grows with downstream distance like it does in turbulent boundary
layers and planar jets. In the former case the dissipation coefficient
$C_{\varepsilon}$ increases with decreasing local Reynolds number
whilst in the latter case it decreases with increasing local Reynolds
number. In all cases these increases and decreases happen according to
the same inverse power-law relation. This was also observed in the DNS
of forced periodic turbulence by \cite{goto&vassilicos2015,
  goto&vassilicos2016} where the local Reynolds number undergoes long
periods of growth followed by long periods of decline.

Following \cite{Townsend}, \cite{george1989} and \cite{dairay2015},
the non-equilibrium dissipation law combined with various self-similar
profiles imply new centreline mean flow and jet width scalings, see
eq. (\ref{eq:scalingNEQ_u_0}) and eq. (\ref{eq:scalingNEQ_delta}). Our
experiments have provided evidence that the profiles of mean flow
velocities $U$ and $V$, Reynolds shear stress $R_{xy}$, turbulent
kinetic energy $K$ and dissipation $\varepsilon$ are indeed
self-similar for streamwise distances downstream of $x=18h$ even
though this is not far enough downstream for the memory of inlet
conditions to fully fade away. The inlet conditions $U_J$ and $h$ are
explicitly present in the non-equilibrium dissipation law which is
found to hold in a region downstream of $x \approx 20h$ that extends
at least as far as $x=140h$ as shown by the data of
\cite{Antonia1980}. An important implication which our experiment
confirms is that the entrainment coefficient $\alpha$ is not constant
but decreases as $(x/h -x_{0}/h)^{-1/3}$ over the streamwise extent
where the non-equilibrium dissipation law holds.

\section*{Acknowledgements}
The authors were supported by ERC Advanced Grant 320560 awarded to
JCV. 
\bibliographystyle{jfm}
 Note the spaces between the initials
\bibliography{jfm-instructions}

\end{document}